\documentclass[manuscript,screen]{acmart}

\AtBeginDocument{%
  \providecommand\BibTeX{{%
    \normalfont B\kern-0.5em{\scshape i\kern-0.25em b}\kern-0.8em\TeX}}}

\setcopyright{acmcopyright}
\copyrightyear{2024}
\acmYear{2024}
\acmDOI{XXXXXXX.XXXXXXX}

\acmConference[ACM TOSEM]{ACM Transactions on Software Engineering and Methodology}{2024}{New York, NY, USA}
\acmISBN{978-1-4503-XXXX-X/18/06}

\usepackage{booktabs}
\usepackage{threeparttable}
\usepackage{makecell}
\usepackage{enumitem}
\usepackage{xspace}
\usepackage{verbatim}
\usepackage{tikz}
\usepackage{colortbl}
\usepackage{hyperref}
\usepackage{collectbox}
\usepackage{multirow}
\usepackage{diagbox}
\usepackage{amsthm}
\usepackage{algpseudocode}
\usepackage[ruled,vlined,linesnumbered]{algorithm2e}

\theoremstyle{definition}
\newtheorem{definition}{Definition}[section]
\usepackage[caption=false]{subfig}

\usepackage{amsmath}

\usepackage{pifont}

\newcommand{\xmark}{\ding{55}}
\newcommand{\tick}{\textcolor{black}{\checkmark}}
\newcommand{\notick}{\textcolor{red}{\xmark}}
\newcommand{\header}[1]{\vspace{-1.5mm}\par\medskip\noindent\textbf{#1:}}
\usepackage{circledsteps}
\newcommand{\totalBugsInDefects}[0]{835\xspace}

\usepackage[nomessages]{fp}
\newcommand{\FPuse}[1]{\FPeval{\result}{#1}{\result}}
\newcommand{\tableRow}[3]{
	#1 & #2 & #3 &
	\ifthenelse{\equal{#2}{0}}{$100$}{$\FPuse{round(#1 / (#1 + #2) * 100, 2)}$}
	&
	\ifthenelse{\equal{#3}{0}}{$100$}{$\FPuse{round(#1 / (#1 + #3) * 100, 2)}$}
	&
	$\FPuse{round(2 * #1 / (2* #1 + #2 + #3) * 100, 2)}$
}

\newcommand{\tableRowBold}[3]{
	#1 & #2 & #3 &
	\ifthenelse{\equal{#2}{0}}{\boldmath$100$}{\boldmath$\FPuse{round(#1 / (#1 + #2) * 100, 2)}$}
	&
	\ifthenelse{\equal{#3}{0}}{\boldmath$100$}{\boldmath$\FPuse{round(#1 / (#1 + #3) * 100, 2)}$}
	&
	\boldmath$\FPuse{round(2 * #1 / (2* #1 + #2 + #3) * 100, 2)}$
}

\usepackage{tcolorbox}
\definecolor{findingsbox-bg-color}{gray}{0.90}
\newtcbox{\findingsbox}{colback=findingsbox-bg-color, boxrule=0.2pt, arc=2pt, boxsep=0pt, left=5pt, right=5pt, top=5pt, bottom=3pt}
\newcommand{\findings}[2] {
	\noindent
	\findingsbox{
		\begin{minipage}{.98\linewidth}
			\textbf{#1}: #2
		\end{minipage}
	}
}

\definecolor{darkgray}{RGB}{90,90,90}
\definecolor{lightgray}{RGB}{210,210,210}

\newcommand\sbar[3]{
	\ifthenelse{\equal{#2}{0} \OR \equal{#2}{1}}{$100$}{$\FPuse{round(#1 / (#1 + #2) * 100, 1)}$} &
	{\color{darkgray}\rule{\dimexpr 0.3cm * #1 / (#1 + #2)}{5pt}}{\color{lightgray}\rule{\dimexpr 0.3cm - (0.3cm * #1 / (#1 + #2))}{5pt}} &
	\ifthenelse{\equal{#3}{0} \OR \equal{#3}{1}}{$100$}{$\FPuse{round(#1 / (#1 + #3) * 100, 1)}$} &
	{\color{darkgray}\rule{\dimexpr 0.3cm * #1 / (#1 + #3)}{5pt}}{\color{lightgray}\rule{\dimexpr 0.3cm - (0.3cm * #1 / (#1 + #3))}{5pt}}
}

\newcommand\sbarOnlyZero[3]{
	\ifthenelse{\equal{#2}{0}}{$100$}{$\FPuse{round(#1 / (#1 + #2) * 100, 1)}$} &
	{\color{darkgray}\rule{\dimexpr 0.3cm * #1 / (#1 + #2)}{5pt}}{\color{lightgray}\rule{\dimexpr 0.3cm - (0.3cm * #1 / (#1 + #2))}{5pt}} &
	\ifthenelse{\equal{#3}{0}}{$100$}{$\FPuse{round(#1 / (#1 + #3) * 100, 1)}$} &
	{\color{darkgray}\rule{\dimexpr 0.3cm * #1 / (#1 + #3)}{5pt}}{\color{lightgray}\rule{\dimexpr 0.3cm - (0.3cm * #1 / (#1 + #3))}{5pt}}
}

\newcommand\pbar[2]{
	$\FPuse{round(#1 /#2 * 100, 1)}$ &
	{\color{darkgray}\rule{\dimexpr 0.5cm * #1 / #2}{5pt}}{\color{lightgray}\rule{\dimexpr 0.5cm - (0.5cm * #1 / #2)}{5pt}}
}
\expandafter\newcommand\csname r@tocindent4\endcsname{4in}
\begin{document}

\title{A Novel Refactoring and Semantic Aware Abstract Syntax Tree Differencing Tool and a Benchmark for Evaluating the Accuracy of Diff Tools}

\author{Pouria Alikhanifard}
\email{po_alikh@encs.concordia.ca}
\author{Nikolaos Tsantalis}
\email{nikolaos.tsantalis@concordia.ca}
\affiliation{%
  \institution{Concordia University}
  \streetaddress{1455 De Maisonneuve Blvd. W.}
  \city{Montreal}
  \state{Quebec}
  \country{Canada}
  \postcode{H3G 1M8}
}

\renewcommand{\shortauthors}{Alikhanifard and Tsantalis}

\begin{abstract} 
	Software undergoes constant changes to support new requirements, address bugs, enhance performance, and ensure maintainability.
	Thus, developers spend a great portion of their workday trying to understand and review the code changes of their teammates.
	Abstract Syntax Tree (AST) diff tools were developed to overcome the limitations of line-based diff tools, which are used by the majority of developers. Despite the notable improvements brought by AST diff tools in understanding complex changes, they still suffer from serious limitations, such as (1) lacking multi-mapping support, (2) matching semantically incompatible AST nodes, (3) ignoring language clues to guide the matching process, (4) lacking refactoring awareness, and (5) lacking commit-level diff support.
	We propose a novel AST diff tool based on RefactoringMiner that resolves all aforementioned limitations.
	First, we improved RefactoringMiner to increase its statement mapping accuracy, and then we developed an algorithm that generates AST diff for a given commit or pull request based on the refactoring instances and pairs of matched program element declarations provided by RefactoringMiner.
	To evaluate the accuracy of our tool and compare it with the state-of-the-art tools, we created the first benchmark of AST node mappings, including 800 bug-fixing commits and 188 refactoring commits.
	Our evaluation showed that our tool achieved a considerably higher precision and recall, especially for refactoring commits, with an execution time that is comparable with that of the faster tools.
\end{abstract}



\maketitle

\section{Introduction}
Source code differencing is an important functionality for both software engineering practitioners and researchers, as software is changing constantly and there is a need to understand how and why it changed.
Software developers spend a significant portion of their workday trying to understand and review the code changes of their teammates.
Based on data collected from over 250 thousand developers \cite{code-time-report}, developers spend 41 minutes per day on code reviewing and understanding code changes, which is almost as much time as they spend on coding (52 minutes per day).
MacLeod et al. \cite{MacLeod:CodeReviewing:2018} surveyed 911 developers at Microsoft and found that the major challenge faced by code reviewers is finding time to perform all the code reviews requested of them, as well as understanding the code’s purpose, the motivations for the change, and how the change was implemented.
Ebert et al. \cite{10.1007/s10664-020-09909-5} found that long or complex code change is the most frequently experienced reason causing confusion in code reviews according to developers.
Regarding research, source code differencing is utilized to extract change patterns that can be used for automated program repair \cite{7476644, 10.1007/s10664-019-09772-z}, automated library migration \cite{10.1145/3477314.3507153, 10.1145/3594264.3594265, 10.1145/3468264.3468571} and automated type/API migration \cite{8812061, 10.1145/3510003.3510115, 10.1145/3368089.3409725}.

The majority of the diff tools used for code reviews including GitHub Pull Request \cite{github-pull-request}, Gerrit \cite{gerrit}, as well as in-house tools used by companies (e.g., the DiffTool used at Meta \cite{10.1145/3540250.3549104}) are applying the Myers algorithm \cite{Myers} at the text line granularity \cite{10.1007/s10664-019-09772-z}.
However, according to Falleri et al. \cite{gumtree} line-based diff has major limitations, as it only computes additions and deletions, it does not consider other kinds of edit actions, such as update and move, and it works at a coarse-grained granularity (i.e., text line) that is not aligned with the source code structure (i.e., the abstract syntax tree).
These limitations make it more difficult for code reviewers to understand the changes intended by the commit authors when inspecting a pull request, especially when the changes are complex and involve refactoring operations, which tend to move code to new locations relatively far from the original code locations (e.g., \textsc{Extract Method} \cite{Fowler:1999}).

To overcome these limitations Abstract Syntax Tree (AST) diff tools have been developed.
AST differencing aims at computing a sequence of edit actions (called an \textit{edit script}) that transform an AST into another \cite{gumtree}. 
The typical edit actions supported by most tools are AST node \textit{updates}, \textit{additions}, \textit{deletions}, and \textit{moves}.
The most prominent AST diff tools are ChangeDistiller \cite{ChangeDistiller}, GumTree \cite{gumtree}, IJM (Iterative Java Matcher) \cite{IJM}, MTDiff \cite{MTDiff}, and CLDiff \cite{ClDiff}.
Despite the notable improvements brought by AST diff in understanding complex code changes, as shown in experiments with professional developers \cite{spike}, AST diff tools have some constraints, which make them generate edit scripts that are not ideal or optimal from the change comprehension point of view,
because they do not reflect the actual changes intended by the commit author.
Fan et al. \cite{10.1109/ICSE43902.2021.00108} proposed a differential testing approach that automatically compares the similarity of mapped program elements by different AST mapping algorithms, and found that GumTree, MTDiff and IJM generate inaccurate mappings for 20\%–29\%, 25\%–36\% and 21\%–30\% of the file revisions, respectively.
Fan et al. concluded that based on their experimental results, the state-of-the-art AST mapping algorithms still need improvements.

\subsection{Motivation}
\label{sec:motivation}
The current AST diff tools are based on constraints that limit their ability to generate accurate edit scripts, i.e., edit actions accurately reflecting the changes intended by the commit author.

\begin{enumerate}[wide,label=\bfseries Constraint \arabic*:, leftmargin=*,labelwidth=!, labelindent=0pt]
	\item ``A given node can only belong to one mapping'' \cite{gumtree}.
	This constraint makes impossible for an AST diff algorithm to generate multi-mappings (i.e., a node from the left AST having more than one corresponding node in the right AST and vice versa). 
	However, in reality developers tend to merge duplicated code fragments into a single one to avoid redundancy (i.e., many-to-one multi-mapping) or duplicate a code fragment in multiple places to reuse it (i.e., one-to-many multi-mapping).
	Currently, there is not even a single AST diff tool in the literature supporting multi-mappings.
	
	\item ``Mappings involve two nodes with identical labels'' \cite{gumtree}.
	Although this constraint seems sensible as it requires two AST nodes to have the same AST type to be matched, it ignores the semantic role that an AST node plays in the program.
	For example, type references, variable identifiers and method call names are three semantically different groups of AST nodes, which are all represented with the same AST type (i.e., \texttt{SimpleName} in the Eclipse JDT parser). However,  it makes no semantic sense to match nodes between these three different groups.
	
	\item ``The AST differencing problem inputs two ASTs and aims at identifying a short edit script $\sigma$ of edit actions (including move) that transforms a first AST (named $t_1$) into a second one (name $t_2$)'' \cite{gumtree}.
	This constraint restricts the AST diff problem to a pair of files, while in reality a commit (or pull request) may include code being moved between different files.
	
	\item ``The algorithm is independent of any language specificity'' \cite{gumtree}.
	Although being language-independent may be considered a desirable feature as it makes the algorithm easily extensible to other programming languages, at the same time it hinders the ability of the algorithm to utilize language-specific information (i.e., language clues), such as method signatures, method calls, field accesses and refactorings, to guide more effectively the AST matching process.
\end{enumerate}

Some of these constraints were never addressed in the literature (i.e., \textbf{Constraint 1}: multi-mappings), while others were only partially addressed with incomplete or non-evaluated solutions.
For instance, IJM~\cite{IJM} is a language-aware solution that attempted to address \textbf{Constraint 2} and \textbf{Constraint 4}.
With respect to Constraint 2, IJM modifies the AST structure by merging the value of name nodes (i.e., \texttt{SimpleName} AST nodes) with their respective parent nodes and deleting the name nodes with redundant information. 
Although this approach reduced the number of semantically incompatible mapping (as shown in our experiments in Section~\ref{sec:RQ2}), the solution is incomplete, as there are other AST node types playing multiple semantic roles that were not covered.
Moreover, modifying the structure of AST or using more fine-grained labels for leaf nodes caused some unwanted side-effects, discussed in Sections~\ref{sec:multi-mappings} and~\ref{sec:RQ6}, respectively.
With respect to Constraint 4, IJM applies \textit{partial matching} by restricting the scope of the matching process to selected parts of the source code (e.g., when matching method declarations with identical signatures). 
Although this approach improved the matching accuracy for method declarations (as shown in our experiments in Section~\ref{sec:program-element-mappings}), the solution is incomplete, 
as more language clues, such as method calls and field accesses, can be utilized to further guide the matching process to more accurate mappings.
With respect to \textbf{Constraint 3}, \textit{Staged Tree Matching} (STM) \cite{StagedTreeMatching} is a GumTree extension that first applies the standard GumTree algorithm on the pairs of files modified in a project. Then, it constructs a project-level AST for each version by connecting the AST roots corresponding to each file to a pseudo-project-root node, and runs another matching phase on the project-level ASTs by considering only the remaining unmatched AST subtrees.
Since the authors of \textit{Staged Tree Matching}~\cite{StagedTreeMatching} did not provide any experimental results regarding its accuracy, we implemented this algorithm and conducted our own experiments in Section~\ref{sec:inter-file-mappings}, which show that STM cannot achieve the same level of accuracy as a \textit{move-refactoring-aware} approach does.

Through extensive experimentation and testing of various AST diff tools with real case studies (i.e., commits) from open-source projects, we compiled a catalogue with their limitations (Section \ref{sec:motivating-examples}) to demonstrate the negative impact of these constraints on the accuracy and comprehensibility of the generated edit scripts.
The \textit{root cause} behind all these limitations is \textit{language independence}.
To overcome each one of these limitations requires some level of language awareness:
\begin{enumerate}[leftmargin=*]
	\item \textbf{Lack of support for multi-mappings (Section \ref{sec:multimapping}):}
	The mechanisms of how code can be merged (i.e., multi-mappings) are language specific and even framework specific.
	For example, in Java,
	duplicated code can be merged 
	by extracting it into a method, 
	by moving it into test fixtures (JUnit5),
	by replacing duplicated tests with a parameterized test (JUnit5), 
	by merging two methods with similar code into a new one, 
	by pulling up duplicated methods within subclasses to their common superclass,
	by moving repeated code within \texttt{if-else-if} branches or \texttt{switch} cases outside of the conditionals (i.e., \textsc{Consolidate Duplicate Conditional Fragments} refactoring).
	Different languages and frameworks provide different mechanisms to merge code and eliminate duplication.
	For example, \textit{mixins} and \textit{object composition} can be used in other programming languages to enable code reuse.
	\item \textbf{Incorrect matching of program declarations (Section~\ref{sec:language-clues}):} 
	Determining the corresponding declarations (i.e., methods, attributes) between two versions of a program requires language specific information.
	For example, in Java, the signature of a method can be uniquely identified based on its name and list of parameter types. However, in weakly typed programming languages, the signature of a function can be determined by its name and the number of parameters.
	Moreover, language specific clues in the code (i.e., method calls, field references) can help guide and speed-up the matching process by allowing to reason about the potential location of moved code.
	For example, 
	if part of a method is found as deleted and in its place there is a call to a newly added method (i.e., implying the potential application of an \textsc{Extract Method} refactoring), the matching process can be guided to match the deleted code with the body of the newly added method without searching for matches in the rest of files.
	However, inferring the function declaration corresponding to a function call also requires language specific information, as some programming languages allow optional arguments, while others do not.
	\item \textbf{Lack of refactoring awareness (Section~\ref{sec:refactoring-awareness}):}
	Refactoring operations are language specific, as the types of refactorings that can be applied to a program depend on the features provided by the programming language.
	For example, certain refactoring types are applicable only in programming languages that support inheritance.
	Moreover, statically-typed languages have a richer and more reliable refactoring support in IDEs compared to dynamically-typed languages~\cite{Unterholzner:2014}, and thus refactoring operations are more prevalent in the former languages.
	\item \textbf{Matching semantically incompatible AST nodes (Section~\ref{sec:semanticallyIncompatibleNodes}):}
	Determining the semantic role of an AST node in a program requires knowing the language specifications, and how the AST parser that is used to parse the program represents each program element. As shown in Section~\ref{sec:RQ2}, some AST types can play multiple different semantic roles in Java programs.
	\item \textbf{Commit-level change ignorance (Section~\ref{sec:commitLevelDiff}):}
	The mechanisms of how code can be moved between files are language specific. 
	Some languages support inheritance and code can be moved by \textsc{Pull Up} or \textsc{Push Down} operations in the inheritance hierarchy, while other languages do not support inheritance.
	Static methods (i.e., methods that do not access the state of an object) have a different move mechanism compared to instance methods (i.e., methods that access the state of an object).
\end{enumerate}

Our work is a language-aware solution attempting to overcome all aforementioned constraints and limitations with the goal of generating highly accurate edit scripts at commit (or pull request) level, and thus reduce the time and effort it takes to perform code reviews and change comprehension tasks, especially in complex change scenarios involving refactorings.

Regarding the evaluation methods, the majority of AST diff tools have been evaluated based on the number of generated mappings and the size of the resulting edit script, without validating whether the mappings and edit actions are correct and reflect the actual changes intended by the commit author.
The assumption behind this evaluation approach, as explained by Falleri et al. \cite{gumtree} is that ``the length of the edit script is a proxy to the cognitive load for a developer to understand the essence of a commit''.
Thus, shorter edit scripts should be easier to understand.
However, in Section \ref{sec:motivating-examples} we demonstrate through real examples from open-source projects that shorter edit scripts may confuse and mislead the code reviewers with edit actions that are inconsistent with the actual changes applied by the commit author.
We believe that the most reliable way to evaluate AST diff tools is based on an established ground truth of AST node mappings, which would make it possible to assess the correctness of the mappings generated by the tools, and thus compute their respective precision and recall.
To this end, we created the first benchmark of AST node mappings with commits from open-source projects.

\subsection{Approach overview}
To overcome all these limitations we propose and implement a refactoring, semantic, and language-aware solution by extending RefactoringMiner \cite{RefactoringMiner1, RefactoringMiner2} to generate AST diff at the commit level.
RefactoringMiner \cite{RefactoringMiner} is a mature tool maintained over a period of 8 years by members of our Refactoring Research Group at Concordia University.
It currently supports the detection of 60 different refactoring types and 40 kinds of API changes and has been established as the tool with the highest accuracy and fastest execution time among competitive tools \cite{SurveyRefactoringDetection, RefDiff2}.
RefactoringMiner has been used in numerous empirical studies to extract refactoring datasets from the commit history of open-source projects.
Moreover, the tool was used in solving various software evolution analysis problems, such as
automatic refactoring-aware commit merging \cite{EllisTSE2023},
automatic decomposition of commits into distinct activities \cite{SmartCommit},
automatic source code comment updating \cite{10.1145/3324884.3416581},
refactoring-aware implementation of the SZZ algorithm for identifying bug-inducing changes \cite{10.1109/ICSE43902.2021.00049},
automatic extraction of complete and concise bug-fixing patches cleaned-up from overlapping refactoring edits \cite{BugBuilder},
and refactoring-aware commit change history extraction for methods and variables \cite{CodeTracker}, just to name a few.
Finally, RefactoringMiner is used by tools visualizing refactoring information in IDEs, such as the RefactorInsight \cite{RefactorInsight} (developed and maintained by JetBrains Research) and Spike \cite{spike} plug-ins for IntelliJ IDEA.

Under the hood, RefactoringMiner applies a statement mapping process \cite{RefactoringMiner2}. The mapped statements and the AST node replacements found within the mapped statements are then used to infer refactoring operations.
Despite its high accuracy in refactoring detection, RefactoringMiner needed some significant and novel extensions (Section~\ref{sec:changes-to-improve-statement-mapping-accuracy}) to be able to generate surgically accurate mappings at statement-level.
Furthermore, we developed an algorithm that generates AST diff (Section~\ref{sec:ast-diff-generation}) using the output of RefactoringMiner (i.e., the detected refactoring instances and the pairs of matched program element declarations).
As explained in Section~\ref{sec:how-we-address-limitations}, the combination of RefactoringMiner's output with our AST diff generation algorithm overcomes all constraints discussed in Section~\ref{sec:motivation}

\subsection{Contributions}
\begin{enumerate}[leftmargin=*]
	\item We extend RefactoringMiner to generate AST diff at the commit level.
	\item We create the first benchmark of AST node mappings from two popular datasets, namely Defects4J \cite{Defects4J}, which includes \totalBugsInDefects bug fixes from 17 open-source projects, and Refactoring Oracle \cite{RefactoringMiner2}, which includes over 11K refactorings and code changes found in 546 commits from 187 open-source projects.
	Both datasets have become the de facto standard for evaluating automated program repair and refactoring mining tools, respectively.
	Our benchmark includes bug-fixing and refactoring commits, to highlight the fact that most AST diff tools perform reasonably well in bug-fixing commits as the changes are small, while their accuracy deteriorates in commits involving larger and more complex changes, such as refactorings.
	\item We propose a novel framework for the evaluation of AST diff tools covering multiple diff quality dimensions, including multi-mappings, semantically compatible mappings, program element mappings, refactoring-related mappings, and inter-file mappings.
	Moreover, we utilize our benchmark to compute precision, recall and F-score, as objective metrics to measure the quality of the AST diff generated by each tool.
	\item We develop an infrastructure to execute multiple AST diff tools, including RefactoringMiner, GumTree 3.0 (greedy and simple), IJM, MTDiff, GumTree 2.1.0, and compute the aforementioned metrics.
	\item We evaluate the precision, recall and execution time of the aforementioned tools on our benchmark.
\end{enumerate}
\section{Motivating examples demonstrating the limitations of current AST diff tools}
\label{sec:motivating-examples}
In this section, we demonstrate the limitations of the current state-of-the-art AST diff tools with illustrative examples found in open-source project commits.
\subsection{Lack of support for multi-mappings}
\label{sec:multimapping}
Multi-mapping is the case where a node from the left AST has more than one corresponding node in the right AST, and vice versa.
None of the current state-of-the-art AST diff tools supports multi-mappings.
However, multi-mappings are quite common in various change scenarios involving duplicated code.
For example, developers tend to extract duplicated code from multiple methods into a newly added method,  eliminate duplicate code within \texttt{if-else-if} or \texttt{switch-case} control structures by merging the duplicated statements and moving them out of the conditionals, eliminate duplicate code within \texttt{catch} blocks handling different exception types by merging the exception handling code into a single \texttt{catch} block using the Union type (i.e., \texttt{catch(Exception1 | Exception2)}).
Of course, the reverse scenarios can also occur, e.g., by inlining a method being called by multiple different methods.
Another example of one-to-many multi-mapping is the common case where a variable declaration with an initialization expression is split to a variable declaration initialized with a default value, and the initialization is moved to a separate variable assignment statement.
Higo et al. \cite{10.5555/3155562.3155630} partially addressed this issue by introducing copy-and-paste as a new kind of edit action.
However, their solution is uni-directional as it supports only scenarios in which developers duplicate existing code, but not scenarios in which developers eliminate duplicate code.
Aalizadeh \cite{Sadegh} studied 346K instances of \textsc{Extract Method} refactorings mined from 132,897 commits of 325 open-source repositories, and found that 41\% of them were intended for reusing the extracted code within the same commit (i.e., avoiding the duplication of pre-existing code), and 25\% of them were intended for eliminating pre-existing duplicate code.

\begin{figure}[ht]
	\centering
	\includegraphics[width=\linewidth]{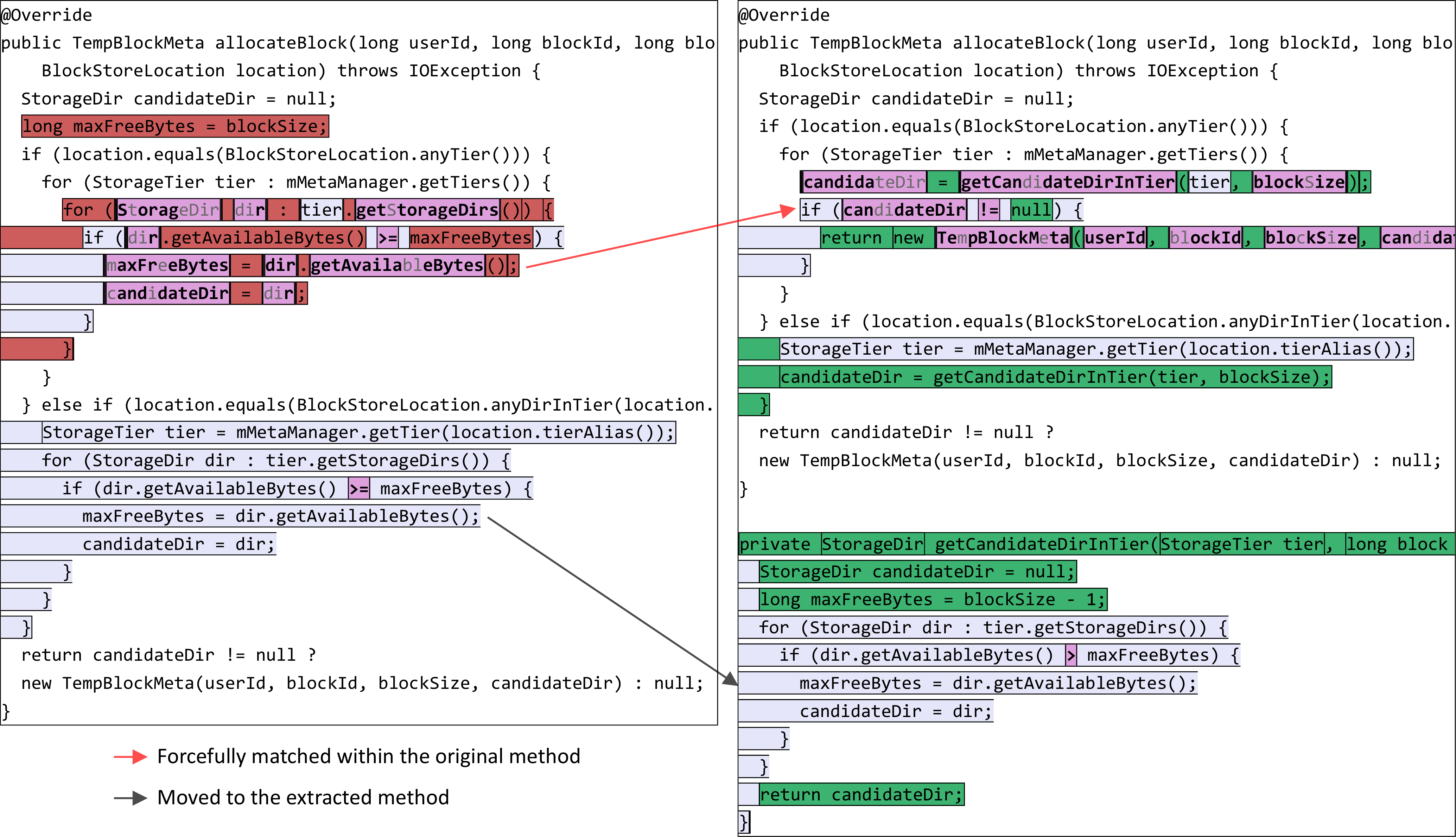}
	\vspace{-7mm}
	\caption{AST diff generated by GumTree 3.0 (greedy) for commit \url{https://github.com/Alluxio/alluxio/commit/9aeefcd}.}
	\label{fig:multi-mappings-GTG}
\end{figure}

Figure \ref{fig:multi-mappings-GTG} shows the AST diff generated by GumTree (greedy) for a commit in which duplicated code within the method \texttt{allocateBlock()} has been extracted to \texttt{getCandidateDirInTier()}.
As GumTree does not support multi-mappings, it matches correctly one of the duplicated code fragments with the code inside the body of the extracted method \texttt{getCandidateDirInTier()}. However, the other duplicated code fragment is forcefully matched with some newly added code in method \texttt{allocateBlock()} on the right side of the diff. This side-effect is a result of the lack of multi-mapping support and can cause great confusion to the code reviewers as it makes it very hard to understand that the actual intention of the commit author was to extract duplicated code into a single reusable method.

\begin{figure}[ht]
	\centering
	\vspace{-3mm}
	\includegraphics[width=\linewidth]{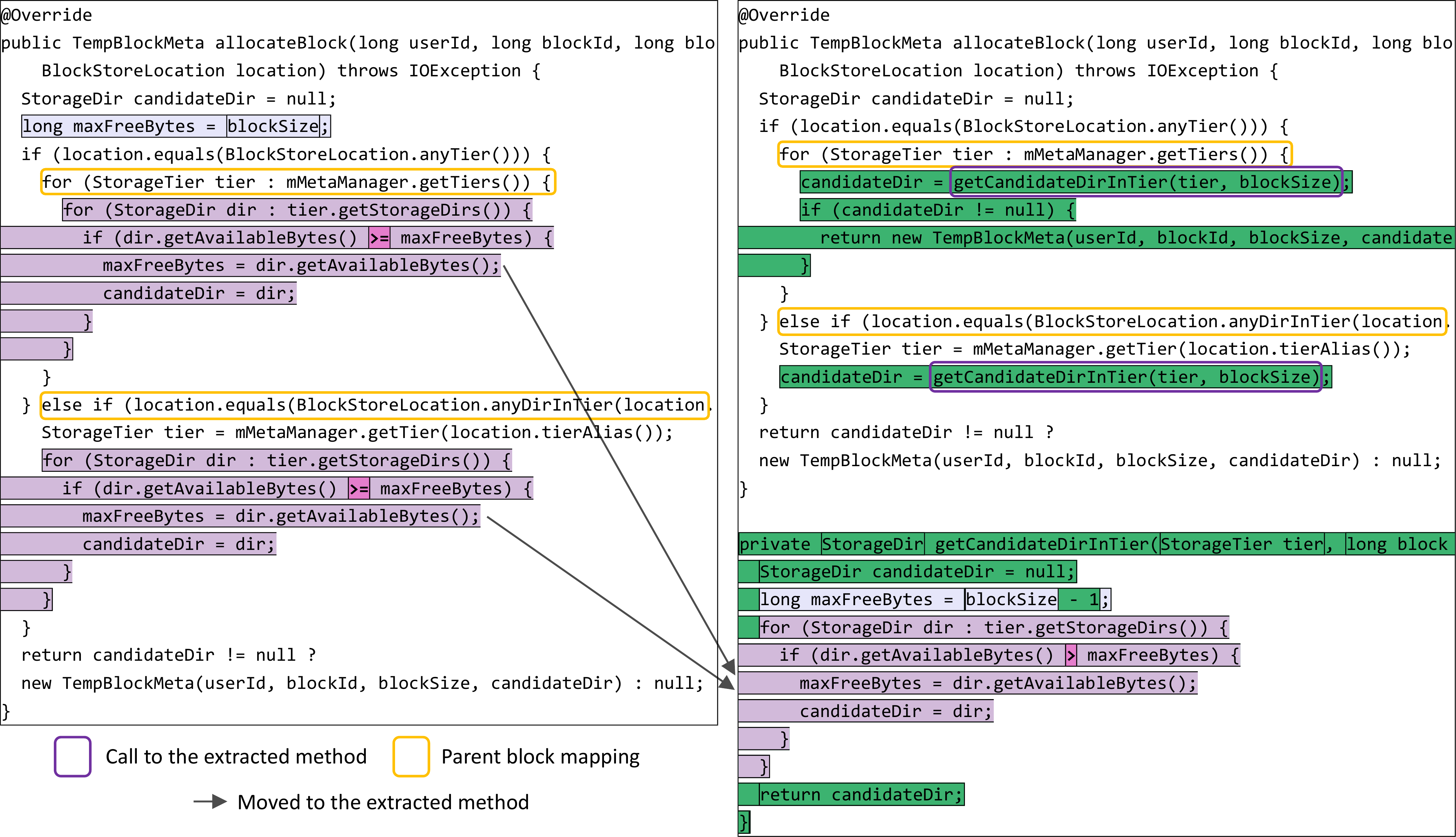}
	\vspace{-7mm}
	\caption{AST diff generated by RefactoringMiner for commit \url{https://github.com/Alluxio/alluxio/commit/9aeefcd}.}
	\label{fig:multi-mappings-RMiner}
	\vspace{-4mm}
\end{figure}


Figure \ref{fig:multi-mappings-RMiner} shows the AST diff generated by RefactoringMiner for the same commit. In contrast to the diff generated by GumTree (Figure \ref{fig:multi-mappings-GTG}), the intention of the commit author to extract duplicated code into a single reusable method is clearly expressed in the diff. Moreover, this AST diff is more aesthetically pleasing, less chaotic, and thus less confusing.
RefactoringMiner was able to discover the multi-mappings, because there exist two calls to the extracted method \texttt{getCandidateDirInTier()} within the body of \texttt{allocateBlock()} on the right side of the diff. By taking advantage of this language clue, it executes twice the process of mapping the statements within the body of \texttt{allocateBlock()} on the left side with the statements within the body of \texttt{getCandidateDirInTier()} on the right side. Moreover, in each execution the location of the call to the extracted method is utilized to restrict the scope of the left-side statement mappings within the parent block mapping under which each call is nested.

\subsection{Incorrect matching of program declarations}
\label{sec:language-clues}
Most AST diff tools are language-agnostic. The premise is that being language-agnostic, it is possible to devise a solution that can support any programming language (silver bullet).
Therefore, GumTree and its extensions try to ``blindly'' find the largest identical (or similar) AST sub-trees in a pair of files without relying on language clues to guide the matching process.
This approach has negative implications on both the accuracy of the mappings and the overall execution time.
For example, if a relatively large portion of the code in the body of a method on the left is extracted to a new method on the right, GumTree erroneously matches the left method declaration with the extracted method declaration, because the left method shares a larger common AST subtree with the extracted method compared to the code remaining in the original method (after extraction) on the right side.
However, this gives the code reviewers the wrong impression that the left method is ``renamed'' to the extracted method, which makes it more difficult to understand the actual change intended by the commit authors (i.e., \textsc{Extract Method} refactoring).

Some examples of language clues that can be used to guide the matching process are method signatures and method calls. 
If a method with the same signature can be found in both the left and right ASTs, the matching process can be narrowed down to the bodies of these two methods without searching for matches in the rest of the files.
If after matching the bodies of the methods, a portion of the left method is found as deleted and in its place in the right method there is a call to a newly added method (i.e., implying the potential application of an \textsc{Extract Method} refactoring), the matching process can be guided to match the deleted code from the left with the body of the newly added method from the right without searching for matches in the rest of the right file.
The only AST diff tool that considers some language clues is IJM~\cite{IJM}, which applies partial matching by restricting the scope of the matching process to selected parts of the source code (e.g., when matching method declarations with identical signatures). However, IJM does not utilize method calls to further guide the matching process.

\begin{figure}[ht]
	\centering
	\includegraphics[width=\linewidth]{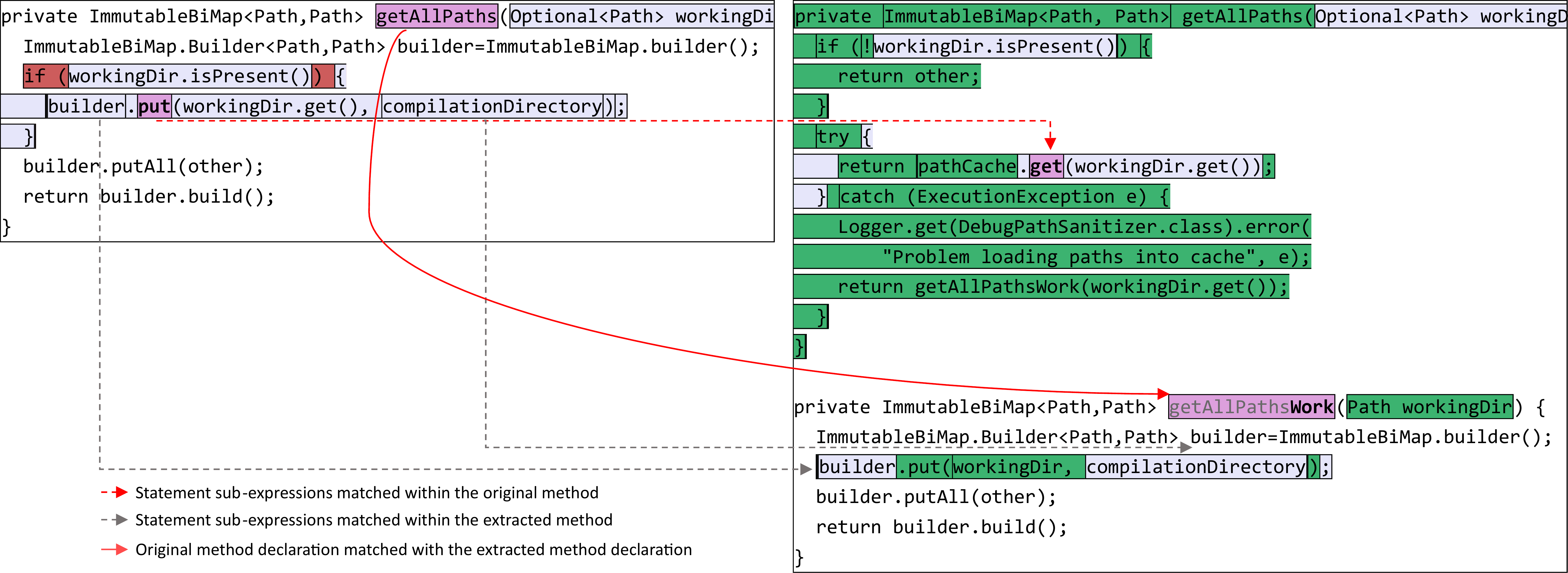}
	\vspace{-7mm}
	\caption{AST diff generated by GumTree 3.0 (greedy) for commit \url{https://github.com/facebook/buck/commit/a1525ac}.}
	\label{fig:language-clues-GTG}
\end{figure}

Figure \ref{fig:language-clues-GTG} shows the AST diff generated by GumTree (greedy) for a commit in which almost the entire body of method \texttt{getAllPaths()} has been extracted to \texttt{getAllPathsWork()}.
The vast majority of AST diff tools have been designed under the assumption that the shortest edit script
is the most beneficial to grasp the nature of a change \cite{gumtree, MTDiff}, as the length of the edit script is
a proxy to the cognitive load for a developer to understand the source code changes.
For this reason, GumTree matches \texttt{getAllPaths()} method declaration with \texttt{getAllPathsWork()} method declaration, as this match minimizes the \textit{move} edit operations (highlighted in gray background color), and results in an overall shorter edit script.
However, the diff visualization gives the code reviewers the wrong impression that \texttt{getAllPaths()} has been renamed to \texttt{getAllPathsWork()},
which makes it almost impossible to understand that the actual change intended by the commit author was to extract method \texttt{getAllPathsWork()} in order to provide the same functionality with a different input parameter type (i.e., \texttt{Path} instead of \texttt{Optional<Path>}).
However, this kind of method extractions are quite common and Silva et al. \cite{WhyWeRefactor} found that the second most common motivation for \textsc{Extract Method} refactoring is to introduce an alternative signature for an existing method (e.g., with additional or different parameters) and make the original method delegate to the extracted one, as happened in this particular example.
\begin{figure}[ht]
	\centering
	\vspace{-4mm}
	\includegraphics[width=\linewidth]{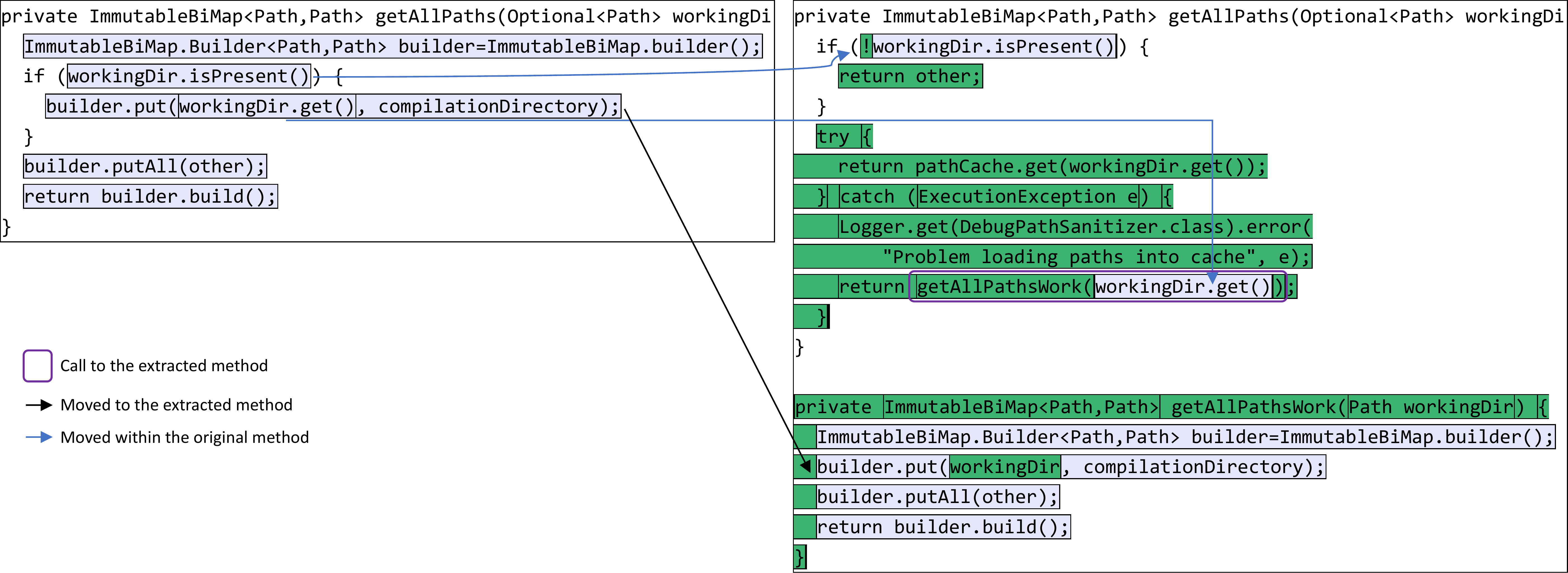}
	\vspace{-7mm}
	\caption{AST diff generated by RefactoringMiner for commit \url{https://github.com/facebook/buck/commit/a1525ac}.}
	\label{fig:language-clues-RMiner}
\end{figure}


By taking advantage of language clues, such as method signatures and method calls, RefactoringMiner generates the AST diff shown in Figure \ref{fig:language-clues-RMiner}.
The edit script corresponding to this diff is actually longer, because the majority of statements within the body of method \texttt{getAllPaths()} are shown as \textit{moved} (highlighted in gray background color) to method \texttt{getAllPathsWork()}. However, this diff depicts more accurately the intention of the commit author to extract method \texttt{getAllPathsWork()} in order to provide the same functionality with a  parameter of type \texttt{Path}.
Moreover, the mapping for sub-expression \texttt{workingDir.get()}, which is matched with the argument passed to the \texttt{getAllPathsWork()} call, accurately reflects the intention of the developer to change the parameter type from \texttt{Optional<Path>} to \texttt{Path}, as \texttt{workingDir.get()} returns the value (i.e., \texttt{Path}) present in \texttt{workingDir} (i.e., \texttt{Optional<Path>}).
This example is the proof that a shorter edit script is not necessarily the best way to convey the intention behind the changes made by a developer, and may actually confuse and mislead the code reviewers.
Moreover, this example demonstrates that AST diff cannot be properly addressed in a language-agnostic manner, as language clues (e.g., method signatures, method calls, field references) are of vital importance for guiding the matching process.

\subsection{Lack of refactoring awareness}
\label{sec:refactoring-awareness}
Refactoring is quite common in software evolution, as it facilitates multiple maintenance activities, such as reusing code, fixing bugs, adding new features, improving code design, improving testability, improving code understandability and readability \cite{WhyWeRefactor, WhyDevelopersRefactor}.
Despite its benefits on software maintainability, refactoring may introduce ``noise'' in the diff, as refactoring-related edits tend to overlap with other non-behavior-preserving changes.
Negara et al. \cite{Negara:2012} found that 46\% of refactored program entities are also edited or further refactored in the same commit, a practice commonly referred to as \textit{floss refactoring} \cite{HowWeRefactor}.

Refactoring information can help to improve the quality of source code diff.
For example, knowing that a variable, field, or method is renamed can help to improve the mapping of sub-expressions within the statements referencing them. 
Knowing that a sub-expression is extracted from a left-side statement and assigned to a newly added right-side variable declaration (i.e., \textsc{Extract Variable} refactoring) can help to properly match the pair of statements from which the variable was extracted, despite any major structural differences caused by the sub-expression extraction.
None of the current state-of-the-art AST diff tools leverages refactoring information in order to improve the accuracy of the generated AST node mappings.

\begin{figure}[ht]
	\centering
	\vspace{-3mm}
	\includegraphics[width=\linewidth]{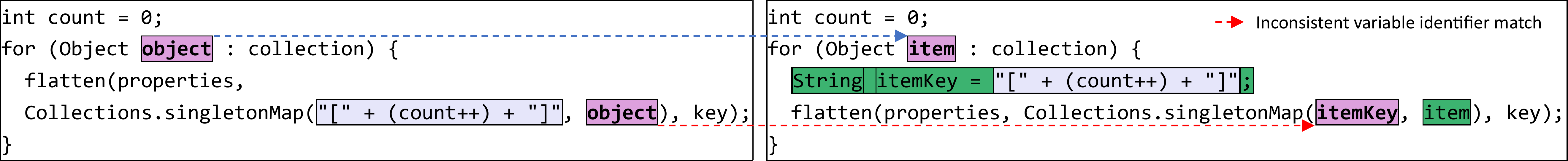}
	\vspace{-7mm}
	\caption{AST diff generated by GumTree 3.0 (simple) for commit \url{https://github.com/spring-projects/spring-boot/commit/20d39f7}.}
	\label{fig:refactoring-aware-GTS}
\end{figure}

Figure \ref{fig:refactoring-aware-GTS} shows the AST diff generated by GumTree (simple) for a piece of code in which two low-level refactorings were applied by the developer. The first refactoring is the renaming of the enhanced-for loop formal parameter \texttt{object} to \texttt{item} (i.e., \textsc{Rename Variable}). The second refactoring is the move of the expression \texttt{"[" + (count++) + "]"} to the initializer of the newly introduced variable \texttt{itemKey} (i.e., \textsc{Extract Variable}).
GumTree correctly matches the \texttt{SimpleName} AST nodes with values \texttt{object} and \texttt{item} in the enhanced-for loop formal parameter, but mismatches the \texttt{SimpleName} AST nodes with values \texttt{object} and \texttt{itemKey} in the argument list of the \texttt{Collections.singletonMap()} invocation.
The reason behind the mismatch is that GumTree tries to find the longest common subsequence with isomorphic structure and \texttt{itemKey} is the first AST node having the same AST type (i.e., \texttt{SimpleName}) with \texttt{object}.
The resulting diff visualization might confuse the code reviewers about whether variable \texttt{object} is actually renamed to \texttt{item}, as there are two inconsistent variable identifier mappings for \texttt{object}.

\begin{figure}[ht]
	\centering
	\vspace{-5mm}
	\includegraphics[width=\linewidth]{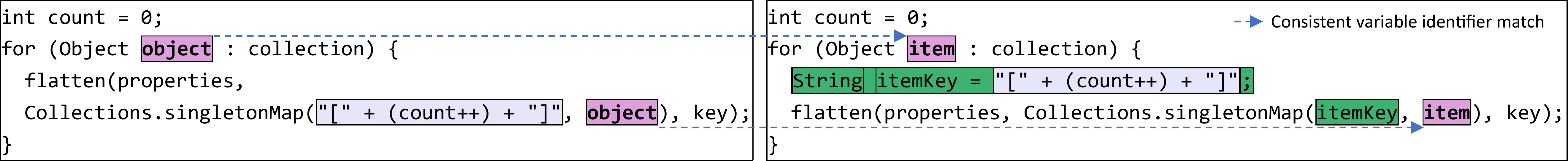}
	\vspace{-7mm}
	\caption{AST diff generated by RefactoringMiner for commit \url{https://github.com/spring-projects/spring-boot/commit/20d39f7}.}
	\label{fig:refactoring-aware-RMiner}
	\vspace{-3mm}
\end{figure}

Figure \ref{fig:refactoring-aware-RMiner} shows the AST diff generated by RefactoringMiner, which takes advantage of the detected \textsc{Rename Variable} and \textsc{Extract Variable} refactorings to optimize the \texttt{SimpleName} AST node mappings within the argument list of the \texttt{Collections.singletonMap()} invocation in a post-processing phase. The resulting diff visualization has no ambiguity regarding the renaming of variable \texttt{object} to \texttt{item}, and thus makes easier the understanding of the actual changes applied by the commit author.

\subsection{Matching semantically incompatible AST nodes (i.e., semantic ignorance)}
\label{sec:semanticallyIncompatibleNodes}
As most AST diff tools are language agnostic \cite{gumtree}, they can match any pair of AST nodes, as long as they have the same AST type, ignoring completely the semantic role these AST nodes play in the program. For example, a method parameter, the formal parameter of an \texttt{enhanced-for} loop, the exception declaration of a \texttt{catch} block, and the formal parameter of a lambda expression, all share the same AST type, which is the \texttt{SingleVariableDeclaration} type in the AST representation of the Eclipse JDT parser.
GumTree can match any combination of AST nodes from the aforementioned semantic groups, just because they have the same AST type. However, such matches make no semantic sense, make the resulting diff look ``unnatural'', may confuse the code reviewers, or even mislead the code reviewers in comprehending the actual changes intended by the commit authors in a wrong way.
Another example of a semantically incompatible match allowed by GumTree is the matching of a type reference (i.e., \texttt{SimpleType} AST node with a nested \texttt{SimpleName} AST node) with a variable reference (i.e., \texttt{SimpleName} AST node),
or the matching of a method name reference (i.e., \texttt{SimpleName} AST node) with a variable reference.
Again, although the AST nodes from the aforementioned semantic groups share the same AST type (\texttt{SimpleName}), they serve a completely different semantic purpose in the program.
In order to prevent the latter semantically incompatible matches, IJM \cite{IJM} modifies the AST structure by merging the value of name nodes (i.e., \texttt{SimpleName} AST nodes) with their respective parent nodes and deleting the name nodes with redundant information.
The last example occurs when there is a change in the block structure of a method (i.e., control flow statements being added or deleted). In such case, GumTree may match the block corresponding to the method body with a control structure block (e.g., the block corresponding to the body of an \texttt{if} or \texttt{for} statement). The block corresponding to the method body has a unique semantic purpose (i.e., enclose the functionality of a method), and thus should be matched only with a method body block.

\begin{figure}[ht]
	\centering
	\vspace{-2mm}
	\includegraphics[width=\linewidth]{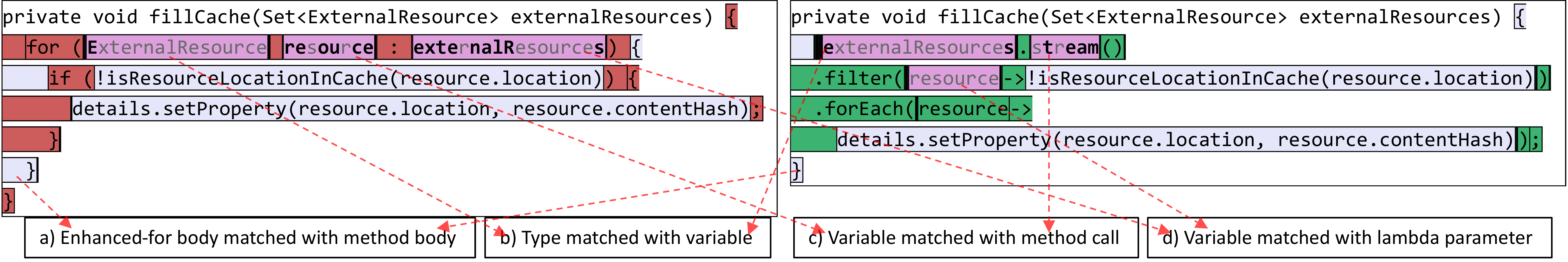}
	\vspace{-7mm}
	\caption{AST diff generated by GumTree 3.0 (greedy) for commit \url{https://github.com/checkstyle/checkstyle/commit/de022d2}.}
	\label{fig:semantic-aware-GTG}
	\vspace{-2mm}
\end{figure}

Figure \ref{fig:semantic-aware-GTG} shows the AST diff generated by GumTree (greedy) for a piece of code in which an enhanced-for loop has been migrated to a semantically equivalent Java Stream API call chain.
Although GumTree correctly matched the condition of the \texttt{if} statement with the lambda expression within the \texttt{filter} method call, and the statement within the body of the \texttt{if} statement with the lambda expression within the \texttt{forEach} method call, it incorrectly matches four semantically incompatible AST nodes:
a) the enhanced-for loop body block with the method body block, b) the type of the enhanced-for loop formal parameter \texttt{ExternalResource} with variable \texttt{externalResources},
c) the variable \texttt{resource} with the method call \texttt{stream}, and d) the variable \texttt{externalResources} with the lambda parameter \texttt{resource}.
Semantically incompatible AST node mappings may cause great confusion to the code reviewers, and in this particular case make it more difficult to assess whether the migration to a new language feature (i.e., Stream API) is behavior-preserving.

\begin{figure}[ht]
	\centering
	\vspace{-5mm}
	\includegraphics[width=\linewidth]{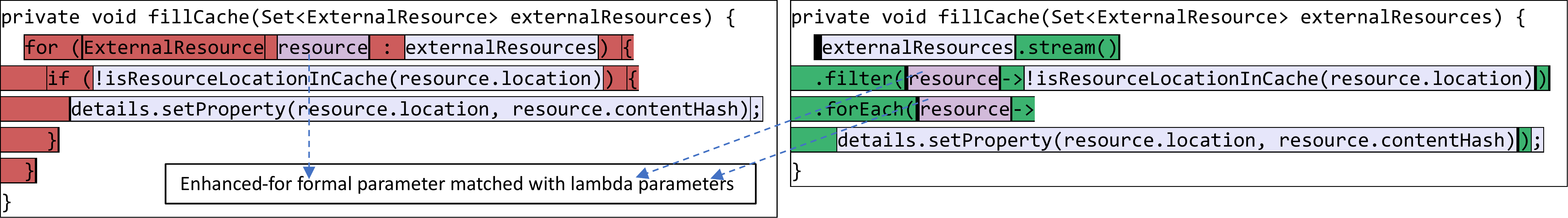}
	\vspace{-7mm}
	\caption{AST diff generated by RefactoringMiner for commit \url{https://github.com/checkstyle/checkstyle/commit/de022d2}.}
	\label{fig:semantic-aware-RefactoringMiner}
	\vspace{-3mm}
\end{figure}

Figure \ref{fig:semantic-aware-RefactoringMiner} shows the AST diff generated by RefactoringMiner, which is fully semantic-aware and shows that the enhanced-for loop formal parameter \texttt{resource} corresponds to the lambda parameters of the \texttt{filter} and \texttt{forEach} method calls (i.e., multi-mapping). 
The resulting diff visualization shows clearly that the applied migration is behavior-preserving and helps the code reviewers to understand (and possibly learn) the transformation mechanism.

\subsection{Lack of diff support at the commit level (i.e., commit-level change ignorance)}
\label{sec:commitLevelDiff}
All current state-of-the-art AST diff tools accept as input a pair of files, and thus cannot detect the move of a code fragment to another file modified in the same commit. This limitation has a further side-effect. It is possible that a code fragment from the left AST that has not been properly marked as moved to another file to be forcefully and incorrectly matched with some newly added code in the right AST.
Such incorrect code fragment matches may mislead code reviewers about the actual code re-organization and design changes applied by the commit authors.
A notable exception is Staged Tree Matching \cite{StagedTreeMatching}, which is a GumTree extension. Staged Tree Matching first applies the standard GumTree algorithm on the pairs of files modified in a project. Then, it constructs a project-level AST for each version by connecting the AST roots corresponding to each file to a pseudo-project-root node, and runs another matching phase on the project-level ASTs by considering only the remaining unmatched AST subtrees. However, this is an Early Research Achievements work and the authors do not provide any experimental results regarding the accuracy and execution time of the proposed approach.
\begin{figure}[ht]
	\centering
	\vspace{-3mm}
	\includegraphics[width=\linewidth]{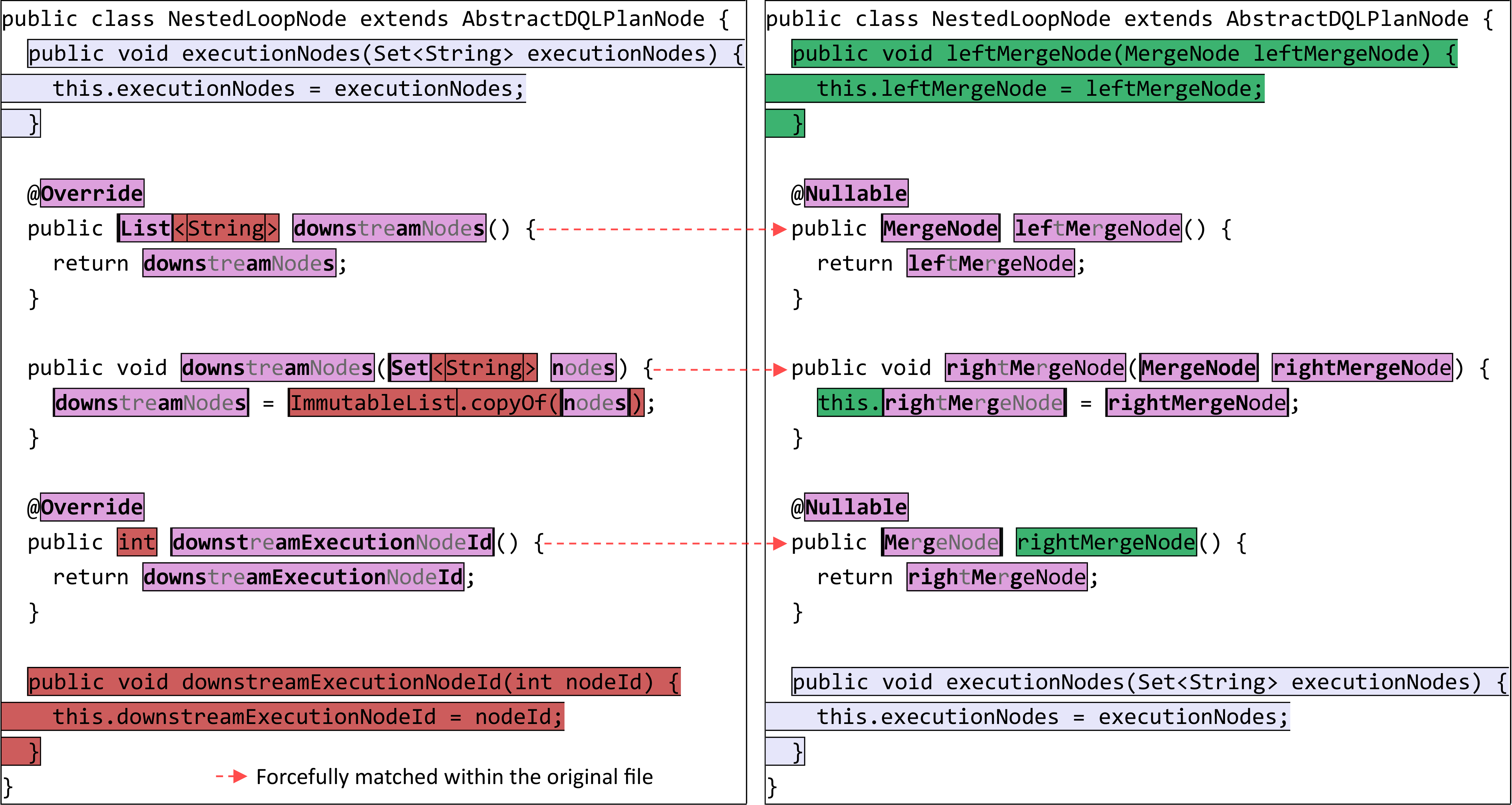}
	\vspace{-7mm}
	\caption{AST diff generated by GumTree 3.0 (greedy) for commit \url{https://github.com/crate/crate/commit/72b5348}.}
	\label{fig:commit-GTG}
\end{figure}
\begin{figure}[ht]
	\centering
	\vspace{-2mm}
	\includegraphics[width=\linewidth]{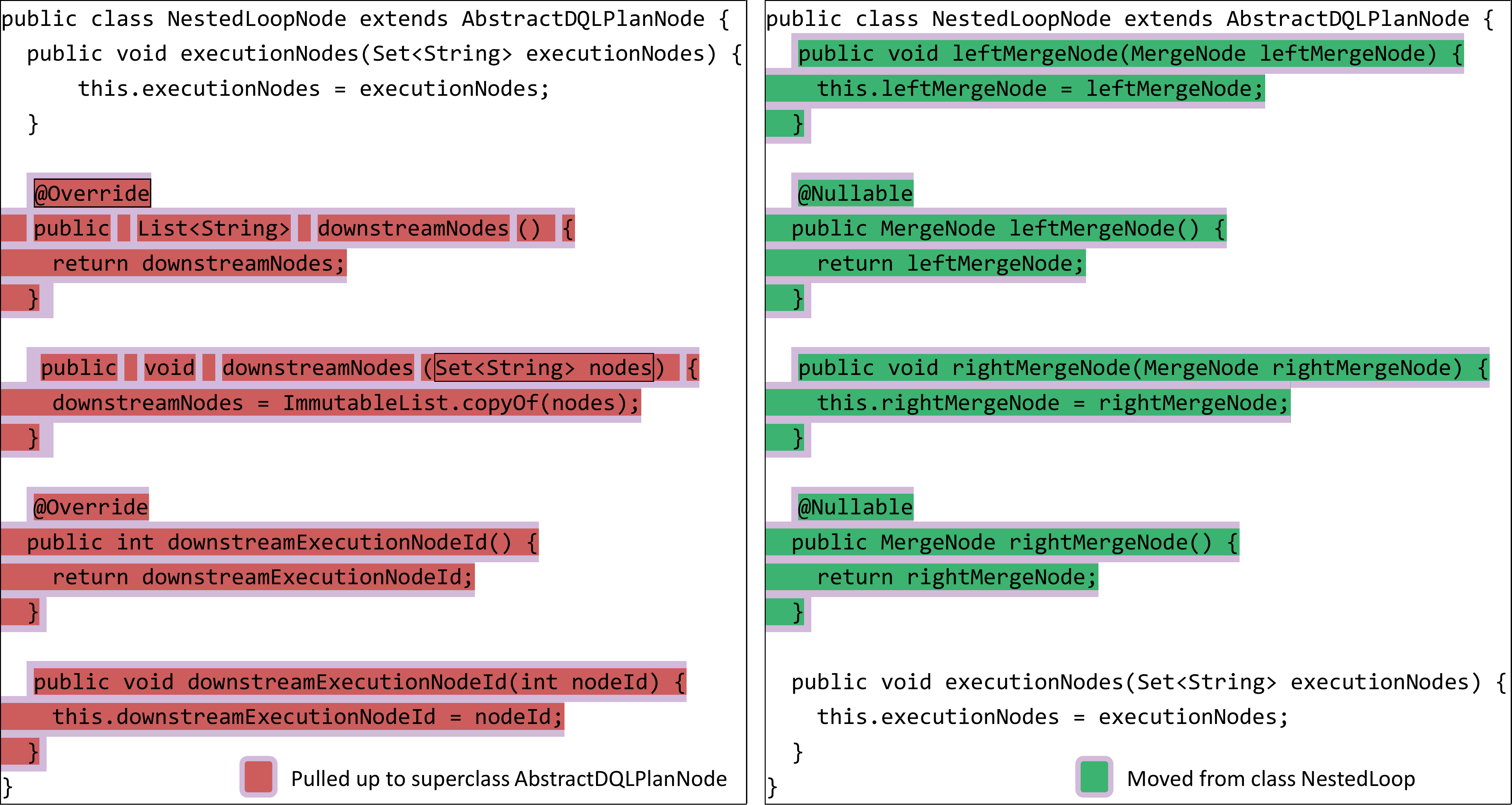}
	\vspace{-7mm}
	\caption{AST diff generated by RefactoringMiner for commit \url{https://github.com/crate/crate/commit/72b5348}.}
	\label{fig:commit-RefactoringMiner}
	\vspace{-5mm}
\end{figure}
Figure \ref{fig:commit-GTG} shows the AST diff generated by GumTree (greedy) for a commit in which some methods are moved to other classes.
As GumTree can perform a diff between pairs of files, it is not aware of the changes taking place in other files modified in the same commit.
As a result GumTree forcefully matches the deleted methods \texttt{downstreamNodes} and \texttt{downstreamExecutionNodeId} (actually pulled up to superclass \texttt{AbstractDQLPlanNode}) to the added methods \texttt{leftMergeNode} and \texttt{rightMergeNode} (actually moved from class \texttt{NestedLoop}).
The resulting diff visualization will definitely mislead the code reviewers to think that the corresponding methods have been renamed.


On the other hand, RefactoringMiner analyzes the changes within all files modified in a commit and is capable of detecting method moves between files.
As shown in Figure \ref{fig:commit-RefactoringMiner}, moved methods are specially highlighted with a purple border, and by hovering over them a tooltip appears showing where each method is moved from/to.
\section{Approach}
In Section~\ref{sec:background}, we provide a quick overview of how RefactoringMiner 2.0 \cite{RefactoringMiner2} works, as background information needed to understand the changes we performed in version 3.0 (Section~\ref{sec:changes-to-improve-statement-mapping-accuracy}) to improve statement mapping accuracy. In Section~\ref{sec:ast-diff-generation}, we explain how we utilize the information provided by RefactoringMiner 3.0 to generate AST diff.

\subsection{Background on how RefactoringMiner 2.0 works}
\label{sec:background}
\begin{figure}[ht]
	\centering
	\includegraphics[width=\linewidth]{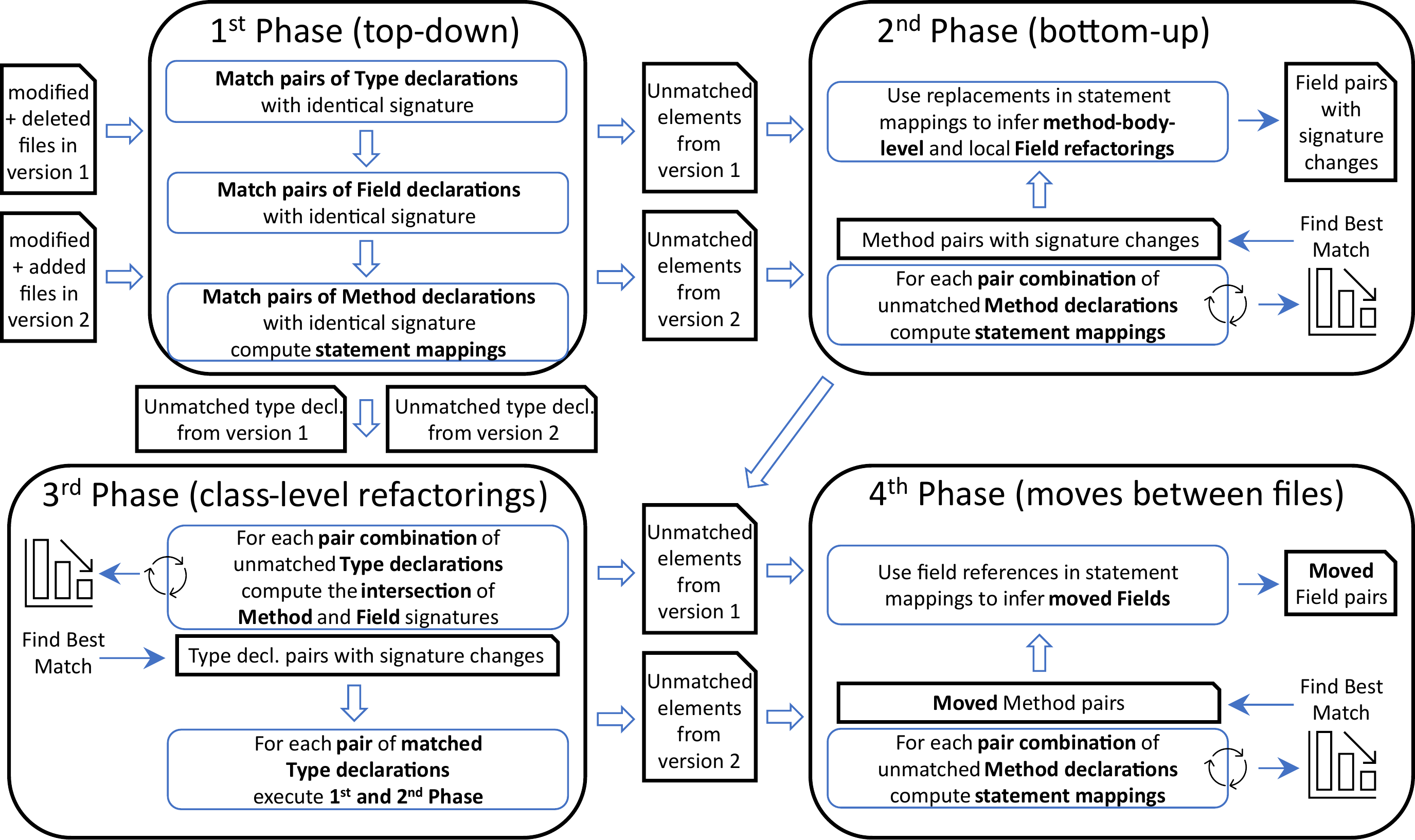}
	\vspace{-7mm}
	\caption{Overview of RefactoringMiner's workflow.}
	\label{fig:workflow}
	\vspace{-3mm}
\end{figure}

Figure \ref{fig:workflow} shows the overall workflow of RefactoringMiner.
Although this workflow remained unchanged, since version 2.0 \cite{RefactoringMiner2}, we decided to include it for the sake of completeness, as it will make it easier for readers who are not familiar with RefactoringMiner to understand better how it works.
RefactoringMiner accepts as input two sets of files from two versions. It provides APIs that can extract this input from a repository commit, from two directories containing Java files, or even from raw file contents provided as strings.
The matching process consists of four phases:
\begin{enumerate}[wide,label=\bfseries Phase \arabic*., leftmargin=*,labelwidth=!, labelindent=0pt]
	\item In the first phase, all program elements (i.e., type, method, field declarations) with identical signatures are getting matched in a top-down fashion, starting from classes having an identical signature (i.e., package and class name), and moving on to the fields and methods declared in them.
	\item For each pair of type declarations that were matched in Phase 1,
	we collect any remaining unmatched methods and fields, and provide them as input to Phase 2, which is responsible for matching program elements with changes in their signatures.
	For every pair combination of unmatched method declarations, we compute their statement mappings, which are then passed to a sorting function that finds the best match among all pairs.
	This step supports the detection of \textit{method signature changes}, such as  renames, addition/deletion of parameters, change of return or parameter types, as well as the detection of \textit{merged} and \textit{split methods}.
	In the next step, the statement mappings within the matched method declarations are utilized to infer \textit{method-body-level} and \textit{local field} refactorings.
	RefactoringMiner does not use a similarity threshold to match a pair of non-identical statements. Instead, it applies all possible syntactically valid AST node replacements within a pair of input statements, and matches these statements only if they become textually identical after the application of replacements.
	These AST node replacements within the statement mappings could be references to renamed/extracted/inlined fields, local variables, or method parameters. We examine the consistency of these replacements throughout the methods of a type declaration to ensure that the detected refactorings are valid.
	\item For every pair combination of type declarations that were not matched in Phase 1, we compute the intersection of methods and fields based on their signatures. Then, a sorting function determines the best match among all pairs. This step supports the detection of \textit{type signature changes}, such as class renames, class moves to another package, as well as the detection of \textit{merged} and \textit{split classes}. For each pair of matched type declarations, we execute Phase 1 and 2, as described above, to compute statement mappings and infer \textit{method-body-level} and \textit{local field} refactorings.
	\item In the last phase, we collect any remaining unmatched methods and fields from all type declarations, and for every pair combination of methods, we compute their statement mappings. Again, a sorting function determines the best match among all pairs.
	This step supports the detection of method moves between previously existing files (if these files are part of the same inheritance hierarchy then the move is actually a \textsc{Pull Up} or \textsc{Push Down Method} refactoring), as well as method moves to new files (i.e,. \textsc{Extract Class} or \textsc{Extract Superclass} refactoring).
	Finally, the statement mappings within the matched moved method declarations are examined for references to unmatched fields in order to infer moved fields along with the moved methods.
\end{enumerate}

\subsection{Changes in RefactoringMiner 3.0 to improve statement mapping accuracy}
\label{sec:changes-to-improve-statement-mapping-accuracy}
\begin{figure}[ht]
	\centering
	\includegraphics[width=\linewidth]{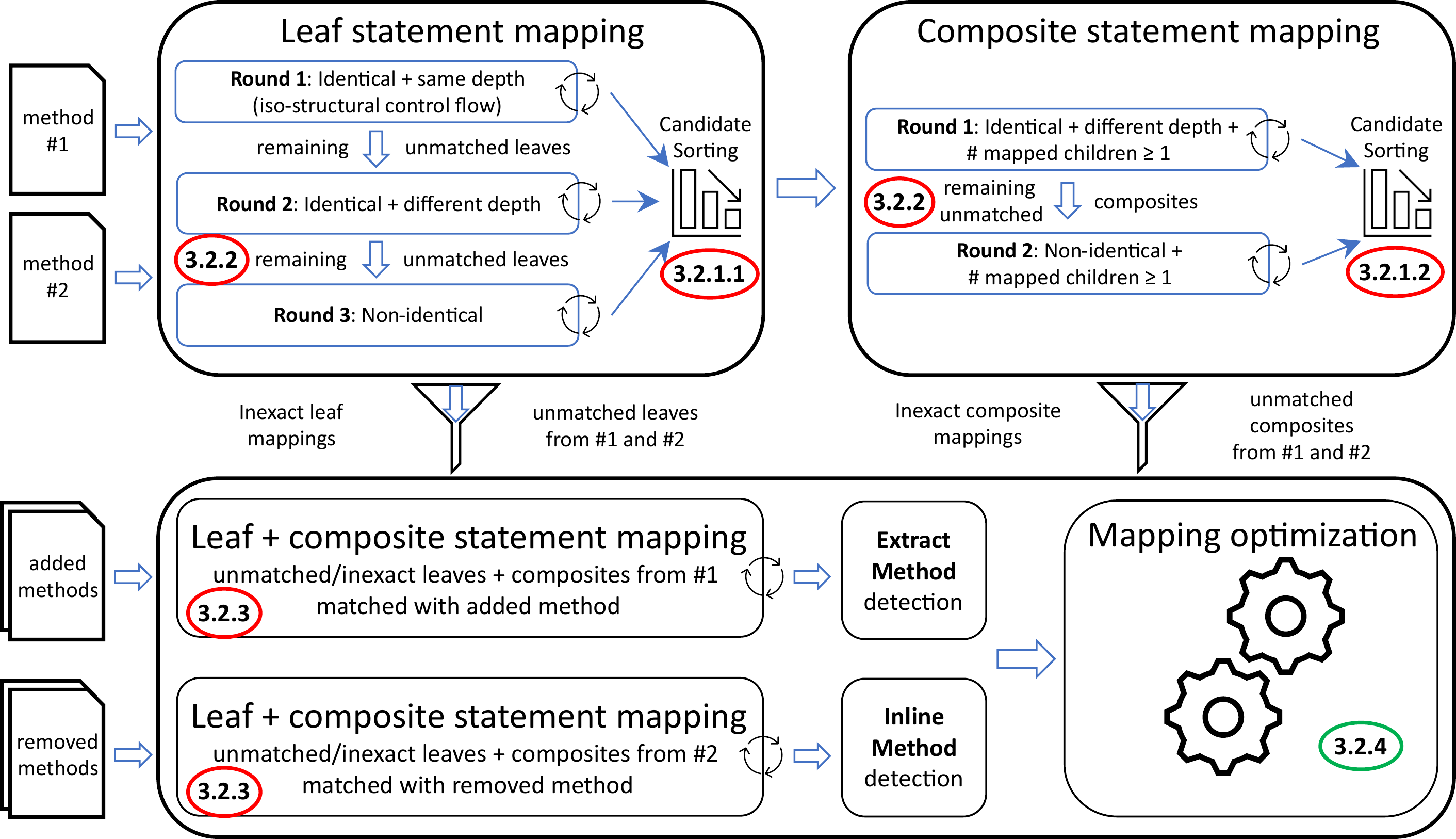}
	\vspace{-7mm}
	\caption{Overview of the statement mapping process highlighting the parts that are added or modified and the corresponding sections explaining these new features in detail. Red-color oval shapes e.g., \Circled[outer color=red]{3.2.2} indicate a change to a previously existing component in RefactoringMiner 2.0. Green-color oval shapes e.g., \Circled[outer color=green]{3.2.4} indicate a new component that did not exist in RefactoringMiner 2.0.}
	\label{fig:overview}
\end{figure}

Figure \ref{fig:overview} shows the core of RefactoringMiner's statement mapping process and highlights the parts that are added or modified to improve the overall accuracy of our AST diff.
This process is executed wherever ``compute statement mappings'' appears in Figure~\ref{fig:workflow}.
In a nutshell, given two code fragments as input (e.g., method \#1 and method \#2),
first the leaf statements (i.e., statements without a body) are matched in three rounds. The first round is executed only if the input code fragments have an iso-structural control flow structure. After the completion of each round, the remaining unmatched leaves are passed to the next round. During the execution of each round, a pair of statements is matched only if the corresponding matching criterion shown in Figure \ref{fig:overview} is met, and if there are multiple candidate mappings for a given statement, a sorting function is used to select the ``best'' one.
Next, the composite statements, i.e., statements with a body containing other nested statements (\texttt{for}, \texttt{while}, \texttt{if}, \texttt{do}, \texttt{switch}, \texttt{synchronized}, \texttt{try}, \texttt{catch}, \texttt{finally}), are matched in two rounds, utilizing the previously established leaf statement mappings.
At the end of this process, there might be some inexact mappings (i.e., non-identical matched statements) and some unmatched statements.
These are then matched with statements of added/removed methods in the same class to detect instances of \textsc{Extract/Inline Method} refactorings (i.e., methods extracted from method \#1 or methods inlined to method \#2).
At the end, a statement mapping optimization process is executed to handle statements that
might be matched two or more times (e.g., once with a statement in the original method, and once with a statement in an
extracted/inlined method). This process determines which one of the duplicate statement
mappings is the ``best'' and discards the other ones.

\subsubsection{Improved sorting of candidate mappings}
\label{sec:candidate-sorting}
Originally, RefactoringMiner had a very simplistic approach for sorting leaf mapping candidates (feature \Circled[outer color=red]{3.2.1.1} in Figure~\ref{fig:overview})
based on 3 criteria~\cite{RefactoringMiner2}.
First, based on the string edit distance of the nodes in ascending order (i.e., more textually similar node pairs rank higher). 
Second, based on the absolute difference of the nodes’ depth in ascending order (i.e., node pairs with more similar depth rank higher). 
Third, based on the absolute difference of the nodes’ index in their parent’s list of children in ascending order (i.e., node pairs with more similar position in their parent’s list of children rank higher).
Composite statement mappings (feature \Circled[outer color=red]{3.2.1.2} in Figure~\ref{fig:overview}) were sorted with the same criteria plus an additional one~\cite{RefactoringMiner2}, namely
based on the ratio of the nodes’ matched children in descending order (i.e., node pairs with more matched children rank higher).

This simplistic ranking was sufficient for the sake of detecting method-level refactorings, because the refactoring detection rules simply require the number of matched statements within a pair of methods to be larger than the number of unmatched statements. As a result, even if the statement mappings are not perfectly matched, the refactoring would be still detected. 
However from our early experiments, we discovered that the resulting mappings were not always accurate, especially in some corner cases involving commonly repeated statements (e.g., \texttt{break}, \texttt{continue}, \texttt{return}, \texttt{throw}) within the body of a method.
As a matter of fact, in 752 out of 988 commits (76\%) in our benchmark (Section~\ref{sec:datasets}), there is at least one case of identical repeated statements.
Therefore, we introduced more fine-grained sorting criteria to handle properly the ties between candidate mappings.
These new criteria include checking whether the statements that precede/follow the statement mappings are identical or not,
and taking into account the edit distance of the statement mappings' parent nodes.
The formal definitions of the sorting criteria for leaf and composite mappings are detailed in sections~\ref{sec:leaf-mapping-sorting-criteria} and ~\ref{sec:composite-mapping-sorting-criteria}, respectively.

Given a mapping $M = (m_1, m_2)$ with $m_1$ being the left-side statement and $m_2$ being the right-side statement,
\begin{definition}
	$\textit{edit-distance}(M)$ is the Levenshtein edit distance between $m_1$ and $m_2$. It should be noted that the bodies of composite statements are not included in the computation of their edit distance.
\end{definition}

\begin{definition}
	$\textit{depth}(m)$ is the number of composite statement parents of statement $m$, until we reach the body of the method declaration containing statement $m$.
\end{definition}

\begin{definition}
	$\textit{parent-edit-distance}(M)$ is an array where index $i$ has the Levenshtein edit distance between the $i$th parent of $m_1$ and the $i$th parent of $m_2$ and $i = 1,..., \textit{depth}(M)$ where $\textit{depth}(M) = min(\textit{depth}(m_1), \textit{depth}(m_2))$.
	At index 1, we compute the edit distance between the direct parents of $m_1$ and $m_2$,
	at index 2, we compute the edit distance between the parents of the direct parents of $m_1$ and $m_2$, and so on, until $i$ becomes equal to $\textit{depth}(M)$.
\end{definition}

\begin{definition}
	$\textit{direct-parent-edit-distance}(M) =
	\textit{parent-edit-distance}(M)[1]$
\end{definition}

\begin{definition}
	$\textit{parent-edit-distance-sum}(M) =
	\sum_{i=1}^{n} \textit{parent-edit-distance}(M)[i]$, where $n = \textit{depth}(M)$
\end{definition}

\begin{definition}
	$\textit{depth-diff}(M) = |\textit{depth}(m_1) - \textit{depth}(m_2)|$
\end{definition}

\begin{definition}
	$\textit{index-diff}(M) = |\textit{index}(m_1) - \textit{index}(m_2)|$, where $\textit{index}(m)$ is $m$'s index in its parent’s statement list
\end{definition}

\paragraph{Sorting criteria for leaf statement mappings}
\label{sec:leaf-mapping-sorting-criteria}
Given two leaf mappings (i.e., pairs of matched leaf statements) $M_a$ and $M_b$, we check the following rules in the order they are listed and $M_a$ is ranked over $M_b$ iff (a rule is applicable if its condition holds for $M_a$ but not for $M_b$):
\begin{enumerate}[leftmargin=*]
	\item $M_a$ is part of a multi-mapping (i.e., a single statement matched to multiple statements or vice versa). If both $M_a$ and $M_b$ are part of a multi-mapping, $M_a$ is ranked first only if it involves more statements.
	\item if $\textit{edit-distance}(M_a) \neq \textit{edit-distance}(M_b)$:
	\begin{enumerate}[leftmargin=*]
		\item $M_a$ becomes textually identical after undoing an overlapping extract/inline variable refactoring.
		\item both statement pairs located right before and right after $M_a$ are textually identical.
		\item the statements in $M_a$ have the same nesting depth (i.e., $\textit{depth-diff}(M_a) = 0$), the same index in their parent's statement list, and the same AST type for their first non-block parent.
		\item if $\textit{depth}(M_a) > 2$ and $\textit{depth}(M_b) > 2$,
		$\textit{parent-edit-distance-sum}(M_a) <
		\textit{parent-edit-distance-sum}(M_b)$
		\item $\textit{edit-distance}(M_a) < \textit{edit-distance}(M_b)$
	\end{enumerate}
	\item if $\textit{edit-distance}(M_a) = \textit{edit-distance}(M_b)$:
	\begin{enumerate}[leftmargin=*]
		\item if $\textit{depth}(M_a) = \textit{depth}(M_b)$,
		$\textit{parent-edit-distance-sum}(M_a) = 0$ and
		$\textit{parent-edit-distance-sum}(M_b) > 0$
		\item if $\textit{depth}(M_a) \neq \textit{depth}(M_b)$,
		$\textit{parent-edit-distance-sum}(M_a) < 
		\textit{parent-edit-distance-sum}(M_b)$
		\item 
		$\textit{depth-diff}(M_a) < \textit{depth-diff}(M_b)$
		\item 
		$\textit{index-diff}(M_a) < \textit{index-diff}(M_b)$
		\item $\textit{direct-parent-edit-distance}(M_a) <
		\textit{direct-parent-edit-distance}(M_b)$
	\end{enumerate}
\end{enumerate}
\paragraph{Sorting criteria for composite statement mappings}
\label{sec:composite-mapping-sorting-criteria}
Given a composite statement mapping $M = (m_1, m_2)$ 
with $m_1$ being the left-side statement and $m_2$ being the right-side statement, 
and $\textit{children}(m)$ being the set of statements nested under $m$,
\begin{definition}
	$\textit{child-match-ratio}(M)$ is the number of matched child pairs nested under $m_1$ and $m_2$, respectively, divided by $max(|\textit{children}(m_1)|, |\textit{children}(m_2)|)$
\end{definition}

\begin{definition}
	$\textit{identical-composite-children}(M)$ is the number of matched composite child pairs nested under $m_1$ and $m_2$, respectively, which are identical (i.e., have an edit distance equal to zero)
	\vspace{-2mm}
\end{definition}

Given two composite statement mappings $M_a$ and $M_b$, we check the following rules in the order they are listed and $M_a$ is ranked over $M_b$ iff (a rule is applicable if its condition holds for $M_a$ but not for $M_b$):

\begin{enumerate}[leftmargin=*]
	\item $\textit{child-match-ratio}(M_a) \geq 2*\textit{child-match-ratio}(M_b)$
	\item if $\textit{edit-distance}(M_a) \neq \textit{edit-distance}(M_b)$:
	\begin{enumerate}[leftmargin=*]
		\item $M_a$ becomes textually identical after undoing an overlapping extract/inline variable refactoring.
		\item $\textit{edit-distance}(M_a) < \textit{edit-distance}(M_b)$
	\end{enumerate}
	\item if $\textit{edit-distance}(M_a) = \textit{edit-distance}(M_b)$:
	\begin{enumerate}[leftmargin=*]
		\item $\textit{identical-composite-children}(M_a) >
		\textit{identical-composite-children}(M_b)$
		\item 
		$\textit{child-match-ratio}(M_a) > \textit{child-match-ratio}(M_b)$
		\item 
		$\textit{depth-diff}(M_a) < \textit{depth-diff}(M_b)$
		\item 
		$\textit{index-diff}(M_a) < \textit{index-diff}(M_b)$
	\end{enumerate}
\end{enumerate}

\subsubsection{Multi-mapping support for duplicated code moved out of or moved into conditionals}
\label{sec:multi-mapping support}
RefactoringMiner was already able to infer multi-mappings for some specific scenarios based on the refactorings it detects.
For example, when some duplicated code is extracted from the same method or different methods, RefactoringMiner detects as many \textsc{Extract Method} refactoring instances as the number of duplicated code fragments.
Each \textsc{Extract Method} instance includes its own mappings between statements on the left side (i.e., one of the duplicated code fragments) and the right side (i.e., the extracted code fragment).
All these statement mappings refer to exactly the same statements on the right side, thus naturally forming \textit{multi-mappings}. \textsc{Inline Method} refactorings can also form multi-mappings, if the inlined method is called more than once.

\begin{figure}[ht]
	\centering
	\includegraphics[width=\linewidth]{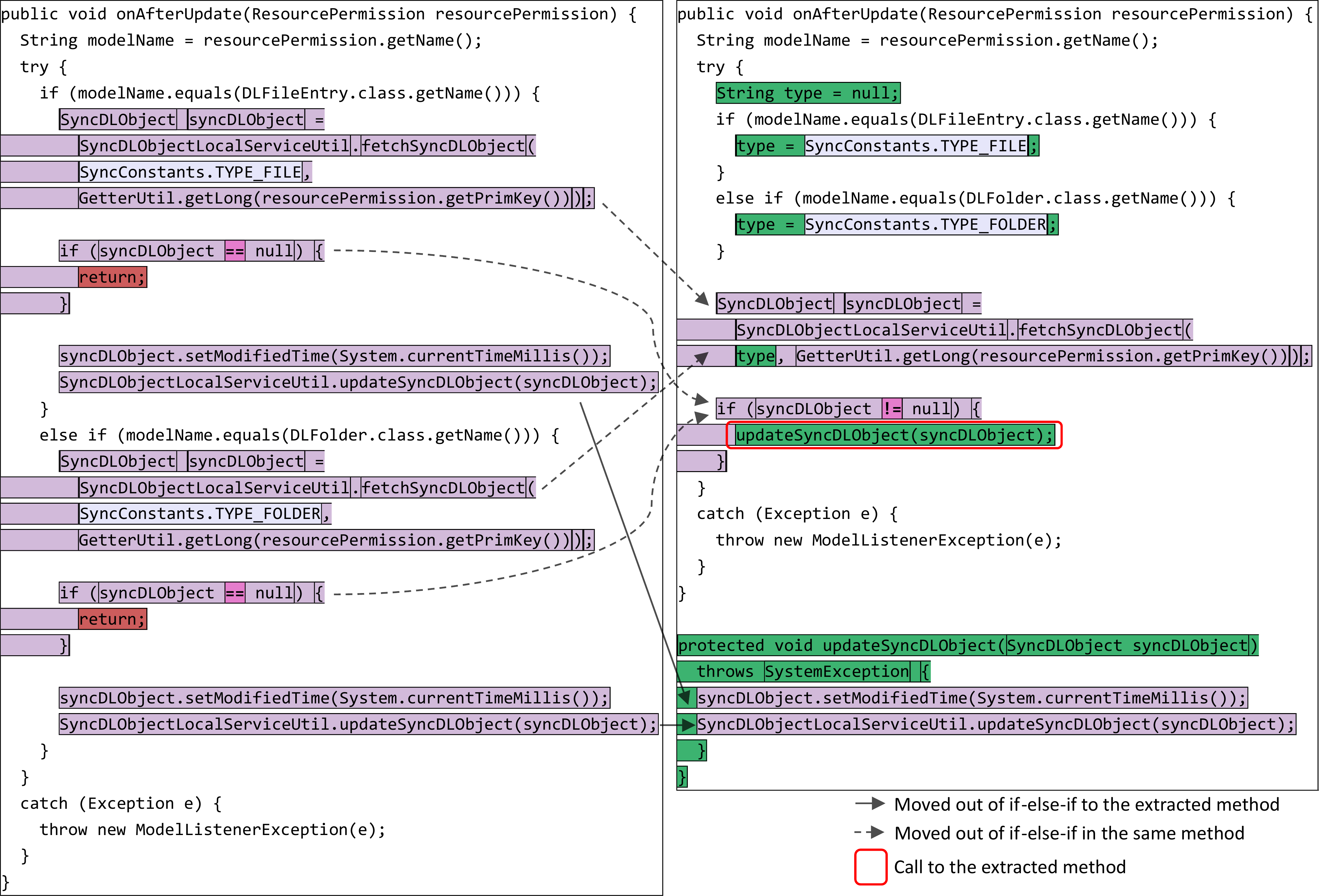}
	\vspace{-7mm}
	\caption{AST diff generated by RefactoringMiner for commit \url{https://github.com/liferay/liferay-plugins/commit/7c7ecf4}.}
	\label{fig:if-else-if-duplication}
	\vspace{-6mm}
\end{figure}

However, there are also scenarios in which developers eliminate (or introduce) duplicated code within the same method. Typically, such duplicated code exists within the branches of conditional statements, such as \texttt{if-else-if} or \texttt{switch-case} statements, where the same logic is repeated within each execution branch with some slight variations.
By applying the \textsc{Consolidate Duplicate Conditional Fragments} refactoring~\cite{Fowler:2018}, it is possible to merge the duplicated logic within each execution branch.
For example, in Figure \ref{fig:if-else-if-duplication}, we can see an \texttt{if-else-if} statement with identical functionality in each branch with the only difference being a \texttt{String} passed as an argument when calling method \texttt{fetchSyncDLObject()}, which creates the \texttt{syncDLObject}.
On the right side, we can see that the developer moved and merged the duplicated statements out of the \texttt{if-else-if} statement, and introduced a variable named \texttt{type} to store the varying value passed as an argument when calling method \texttt{fetchSyncDLObject()}. The original \texttt{if-else-if} statement still remains in the method just to assign the proper value to variable \texttt{type}.
Moreover, we can see that some of the duplicated statements are further extracted to method \texttt{updateSyncDLObject()}, which is called inside the body of statement \texttt{if(syncDLObject != null)}, which was also moved and merged out of the \texttt{if-else-if} statement, and its condition was inverted.

To support such duplication elimination/introduction scenarios, we developed a simple algorithm (feature \Circled[outer color=red]{3.2.2} in Figure~\ref{fig:overview}) that is activated whenever we have \textit{many-to-one} (i.e., many statements on the left side are matched with a single statement on the right side) or \textit{one-to-many} (i.e., one statement on the left side is matched with many statements on the right side) candidate statement mappings. Before passing these candidate mappings to the sorting function to select the ``best'' one, we execute our algorithm to check if we should keep all of them instead.
Assuming we have many-to-one candidate mappings, the algorithm checks if the left-side statements in the candidate mappings are nested under the same \texttt{if-else-if} or \texttt{switch-case} statement, and if each one of them belongs to a separate execution branch. If this is true, next we check if the right-side statement is placed in a shallower depth in the method's nesting structure compared to the left-side statements.
Usually, the new nesting depth of the right-side statement becomes equal to the nesting depth of the \texttt{if-else-if} or \texttt{switch-case} statement from which it was moved out, as in the example of Figure \ref{fig:if-else-if-duplication}.
If all conditions checked by our algorithm are met, then we keep all many-to-one candidate mappings and skip the execution of the sorting function.
A similar algorithm is applied for one-to-many candidate mappings.

\subsubsection{Statement mapping scope based on call sites}
\label{sec:scope-call-sites}
This is a language-aware feature (feature \Circled[outer color=red]{3.2.3} in Figure~\ref{fig:overview}) that utilizes the location of method calls in order to restrict the scope of the statement mapping process for 
accurately matching the statements within extracted and inlined methods.
This feature can be applied in any programming paradigm that involves function calls (e.g., functional, procedural, object-oriented), as long as the program has a control flow and it is possible to bind function calls to declarations.
The idea comes from the observation that developers tend to extract consecutive statements, which are located within the body of a control statement in the origin method.
For example, in Figure~\ref{fig:multi-mappings-RMiner}, the developer extracted duplicated code located within the body of a \texttt{for} and an \texttt{if} statement, respectively. As we can see in the right side of Figure~\ref{fig:multi-mappings-RMiner}, after the refactoring, two method calls to the extracted method \texttt{getCandidateDirInTier} appear in the locations of the duplicated code fragments within the body of the same \texttt{for} and \texttt{if} statement, respectively.
By utilizing the location and the number of calls to the extracted method, we can achieve two things.
First, the number of calls to the extracted method determines how many times the statement mapping process should be executed. Since there are two method calls to the extracted method, we should match the remaining unmatched statements in the origin method with the statements of the extracted method two times.
Second, the location of the calls helps to narrow down the scope of the statement mapping process in each execution.
In the first execution, we include only the unmatched statements within the body of the \texttt{for} loop in the origin method, as the \texttt{for} loop corresponds to the parent mapping of the first call to the extracted method.
In the second execution, we include only the unmatched statements within the body of the \texttt{if} statement in the origin method, as the \texttt{if} statement corresponds to the parent mapping of the second call to the extracted method.
Following this approach, we are able to obtain the multi-mappings shown in Figure~\ref{fig:multi-mappings-RMiner}, without mixing the mappings corresponding to each duplicated code region.

In the alternative scenario, where the duplicated code fragments are located within the method body scope, then the calls to the extracted method would be also located within the method body scope. In this scenario, it is not possible to use the parent mapping to restrict the unmatched statements included in each execution of the statement mapping process, as both duplicated code fragments would have the same parent mapping (i.e., the origin method body block).
For this alternative scenario, in each execution we form a region expanding from the first to the last identical statement mapping between the remaining unmatched statements in the origin method and the statements of the extracted method.
In each subsequent execution of the statement mapping process, we exclude the statements that were matched within a region in all previous executions.
With this approach, it is possible to obtain accurate multi-mappings, even when the duplicated code fragments follow each other directly, without any intermediate non-extracted code between them.
Some examples of this scenario can be found in our AST diff gallery \cite{diff-gallery-duplicated-code-12, diff-gallery-duplicated-code-36}.

\subsubsection{Statement mapping optimization}
\label{sec:statement-mapping-optimization}
\vspace{-2mm}
This is a completely new component added in the statement mapping process (feature \Circled[outer color=green]{3.2.4} in Figure~\ref{fig:overview}), which is executed right after the detection of \textsc{Extract Method} and \textsc{Inline Method} refactorings extracted from or inlined to a given pair of matched methods (e.g., method \#1 and method \#2 in Figure \ref{fig:overview}).
This component first detects statements that participate in more than one mappings.
For example, a statement in method \#1 that is matched with a statement in method \#2, and is also matched with a statement in an added method (\textsc{Extract Method} scenario),
or a statement in method \#2 that is matched with a statement in method \#1, and is also matched with a statement in a removed method (\textsc{Inline Method} scenario).
Assuming we have a set of such mappings, we check the following rules in the order they appear (each rule is applicable, only if there is at least one mapping in the set not satisfying the condition of the rule):
\begin{enumerate}[leftmargin=*]
	\vspace{-1.5mm}
	\item If a mapping in the set includes a call to an extracted method or an inlined method, then this mapping is discarded, and the others are kept.
	\item If one or more mappings in the set do not have their parent statements mapped, then these mappings are discarded, and the others are kept.
	\item If one or more mappings in the set belong to a nested extracted method (i.e., a method extracted from another extracted method) or a nested inlined method (i.e., a method inlined to another inlined method), then these mappings are discarded, and the others are kept.
	\item If one or more mappings are not textually identical (i.e., the statements were matched by replacing AST sub-expressions that were different), then these mappings are discarded, and the others are kept.
	\item We compute the minimum edit distance (i.e., the levenshtein distance between the two statements of a mapping) among the mappings in the set, and discard the mappings having an edit-distance larger than the minimum one.
	\vspace{-1.5mm}
\end{enumerate}
The same optimization process is applied in the case of \textsc{Extract and Move Method} and \textsc{Move and Inline Method} refactorings, where the extracted/inlined code is moved to/from another file.

\subsubsection{Semantic and refactoring-aware import declaration diff}
\label{sec:import-declaration-diff}
A good quality diff of import declaration statements should help code reviewers to quickly understand about library migrations \cite{MigrationMiner} and type/API migrations \cite{10.1145/3368089.3409725} that took place in the file.
To this end, we introduce the first semantic and refactoring-aware import declaration AST diff in the literature, which offers three novel features, namely matching based on package/class refactorings, matching based on local type changes in the file, and grouping/un-grouping based on \textit{import-on-demand} declarations (i.e., \texttt{import package.*}).

\begin{figure}[ht]
	\centering
	\vspace{-3mm}
	\includegraphics[width=\linewidth]{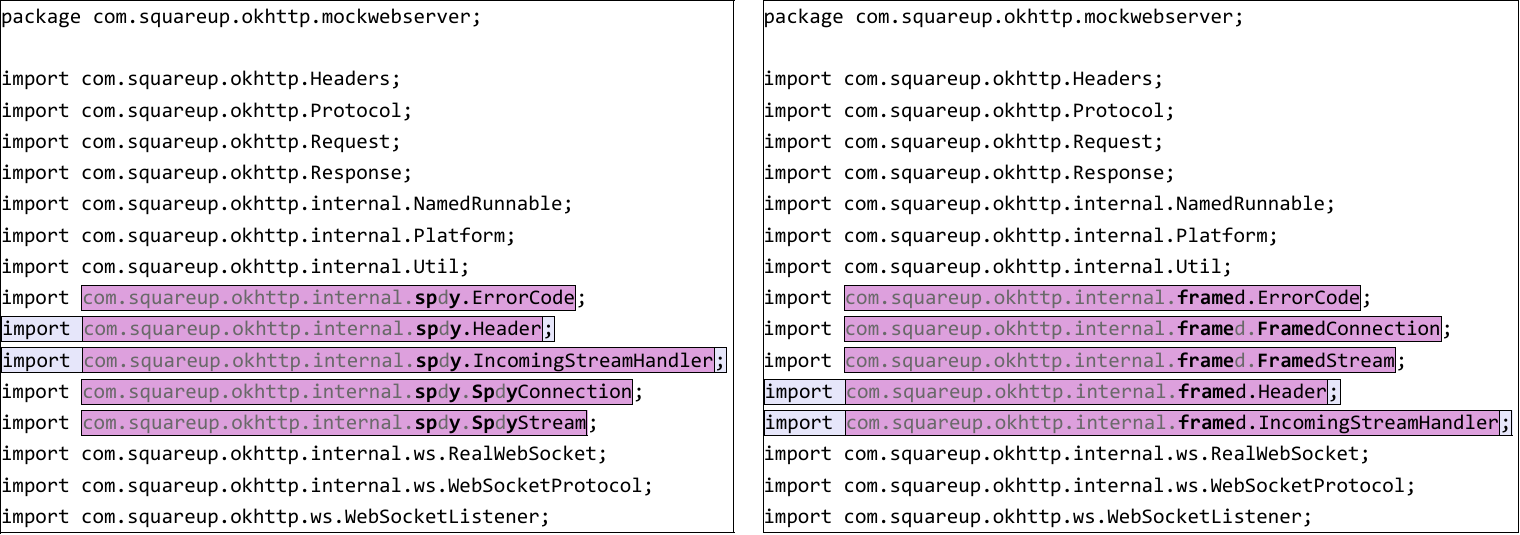}
	\vspace{-7mm}
	\caption{AST diff generated by RefactoringMiner for commit \url{https://github.com/square/okhttp/commit/c753d2e}.}
	\label{fig:imports-with-class-moves-renames}
	\vspace{-4mm}
\end{figure}

After completing the detection of refactorings at commit-level, RefactoringMiner is aware of changes in the structure and organization of the project, as reflected by package-level refactorings (e.g., \textsc{Rename Package}, \textsc{Move Package}) and class-level refactorings (e.g., \textsc{Rename Class}, \textsc{Move Class}).
In a post-processing step, RefactoringMiner goes through each modified Java file in the commit, and matches pairs of previously unmatched (i.e., removed and added) import declarations whose types correspond to the fully-qualified names of moved/renamed classes.
In Figure \ref{fig:imports-with-class-moves-renames}, we can see import declaration statements matched based on \textsc{Rename Class} and \textsc{Move Class} refactoring information. All five type declarations referenced in the import statements have been moved from package \texttt{spdy} to \texttt{framed}, and additionally classes \texttt{SpdyConnection} and \texttt{SpdyStream} have been renamed to \texttt{FramedConnection} and \texttt{FramedStream}, respectively.
RefactoringMiner is able to correctly match the updated import declarations, despite the re-ordering after applying alphabetical sorting.

\begin{figure}[ht]
	\centering
	\includegraphics[width=\linewidth]{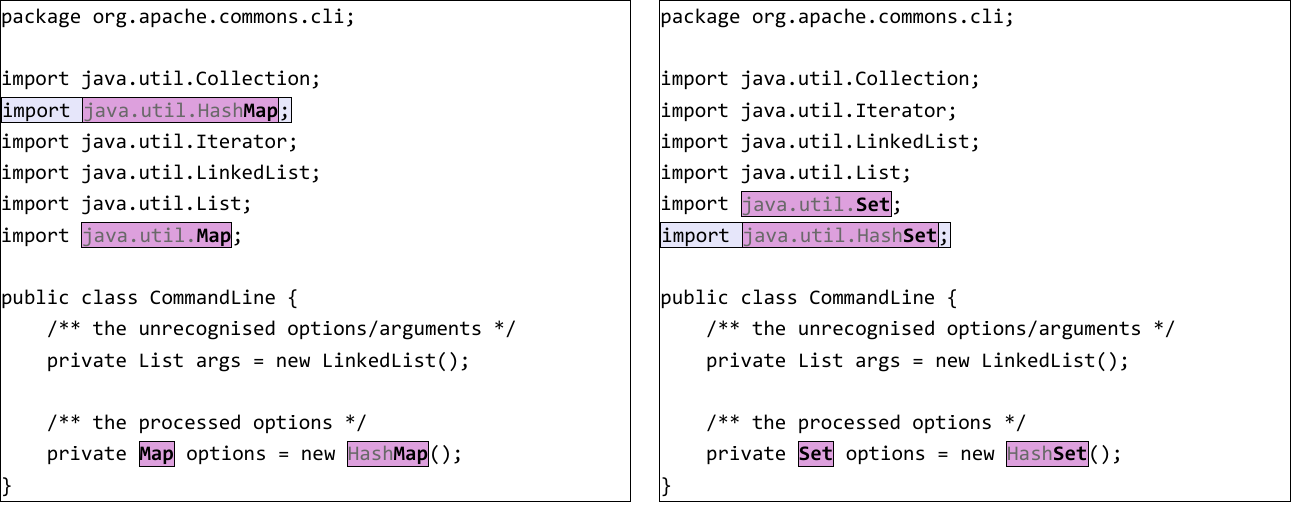}
	\vspace{-7mm}
	\caption{AST diff generated by RefactoringMiner for Cli-1.}
	\label{fig:imports-with-type-changes}
	\vspace{-4mm}
\end{figure}

\begin{figure}[ht]
	\centering
	\includegraphics[width=\linewidth]{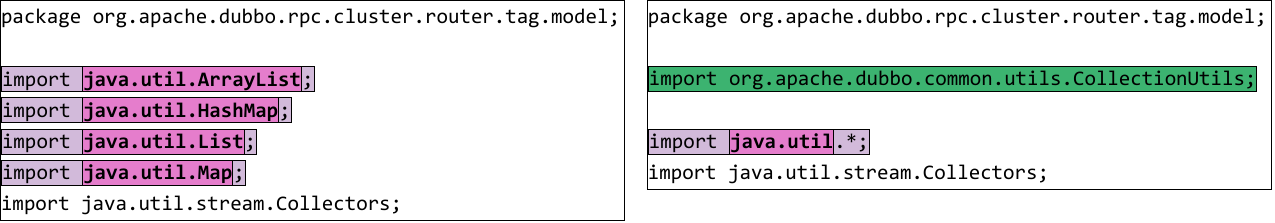}
	\vspace{-7mm}
	\caption{AST diff generated by RefactoringMiner for commit \url{https://github.com/apache/dubbo/commit/500434a}.}
	\label{fig:imports-with-grouping}
	\vspace{-4mm}
\end{figure}
The type changes detected by RefactoringMiner within the scope of a Java file are also used to match pairs of previously unmatched (i.e., removed and added) import declarations whose types correspond to the changed type references.
In Figure \ref{fig:imports-with-type-changes}, we can observe that the developer changed the type of field \texttt{options} from \texttt{Map} to \texttt{Set} and the initializer of field \texttt{options} from \texttt{HashMap} to \texttt{HashSet}. The modified import declaration statements have been matched accordingly to reflect the collection migration from \texttt{Map} to \texttt{Set}.
It should be noted that only the \textit{consistent} type changes within the scope of a Java file are taken into account.
For example, if \texttt{CommandLine.java} included a type change from \texttt{HashMap} to \texttt{TreeSet} in another part of the code, then the import declaration for \texttt{java.util.HashMap} would be reported as deleted, as the matching import declaration on the right-side is ambiguous.
Finally, RefactoringMiner can match a set of \textit{single-type-import} declarations that are replaced with a \textit{type-import-on-demand} declaration and vice-versa by introducing multi-mappings, as shown in Figure \ref{fig:imports-with-grouping}. The grouping/un-grouping of import declarations is based on their common package prefix (e.g., \texttt{java.util.} in the example of Figure \ref{fig:imports-with-grouping}).

\subsection{AST diff generation}
\label{sec:ast-diff-generation}
\begin{figure}[ht]
	\centering
	\vspace{-3mm}
	\includegraphics[width=\linewidth]{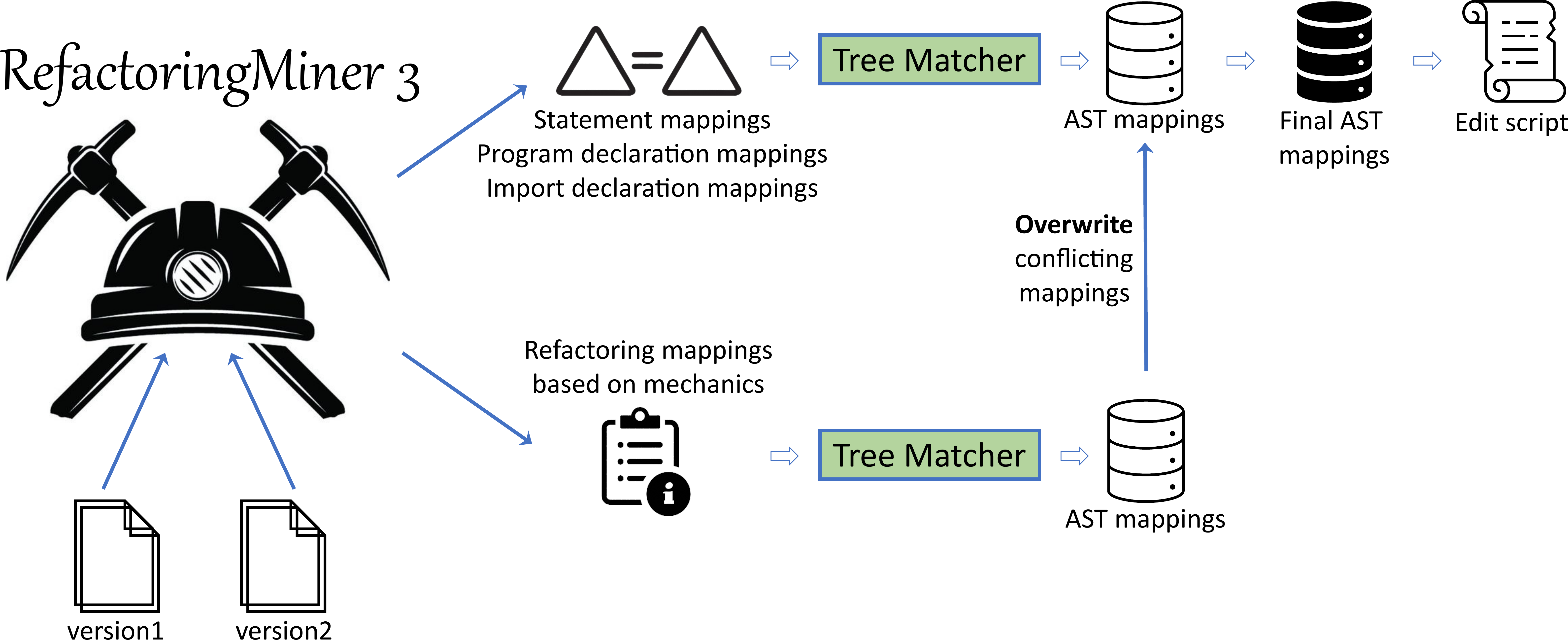}
	\vspace{-7mm}
	\caption{AST diff generation process.}
	\label{fig:ASTdiff-generation}
\end{figure}

As shown in Figure~\ref{fig:ASTdiff-generation}, we utilize the following information provided by RefactoringMiner to generate the AST diff:
\begin{itemize}[leftmargin=*]
	\vspace{-1mm}
	\item Pairs of matched declarations (i.e., import, class/enum, method, field, \texttt{enum}-constant declarations, initializer blocks)
	\item Statement-level mappings
	\item Refactoring-specific mappings by utilizing the mechanics of each supported refactoring type.
	\vspace{-1mm}
\end{itemize}

Table~\ref{table:refactorings} shows the kind(s) of AST node mappings that each refactoring type contributes to.
More specifically, a refactoring may contribute declaration mappings (i.e., a pair of matched declarations with changed signature), multi-mappings (i.e., duplicated statements being merged), intra-file-move mappings (i.e., statements moved within the same file), inter-file-move mappings (i.e., statements moved between different files), or sub-expression mappings (i.e., expressions matched within larger statements, refer to Section~\ref{sec:refactoring-mechanics} for more details).

\begin{table}[ht]
	\vspace{-3mm}
	\caption{Refactoring types and their contributions in AST node mappings}
	\vspace{-4mm}
	\centering
	\setlength{\tabcolsep}{5pt}
	\begin{threeparttable}
	\begin{tabular}{ll}
		\toprule
		\textbf{Refactoring Type} & \textbf{Kinds of AST node mappings} \\
		\toprule
		\textsc{Rename Class/Method/Field} & declaration\\
		\textsc{Move and Rename Class/Method/Field} & declaration, inter-file-move\\
		\textsc{Move/Pull Up/Push Down Method/Field} & inter-file-move\\
		\textsc{Merge/Split/Extract Class} & inter-file-move\\
		\textsc{Extract and Move Method}, \textsc{Move and Inline Method} & multi\tnote{$\dagger$} , inter-file-move, sub-expression\\
		\textsc{Extract/Inline Method} & multi\tnote{$\dagger$} , intra-file-move, sub-expression\\
		\textsc{Merge/Split Method}, \textsc{Parameterize Test} & multi\tnote{$\dagger$} , intra-file-move\\
		\textsc{Move Code Between Methods} & intra-file-move\\
		\textsc{Merge/Split/Invert Conditional} & sub-expression\\
		\textsc{Merge Catch} & multi, sub-expression\\
		\textsc{Replace Loop with Pipeline}, \textsc{Replace Pipeline with Loop} & sub-expression\\
		\textsc{Replace Anonymous with Class} & inter-file-move\\
		\textsc{Replace Anonymous with Lambda} & intra-file-move, sub-expression\\
		\textsc{Extract/Inline Variable/Field} & sub-expression\\
		\bottomrule
	\end{tabular}
	\begin{tablenotes}
		\item[$\dagger$] only if the refactoring involves duplicated fragments being merged, or single  fragment being duplicated
	\end{tablenotes}
	\end{threeparttable}
	\label{table:refactorings}
	\vspace{-5mm}
\end{table}

\subsubsection{Use of refactoring mechanics to infer sub-expression mappings}
\label{sec:refactoring-mechanics}
As explained before, RefactoringMiner generates mappings at the statement level.
Moreover, for some specific refactoring types listed in Table~\ref{table:refactorings}, it generates sub-expression mappings collected by utilizing the refactoring mechanics~\cite{Fowler:1999}:
\begin{itemize}[leftmargin=*]
	\item \header{Extract Method} This refactoring provides mappings between arguments passed to the extracted method call sites on the right side of the diff and sub-expressions within the statements being extracted on the left side of the diff.
	For example, in Figure~\ref{fig:language-clues-RMiner} there is a mapping between the sub-expression \texttt{workingDir.get()} from statement \texttt{builder.put(\textbf{workingDir.get()}, compilationDirectory);} on the left side of the diff with the argument passed to the extracted method call site \texttt{return getAllPathsWork(\textbf{workingDir.get()});} on the right side of the diff.
	This mapping is established because the parameter \texttt{workingDir} is used in the place of the original sub-expression in statement \texttt{builder.put(\textbf{workingDir}, compilationDirectory);} within the body of the extracted method, and RefactoringMiner matched the pair of \texttt{builder.put()} statements by replacing \texttt{\textbf{workingDir.get()}} with \texttt{\textbf{workingDir}}.
	By utilizing the \textsc{Extract Method} mechanics, we know that the argument \texttt{workingDir.get()} passed to the extracted method corresponds to parameter \texttt{workingDir}, and thus we can infer the aforementioned sub-expression mapping.
	\item \header{Inline Method} This refactoring provides mappings between arguments passed to the inlined method call sites on the left side of the diff and sub-expressions within the statements being inlined on the right side of the diff.
	\item \header{Extract Variable/Field} These refactorings provide mappings between the sub-expression(s) being extracted from statements on the left side of the diff and the initializer of the extracted variable/field declaration on the right side of the diff.
	For example, in Figure~\ref{fig:refactoring-aware-RMiner} there is a mapping between the sub-expression \texttt{"[" + (count++) + "]"} on the left side of the diff with the initializer of the newly introduced variable \texttt{itemKey} on the right side of the diff.
	This mapping is established because the variable \texttt{itemKey} is used in the place of the original sub-expression in statement \texttt{flatten(properties, Collections.singletonMap(\textbf{itemKey}, item), key);} on the right side of the diff, and RefactoringMiner matched the pair of \texttt{flatten()} statements by replacing \texttt{\textbf{"[" + (count++) + "]"}} with \texttt{\textbf{itemKey}}.
	It should be emphasized that the longer (in number of characters) the extracted expression is, the lower the textual similarity between the statement originally containing the expression and the statement referring to the extracted variable becomes.
	For this reason, we gave high priority to the rule 2(a) in Section~\ref{sec:leaf-mapping-sorting-criteria}, which checks whether a mapping ``becomes textually identical after undoing an overlapping extract/inline variable refactoring''.
	\item \header{Inline Variable/Field} These refactorings provide mappings between the sub-expression(s) being inlined to statements on the right side of the diff and the initializer of the inlined variable/field declaration on the left side of the diff.
	\item \header{Replace Loop with Pipeline} This refactoring appears in the second edition of Fowler's book~\cite{Fowler:2018} and is essentially a migration of traditional loops and conditionals to the Java Stream APIs~\cite{Fowler:pipelines}, as shown in Figure~\ref{fig:semantic-aware-RefactoringMiner}.
	RefactoringMiner generates mappings between the sub-expressions within the loops and conditionals on the left side of the diff and the lambda expressions appearing within Stream API calls on the right side of the diff.
	For example, in Figure~\ref{fig:semantic-aware-RefactoringMiner} there are four sub-expression mappings established based on the refactoring mechanics.
	The collection \texttt{externalResources} being iterated by the \texttt{enhanced-for} loop on the left side of the diff becomes the invoker of the \texttt{stream()} method invocation on the right side of the diff.
	The condition \texttt{!isResourceLocationInCache(resource.location)} of the \texttt{if} statement on the left side of the diff becomes the lambda expression body of the \texttt{filter()} method invocation on the right side of the diff.
	The statement \texttt{details.setProperty(resource.location, resource.contentHash);} within the body of the \texttt{if} statement on the left side of the diff becomes the lambda expression body of the \texttt{forEach()} method invocation on the right side of the diff.
	Finally, the \texttt{enhanced-for} loop formal parameter \texttt{resource} on the left side of the diff is matched twice (i.e., multi-mapping) with the lambda expression parameters of the \texttt{filter()} and \texttt{forEach()} method invocations on the right side of the diff.
	\item \header{Replace Pipeline with Loop} This is the reverse refactoring of the previously discussed one, and the sub-expression mappings are established in the reverse way.
	\item \header{Split/Merge Conditionals} These refactorings capture the cases where a conditional (typically \texttt{if} statement) is split to two or more conditionals, and cases where two or more conditionals are merged into a single one, respectively.
	Typically, these \texttt{if} statements have complex conditional expressions combined with \texttt{\&\&} and \texttt{||} operators, which are either split into separate conditional sub-expressions, or merged into a single conditional super-expression.
	Before the merge and after the split, typically these \texttt{if} statements are nested within each other when the operator is \texttt{\&\&}, and are sequentially executed when the operator is \texttt{||}.
	Both refactorings provide mappings between the split/merged conditional sub-expressions.
	\item \header{Merge Catch} This refactoring captures the cases where multiple \texttt{catch} blocks handling different exception types and having identical or similar exception handling code in their bodies, are merged into a single \texttt{catch} block that handles all exception types using the Union type (i.e., \texttt{catch(Exception1 | Exception2)}).
	The refactoring provides multi-mappings (many-to-one mappings) for the exception handling statements being merged, as well as sub-expression mappings for the exception types being moved from each individual \texttt{catch} to the Union type.
\end{itemize}

\subsubsection{From RefactoringMiner output to AST diff}
\label{sec:bottom-up-phase}

Algorithm~\ref{algo:ASTDiffWrapper} shows the process we follow to generate AST node mappings using as input the list of program element declaration pairs $\mathcal{P}$ and refactoring instances $\mathcal{R}$ detected by RefactoringMiner, as explained in Sections \ref{sec:background} and \ref{sec:changes-to-improve-statement-mapping-accuracy}.
Each pair $p \in \mathcal{P}$ of program element declarations may additionally include a set of mappings $p.mappings$.
In particular, if $p_1$ and $p_2$ are method declarations, $p.mappings$ contains the pairs of statements matched within the method bodies of $p_1$ and $p_2$.
If $p_1$ and $p_2$ are field or \texttt{enum} constant declarations $p.mappings$ contains the pairs of statements matched within the initializers of $p_1$ and $p_2$.
Each refactoring instance $r \in \mathcal{R}$ includes a set of mappings $r.mappings$, as explained in Table~\ref{table:refactorings}.

Lines 2-7 generate the initial AST node mappings in mapping set $\mathcal{M}$ from the list of program element declaration pairs $\mathcal{P}$.
Lines 8-12 generate AST node mappings in mapping set $\mathcal{A}$ from the list of refactorings $\mathcal{R}$.
Finally, lines 13-16 examine if there exists a mapping in $\mathcal{A}$ that has a conflict with a mapping in $\mathcal{M}$.
Two mappings are considered conflicting, if they have the same left AST subtree, but have a different right AST subtree, and vice versa.
In such a case, the conflicting mapping is removed from $\mathcal{M}$ and is replaced with the mapping from $\mathcal{A}$.
If there is no conflict, the mapping from $\mathcal{A}$ is added to $\mathcal{M}$, anyways.
In this way, we give priority to all mappings established from refactorings, as their correctness is guaranteed by the refactoring mechanics.
In other words, this last step makes our algorithm refactoring-aware.

\begin{algorithm}[H]
	\label{algo:ASTDiffWrapper}
	\caption{Generate AST node mappings from RefactoringMiner output}
	\DontPrintSemicolon
	\SetAlgoLined
	\SetKwInOut{Input}{Input}
	\SetKwInOut{Output}{Output}
	
	\Input{List of Refactorings $\mathcal{R}$, List of pairs of mapped program elements $\mathcal{P}$}
	\Output{Mapping Set $\mathcal{M}$ in which every item is a pair of  subtrees}
	
	$\mathcal{M} \leftarrow \{\}\;$ $\mathcal{A} \leftarrow \{\}$
	\\
	\ForEach{$p=(p_1, p_2)$ in $\mathcal{P}$}{    
		$\mathcal{M} \gets \mathcal{M} \cup \{(p_1, p_2)\}$\\
		\ForEach{$(m_1, m_2)$ in $p.mappings$}{
			$\mathcal{M} \gets \mathcal{M} \cup TreeMatcher(m_1, m_2)$\;
		}
	}
	\ForEach{r in $\mathcal{R}$} { 
		\ForEach{$(m_1, m_2)$ in $r.mappings$ \label{comment_line}}
		{
			$\mathcal{A} \gets \mathcal{A} \cup TreeMatcher(m_1, m_2)$
		}
	}
	
	\ForEach{$(m_1, m_2)$ in $\mathcal{A}$}{
		$\mathcal{M} \gets \mathcal{M} \setminus \{ (x, y) \in \mathcal{M} \mid x = m_1 \lor y = m_2\}$\\
		$\mathcal{M} \gets \mathcal{M} \cup \{(m_1, m_2)\}$\;
	}
\end{algorithm}
Algorithm~\ref{algo:ASTDiffWrapper} calls $TreeMatcher$ (Algorithm~\ref{algo:SubTreeMatcher}) in lines 5 and 10.
Algorithm~\ref{algo:SubTreeMatcher} takes as input a pair of statements or sub-expressions, as provided by RefactoringMiner, and generates fine-grained AST node mappings by matching the leaf nodes within the input statements.
In line 1, we execute the GumTree 3.0 (simple) algorithm to obtain the initial set of fine-grained mappings in $\mathcal{M}$.
By default, GumTree employs a threshold for the hyperparameter $minHeight$
(i.e., length of the longest path from one leaf to the root of the subtree) equal to 2.
This is to avoid matching remaining leaf expressions with height 1 (e.g., \texttt{SimpleName} nodes), which coincidentally have the same value. 
However, since we are applying the algorithm on a single pair of statements (i.e., subtrees with rather shallow depth), we configure a lower threshold for $minHeight$ equal to 1,
and modify the top-down phase of GumTree (see Section~\ref{sec:related-work} for more details)
by adding a condition to prevent the matching of \texttt{SimpleName} nodes with parents having different AST types (i.e., avoid semantically incompatible mappings for \texttt{SimpleName} nodes).

Lines 2-12 optimize the mappings generated by GumTree, as
the lower threshold for $minHeight$ may result in pairs of matched leaf nodes (i.e., \texttt{SimpleName} and \texttt{InfixOperator}) with identical labels, whose parent nodes have not been matched accordingly.
Given a pair of matched leaf nodes $(t_a, t_b)$, if $\mathcal{M}$ does not contain a mapping $(p_a, p_b)$ for the actual parents of $t_a$ and $t_b$ (line 7), then all mismatches for the parent nodes $p_a$ and $p_b$ are removed from $\mathcal{M}$ (line 8), and the pair of actual parent nodes $(p_a, p_b)$ is added to $\mathcal{M}$ (line 9).
Table~\ref{table:sub-expression-matching} includes some illustrative examples involving method invocations and infix expressions, which visually demonstrate the effect of the aforementioned mapping optimization.

\begin{algorithm}[H]
	\label{algo:SubTreeMatcher}
	\caption{Tree Matcher}
	\DontPrintSemicolon
	\SetAlgoLined
	\SetKwInOut{Input}{Input}
	\SetKwInOut{Output}{Output}
	\Input{Two subtrees $st_1$ and $st_2$}
	\Output{Mapping set $\mathcal{M}$}
	
	$\mathcal{M} \gets GumTreeSimple(st_1, st_2)$ \tcp*{Running GumTree 3.0 Simple with minHeight = 1} 
	\ForEach{$(t_a, t_b)$ in $\mathcal{M}$}{ 
		{
			$p_a \gets parent(t_a)$, 
			$p_b \gets parent(t_b)$\\
			$c1 \gets type(t_a) = \texttt{SimpleName} \land type(t_b) = \texttt{SimpleName} \land type(p_a) = \texttt{Invocation} \land type(p_b) = \texttt{Invocation}$\\
			$c2 \gets type(t_a) = \texttt{Operator} \land type(t_b) = \texttt{Operator} \land type(p_a) = \texttt{InfixExpr} \land type(p_b) = \texttt{InfixExpr}$\\
			\If{$c1 \lor c2$}	
			{ 
				\If{$\neg \exists (m_1, m_2) \in \mathcal{M} \mid m_1 = p_a  \land m_2 = p_b$}
				{
					$\mathcal{M} \gets \mathcal{M} \setminus \{ (m_1, m_2) \in \mathcal{M} \mid m_1 = p_a \lor  m_2 = p_b \}$\\
					$\mathcal{M} \gets \mathcal{M} \cup \{(p_a, p_b)\}$\;
				}
			}
		}
	}
\end{algorithm}
\begin{table}[ht]
	\caption{Improvements in sub-expression matching.\\(RM = RefactoringMiner, GTG = GumTree Greedy, GTS = GumTree Simple, GT = GumTree Simple or Greedy)}
	\vspace{-4mm}
	\centering
	\setlength{\tabcolsep}{1pt}
	\renewcommand{\arraystretch}{0.5}
	\begin{tabular}{|c|c|}
		\hline
		\textbf{Tool} & \textbf{Diff}\\
		\noalign{\hrule height 1pt}
		& Example 1: nested invocations at the argument level\\\hline
		RM & \includegraphics[width=0.92\linewidth]{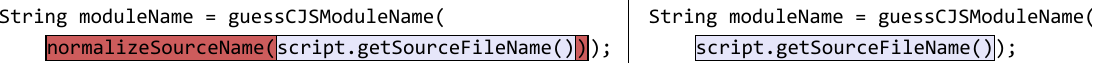}\\
		\hline
		GT & \includegraphics[width=0.92\linewidth]{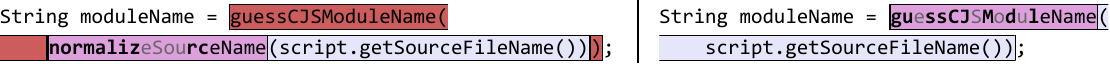}\\
		\noalign{\hrule height 1pt}
		& Example 2: nested invocations at the invoker level\\\hline
		RM & \includegraphics[width=0.92\linewidth]{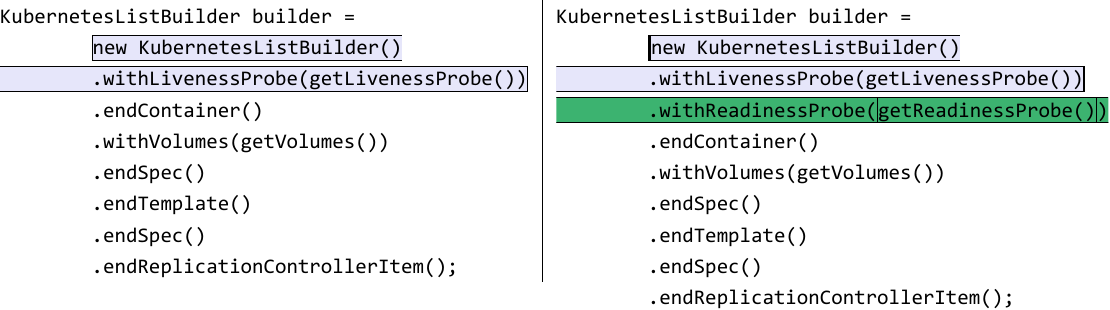}\\
		\hline
		GTS & \includegraphics[width=0.92\linewidth]{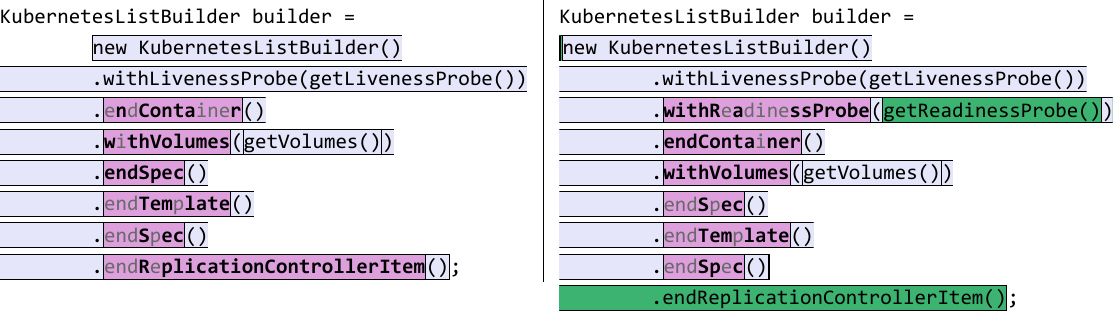}\\
		\hline
		GTG & \includegraphics[width=0.92\linewidth]{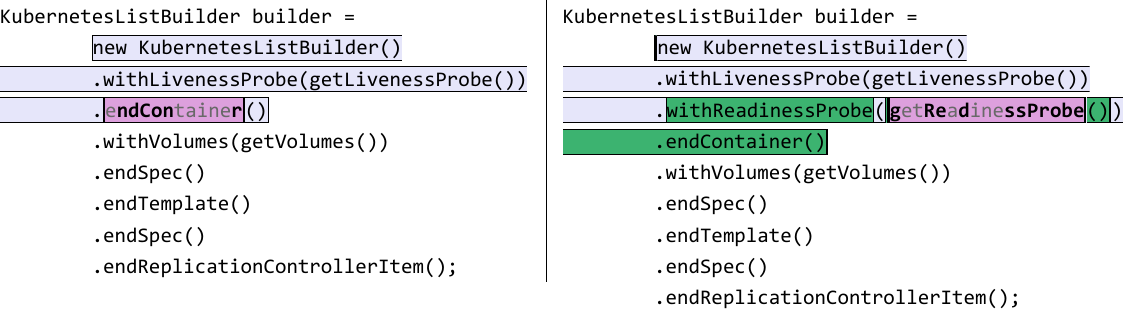}\\
		\noalign{\hrule height 1pt}
		& Example 3: nested infix expressions\\\hline
		RM & \includegraphics[width=0.92\linewidth]{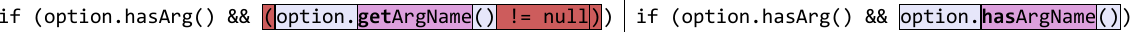}\\
		\hline
		GT & \includegraphics[width=0.92\linewidth]{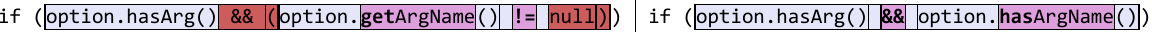}\\
		\noalign{\hrule height 1pt}
	\end{tabular}
	\label{table:sub-expression-matching}
	\vspace{-5mm}
\end{table}

\subsection{How our approach addresses the limitations of AST diff tools}
\label{sec:how-we-address-limitations}
\begin{enumerate}[leftmargin=*]
	\item \header{Lack of support for multi-mappings}
	Although Algorithm \ref{algo:SubTreeMatcher} does not generate multi-mappings by itself, Algorithm \ref{algo:ASTDiffWrapper} aggregates the mappings coming from different refactoring instances and program element declaration pairs (as detected by RefactoringMiner) without discarding new mappings for already mapped subtrees.
	\item \header{Incorrect matching of program declarations}
	RefactoringMiner is utilizing the program structure, the program element declaration types, and language clues (e.g., method calls, method signatures, field references) to reason about and guide the matching process on specific parts of the ASTs, instead of trying to randomly match code fragments between the left and right ASTs based on their similarity.
	\item \header{Lack of refactoring awareness}
	Algorithm \ref{algo:ASTDiffWrapper} (lines 13-16) prioritizes mappings established from refactorings, as their correctness is guaranteed by the refactoring mechanics.
	\item \header{Matching semantically incompatible AST nodes}
	Algorithm \ref{algo:SubTreeMatcher} modifies the top-down phase of GumTree by adding a condition to prevent the matching of \texttt{SimpleName} nodes with parents having different AST types.
	Moreover, RefactoringMiner is guiding the matching process towards semantically compatible mappings by avoiding to match subtrees from program element declarations of a different kind. 
	\item \header{Lack of diff support at the commit level} RefactoringMiner detects refactorings that contribute inter-file-move mappings (Table \ref{table:refactorings}).
\end{enumerate}
\section{Benchmark}
In this section, we describe the process we followed to create a publicly available benchmark \cite{DiffBenchmark} of AST node mappings.
\subsection{Datasets}
\label{sec:datasets}
We relied on two popular datasets, namely Defects4J \cite{Defects4J}, which includes \totalBugsInDefects bug fixes from 17 open-source projects, and Refactoring Oracle \cite{RefactoringMiner2}, which includes over 11K refactorings and code changes found in 546 commits from 187 open-source projects.
Both datasets have been extensively validated over a long period of time.
Defects4J has 10 releases between 2015-2020 and the Refactoring Oracle has been constantly updated between 2018-2023 by incorporating results from more tools and supporting more refactoring types.
Both datasets have become the de facto standard for evaluating automated program repair and refactoring mining tools, respectively.
The reason we selected these two datasets to build our benchmark is because the characteristics of the changes they include pose different levels of challenge for AST diff tools.
Bug fixing commits are a low barrier for assessing the accuracy AST diff tools, as most changes are small (a few lines of code), local (typically located within a single method) and rather simple and straightforward.
On the other hand, refactoring commits are more challenging for AST diff tools, as most changes are big (many lines of code), dispersed (located within different methods and files) and rather complex (overlapping changes and refactorings).
Table \ref{table:benchmarks} shows the characteristics of the two datasets. We can clearly observe that the Refactoring dataset has a higher percentage of cases with multi-mappings, 8 times higher average churn and a higher percentage of challenging cases compared to Defects4J (as explained in Section~\ref{sec:benchmark-structure} \textit{challenge} expresses how challenging the construction of the ground-truth diff is).

\begin{table}[ht]
    \caption{Benchmark characteristics}
    \vspace{-4mm}
    \centering
    \setlength{\tabcolsep}{5pt}
    \begin{tabular}{lrr}
    \toprule
     \textbf{Property} & \textbf{Defects4J} & \textbf{Refactoring Oracle} \\
     \toprule
		\# cases & 800 & 188\\
		\# total number of diff files & 996 & 287 \\
		\# cases with multi-mapping (\%) & 25 (3.1\%) & 52 (27.6\%) \\
		\# cases with refactoring (\%) & 107 (13.4\%) & 188 (100\%) \\
		average churn & 0.011 & 0.086 \\
		\# \textit{low} challenge cases (\%) & 732 (91.5\%) & 70 (37.2\%) \\
		\# \textit{medium} challenge cases (\%) & 65 (8.1\%) & 87 (46.2\%) \\
		\# \textit{high} challenge cases (\%) & 3 (0.4\%) & 31 (16.4\%) \\
     \bottomrule
    \end{tabular}
    \label{table:benchmarks}
\end{table}
Since generating a benchmark of AST node mappings is a rather tedious and challenging task, and both datasets are large, we included from the refactoring oracle only the commits that have 1 or 2 modified files, and from Defects4J 800 out of the \totalBugsInDefects total bug fixes.

\subsection{Tools}
\label{sec:tools}
We did our best effort to include as many AST diff tools as possible in order to do a comprehensive analysis of their strengths and weaknesses.
The criteria we set to select an AST diff tool are the following:
\begin{itemize}[leftmargin=*]
	\item The tool should be publicly available.
	\item The tool should be executable and provide an API, so that it can programatically executed.
	\item The tool should be able to report at minimum AST mappings at statement-level.
\end{itemize}

The tools that were considered and eventually selected based on our criteria are shown in Table \ref{table:selected-tools}. It should be noted that CLDiff~\cite{ClDiff} was not included in our experiments, because it uses GumTree (without any changes in its implementation) to obtain AST mappings and edit actions, and generates concise code differences by grouping fine-grained code differences at or above the statement level.
Since we already compare with many different versions of GumTree, adding CLDiff in the list of evaluated tools is deemed redundant.
Moreover, the higher-level groups of code differences generated by CLDiff are not compatible with the fine-grained differences generated by the other tools, and thus a direct comparison is not feasible.
\begin{table}[ht]
    \vspace{-2mm}
    \caption{Considered and selected AST diff tools}
    \vspace{-4mm}
    \centering
    \setlength{\tabcolsep}{2pt}
    \begin{tabular}{lccc}
    \toprule
     \textbf{Tool} & \textbf{Availability} & \textbf{API} & \textbf{AST mappings} \\
     \toprule
     RefactoringMiner 3.0 & \href{https://github.com/tsantalis/RefactoringMiner}{\tick} & \href{https://github.com/tsantalis/RefactoringMiner#ast-diff-api-usage-guidelines}{\tick} & \tick \\\hline
     GumTree 3.0 (greedy) \cite{martinez2022hyperparameter} & \href{https://github.com/GumTreeDiff/gumtree}{\tick} & \href{https://github.com/GumTreeDiff/gumtree/wiki/GumTree-API}{\tick} & \tick \\\hline
     GumTree 3.0 (simple) \cite{martinez2022hyperparameter, GumtreeSimple:2024} & \href{https://github.com/GumTreeDiff/gumtree}{\tick} & \href{https://github.com/GumTreeDiff/gumtree/wiki/GumTree-API}{\tick} & \tick \\\hline
     GumTree 2.1.0  \cite{gumtree} & \href{https://github.com/GumTreeDiff/gumtree/releases/tag/v2.1.0}{\tick} & \href{https://github.com/GumTreeDiff/gumtree/wiki/GumTree-API}{\tick} & \tick \\\hline
     IJM \cite{IJM} & \href{https://github.com/VeitFrick/IJM}{\tick} & \href{https://github.com/VeitFrick/IJM#running}{\tick} & \tick \\\hline
     MTDiff \cite{MTDiff} & \tick & \tick & \tick \\\hline
     srcDiff \cite{10.1002/smr.2226} & \notick & \notick & \tick \\
     \bottomrule
    \end{tabular}
    \label{table:selected-tools}
    \vspace{-3mm}
\end{table}

\begin{figure}[ht]
	\centering
	\vspace{-3mm}
	\subfloat[AST generator \texttt{gumtree.gen.jdt-2.1.0}]{\label{fig:2.1} \includegraphics[width=\linewidth]{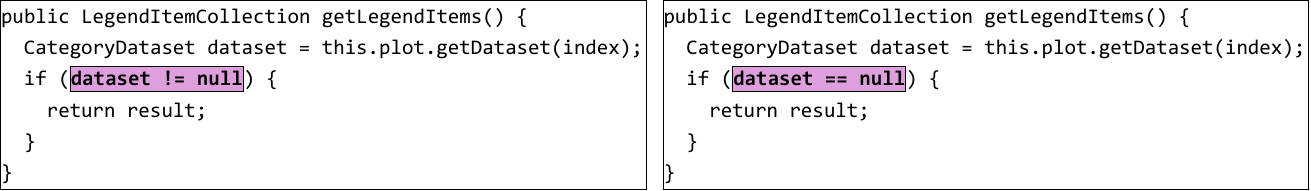}}
	
	\vspace{-3mm}
	\subfloat[AST generator \texttt{gumtree.gen.jdt-3.0.0}]{\label{fig:3.0}
		\includegraphics[width=\linewidth]{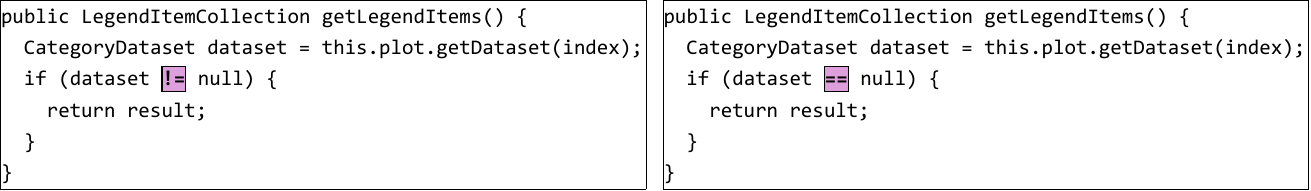}}
	\vspace{-4mm}
	\caption{AST diff with two different AST generators.}
	\vspace{-5mm}
\end{figure}
The implementation of MTDiff is included in the IJM repository~\cite{IJM-repository}. Moreover, the API of GumTree changed significantly in version 3.0, and thus it is no longer compatible with the 2.1 version.
To avoid problems with class name clashes, 
we had to create a shaded version of IJM by moving its class files into a folder named ``\texttt{shaded}'' and referencing the classes corresponding to IJM, MTDiff, GumTree 2.1 by adding the ``\texttt{shaded.}'' prefix in their qualified name.
RefactoringMiner is 100\% compatible with GumTree 3.0, as it generates the AST diff in objects extending the GumTree 3.0 APIs.
Regarding the AST structure, RefactoringMiner and GumTree 3.0 utilize the library \texttt{gumtree.gen.jdt-3.0.0}, which wraps around the Eclipse JDT Generator to generate ASTs. In contrast, the other tools rely on an older version of this tree generation library, \texttt{gumtree.gen.jdt-2.1.0}.
Consequently, there are differences in the generated AST structures among these tools. 
For instance, in version 2.1, the \texttt{InfixExpression} node includes the operator as a label, while in version 3.0, there is an additional child node to store the operator.
The implications of these differences are illustrated in Figures \ref{fig:2.1} and \ref{fig:3.0}. Clearly, the tree structure in version 2.1 hinders the algorithm's ability to generate fine-grained edit actions.
Due to these differences in the AST structure produced by the generators 2.1 and 3.0, we can compute the precision and recall of the tools compatible with GumTree 2.1 only at statement level.

\subsection{Structure and format}
\label{sec:benchmark-structure}
There are two possible ways to store the output of AST diff tools. The first is in the form of the AST node mappings found between two ASTs, and the second is in the form of a sequence of edit actions (e.g., insert, delete, update, move) that transforms one AST into the other. 
We decided to store the ground truth in the form of AST node mappings,
as not all tools support the same kinds of edit actions, and thus their edit scripts are not 100\% compatible. For example, Higo et al. \cite{10.5555/3155562.3155630} introduced a new kind of edit action, named \textit{copy-and-paste}, which is not supported by other AST diff tools. Moreover, in our work we introduced new kinds of edit actions to support multi-mappings and multi-moves, which are not supported by other AST diff tools.
We believe in this way our benchmark will be more usable in the future and facilitate future research, regardless of the new edit actions researchers might come up with.

\begin{figure}[ht]
	\centering
	\includegraphics[width=0.9\linewidth]{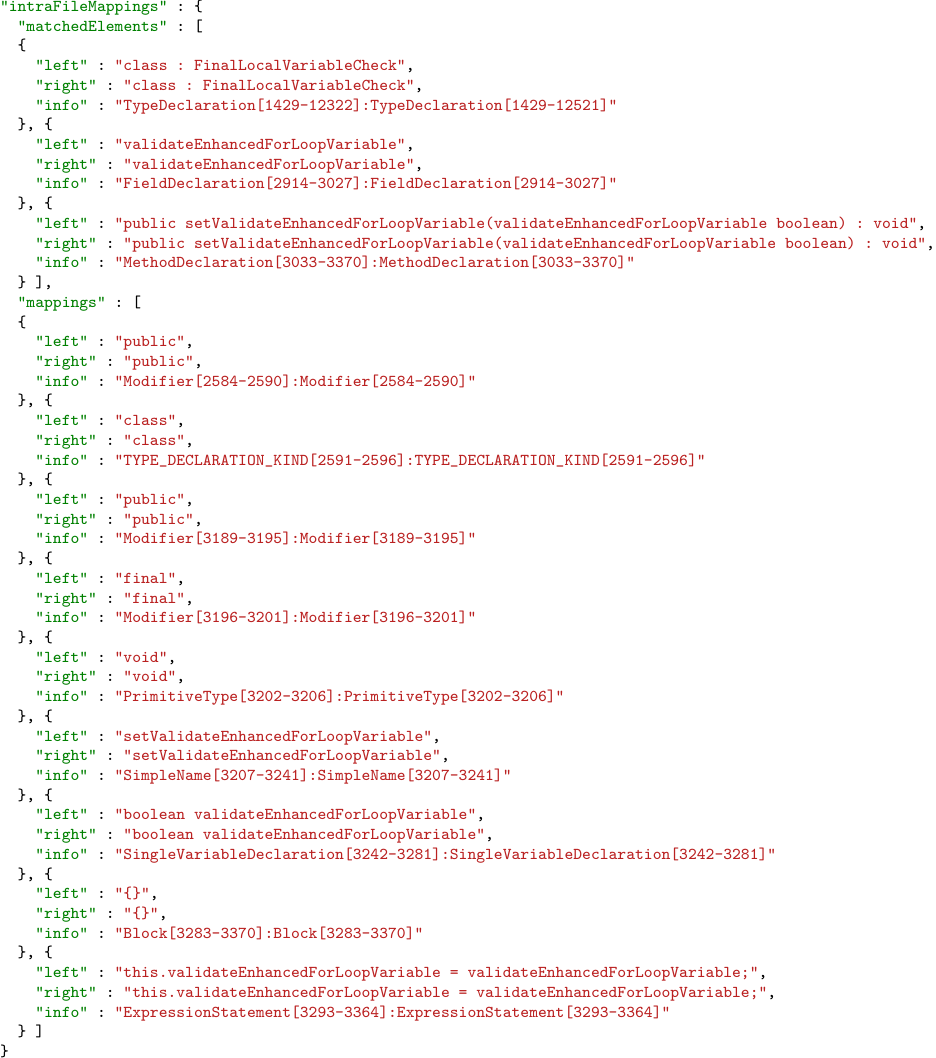}
	\vspace{-2mm}
	\caption{A sample of human-readable AST diff with intra-file mappings}
	\label{listing:human-readable-diff-intraFileMappings}
	\vspace{-4mm}
\end{figure}


Our benchmark is structured as follows. For each commit from the Refactoring dataset, we create a directory named as \texttt{organization/project/commitSHA-1}. For each bug from the Defects4J dataset, we create a directory named as \texttt{ProjectName-BugId}.
Within each directory, there is a separate json file storing mappings between program elements and AST nodes for each modified compilation unit (i.e., Java file) in the diff. Each json file has a mandatory section for \textit{intra-file mappings} (i.e., mappings within the file) and an optional section for \textit{inter-file mappings} (i.e., mappings in different files) in case some code is moved from/to this file.

As shown in Figure \ref{listing:human-readable-diff-intraFileMappings}, \textit{intra-file mappings} section includes two sub-sections.
In the first section, named ``matchedElements'', we store pairs of matched program element declarations, such as type, method, field and enum declarations.
In the second section, named ``mappings'', we store pairs of AST statements or expressions.
Each entry in the json file has three properties, namely ``left'', ``right'' and ``info''.
The properties ``left'' and ``right'' include a string representation of the matched element on the left and right side of the diff, respectively.
If the element is a declaration, the string representation is a short signature of the declaration, including its name, and parameters (only for method declarations).
If the element is an AST statement/expression, the string representation is the actual code corresponding to the statement/expression in pretty-printed format.
The property ``info'' encodes AST type and location information in the format
\texttt{LeftASTType[LeftStartOffset-LeftEndOffset]:RightASTType[RightStartOffset-RightEndOffset]}.
We decided to use this format for two main reasons. First, it is human-readable, which makes easy to understand and manually edit the file if needed.
Second, it facilitates the evaluation framework we propose for assessing the quality of the diff generated by each tool, which covers different quality dimensions, such as statement mapping accuracy, program element mapping accuracy and semantically incompatible mappings.
For instance, based on the ``matchedElements'' section, we can compute the program element mapping accuracy by comparing the pairs of program elements listed in this section with the pairs of program elements matched by an AST diff tool. 
Based on the ``mappings'' section, we can compute the statement mapping accuracy by comparing the mappings listed in this section with the mappings generated by an AST diff tool.
Finally, based on the ``info'' property, we can compute the semantically incompatible mappings by looking for mappings generated by an AST diff tool that match only the left or right side of a ground-truth mapping.

The \textit{inter-file mappings} section
has the same format as the \textit{intra-file mappings} section, with the only difference being that the mappings are grouped based on the origin or destination file, as it is possible to have code moved from/to multiple different files.
Finally, we include the following metadata at the commit level (Refactoring dataset) and bug level (Defects4J dataset) to provide an overview of the commit changes, and facilitate future research.
\begin{itemize}[leftmargin=*]
	\item \textbf{Multi-mappings}: This is a boolean variable (\textit{true}/\textit{false}) to inform about the presence of AST node multi-mappings in the diff (i.e., one-to-many, many-to-one, many-to-many). This is a novel feature in AST diff supported only by RefactoringMiner.
	\item \textbf{Left-side churn}: This is a ratio variable defined as $\frac{\text{removed lines} + \text{modified lines}}{\text{total lines}}$ in the left-side files of the diff.
	\item \textbf{Right-side churn}: This is a ratio variable defined as $\frac{\text{added lines} + \text{modified lines}}{\text{total lines}}$ in the right-side files of the diff.
	\item \textbf{Challenge level}: This a categorical variable that can take three possible values, namely \textit{low}, \textit{medium}, \textit{high} to express how challenging the construction of the ground-truth diff is, as determined by the time taken to construct it.
	The construction time depends on the number and complexity of the changes to be inspected, and the number of missing or incorrect mappings generated by the tools, which should be manually fixed in the ground-truth diff.
	\item \textbf{Comments}: This is free-form text explaining the reason the diff is challenging, along with some arguments justifying some debatable mappings.
\end{itemize}

\subsection{Construction process}
\begin{figure}[ht]
	\vspace{-3mm}
	\centering
	\includegraphics[width=\linewidth]{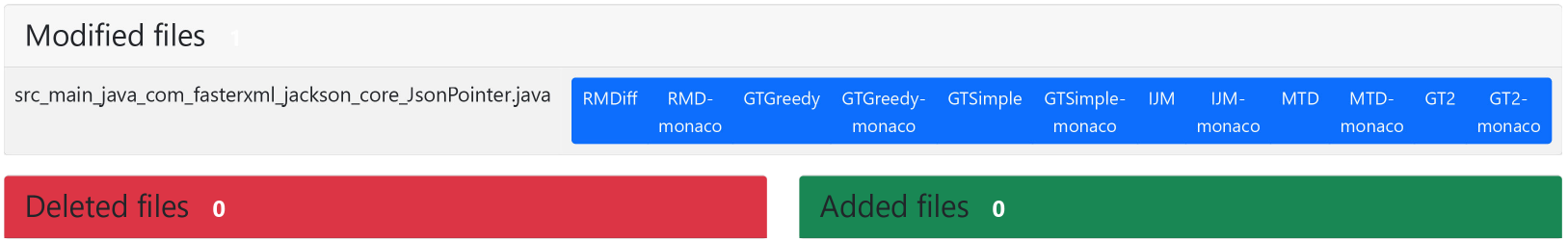}
	\vspace{-7mm}
	\caption{AST diff output for all selected AST diff tools (Defects4J dataset, \texttt{JacksonCore} project, bugID \texttt{6}).}
	\label{fig:diff-compare}
	\vspace{-2mm}
\end{figure}

The process we followed to construct our benchmark for assessing the accuracy of AST diff tools is as follows:
\header{Step 1 - Diff tools execution} We execute all selected tools listed in Table \ref{table:selected-tools}.
We extended the \textit{directory diff viewer} provided by GumTree to include the diff results of all selected AST diff tools visualized with the \textit{classic} and \textit{monaco file diff viewers}, as shown in Figure \ref{fig:diff-compare}.
Moreover, we designed factory APIs that can process input from local files (suitable for the Defects4J dataset), as well as GitHub commits (suitable for the Refactoring dataset).

\header{Step 2 - Diff inspection and validation}
We inspect the diff generated by each tool following the diff assessment criteria listed in Section \ref{sec:diff-assessment-criteria} to assess the correctness of the AST mappings in the diff.
If there are no debatable mappings, only the first author of the paper is involved in the inspection.
If there is at least one debatable mapping that the first author is not sure about, both paper authors are involved in the inspection and try to reach a consensus using arguments justifying their opinions. In the vast majority of the examined cases with debatable mappings the two co-authors were able to reach a consensus. In the rare cases that this was not possible, the co-authors asked the opinion of the current members of the Refactoring research group at Concordia University (listed in the Acknowledgments Section), and the debate was resolved with a majority vote.
As a matter of fact, there were 24 debatable cases in the Defects4J dataset and 9 debatable cases in the Refactoring dataset, which constitute around 3\% of the examined cases.
Overall, the inspection and validation process lasted 6 person-months.
\header{Step 3 - Ground truth generation}
There are three possible outcomes after the inspection and validation of a diff:
\begin{itemize}[leftmargin=*]
	\item At least one of the tools generates the optimal diff with all mappings assessed as correct and without any missing mappings. In that case, we use one of these tools to generate the ground-truth in the human-readable diff format.
	\item A tool generates a part of the optimal diff, while another tool generates the remaining part of the optimal diff. In that case, we combine the optimal diff parts from multiple tools to generate the ground-truth in the human-readable diff format.
	\item There are some AST mappings that all tools matched them incorrectly or missed them. In that case, we programatically add the correct mappings or missing mappings to generate the ground-truth in the human-readable diff format.
\end{itemize}

\subsection{Diff assessment criteria}
\label{sec:diff-assessment-criteria}
\begin{figure}[ht]
	\centering
	\includegraphics[width=\linewidth]{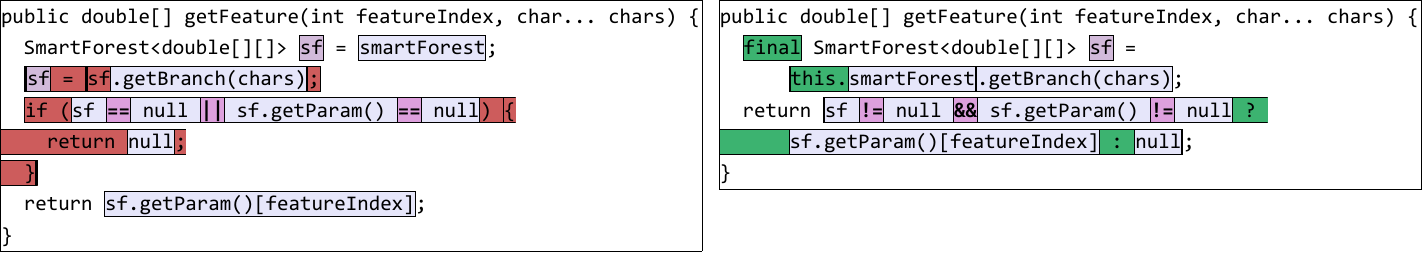}
	\vspace{-7mm}
	\caption{AST diff generated by RefactoringMiner for commit \url{https://github.com/NLPchina/ansj_seg/commit/913704e}.}
	\label{fig:control-flow-restructuring}
\end{figure}

\begin{figure}[ht]
	\centering
	\includegraphics[width=\linewidth]{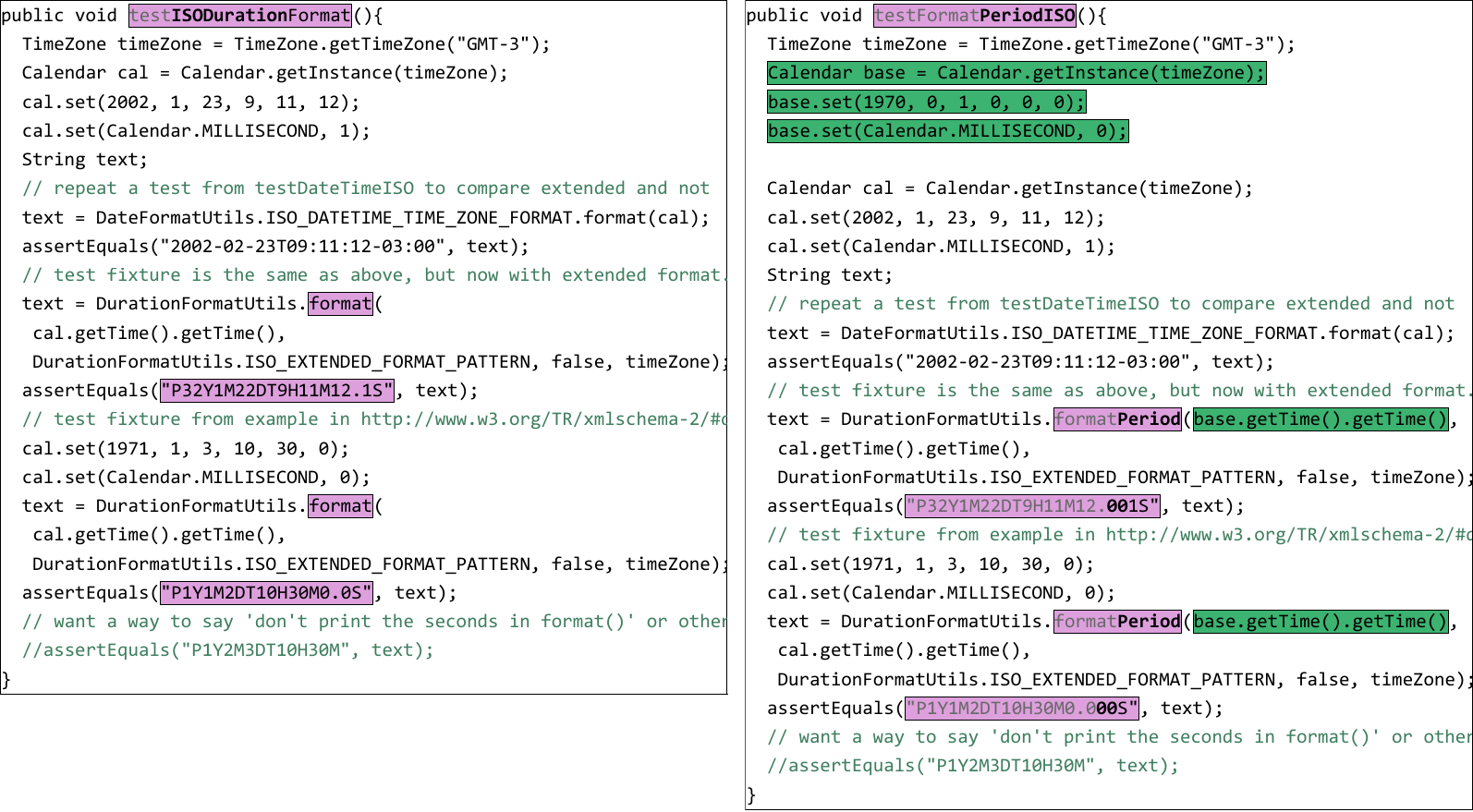}
	\vspace{-7mm}
	\caption{AST diff generated by RefactoringMiner for commit \url{https://github.com/apache/commons-lang/commit/5111ae7}.}
	\label{fig:inline-comments}
	\vspace{-3mm}
\end{figure}

When inspecting the AST diff generated by the tools to construct the ground-truth diff, we check the following criteria (properties) to assess the correctness of the mappings:
\begin{enumerate}[leftmargin=*]
	\item \textbf{Statement continuity}: Consecutive mapped statements should remain in the same relative execution order. In many cases, there exist multiple identical statements (e.g., \texttt{break}, \texttt{continue}, \texttt{return}) within the same method. This introduces multiple potential matching candidates for approaches that generate mappings at statement level, such as RefactoringMiner. With the \textit{statement continuity} criterion, we check whether statements with multiple matching candidates have the same corresponding statements around them (i.e., before and after) on the left and right side of the diff.
	\item \textbf{Compliance with developer actions and intent}: The diff should comply with the changes the developer (i.e., commit author) actually performed. The intention of the developer is often expressed in the commit message or pull request discussions. By reading the commit message of the example discussed in Section \ref{sec:language-clues}, it becomes clear that the intention of the developer was not to rename method \texttt{getAllPaths} to \texttt{getAllPathsWork}, as expressed in the diff shown in Figure \ref{fig:language-clues-GTG}, but rather to enhance the functionality of the already existing method \texttt{getAllPaths} to ``cache the values returned by getAllPaths to avoid the multiple ImmutableBiMap objects being created for the same working directory'', as expressed in the diff shown in Figure \ref{fig:language-clues-RMiner} with the new functionality of \texttt{getAllPaths} being highlighted in green color and the original functionality being extracted into method \texttt{getAllPathsWork}.
	\item \textbf{Aesthetic appeal} or \textbf{Visual appeal} has been commonly used in computer graphics research as a subjective measure to evaluate the quality of a visualization \cite{https://doi.org/10.1111/cgf.14521, https://doi.org/10.1111/cgf.14836}. An AST diff can be considered as a visualization of the changes that took place on the AST of a piece of source code. Thus, a good quality diff should be aesthetically appealing, intuitive, natural, clear and easy-to-understand.
	Incorrect mappings introduce visual noise or clutter in the diff visualization, and make it look unappealing, counter-intuitive, unnatural, ambiguous and confusing, as in the examples shown in Figures \ref{fig:multi-mappings-GTG}, \ref{fig:semantic-aware-GTG} and \ref{fig:commit-GTG}.
	\item \textbf{Consistent control flow restructuring}: In the case of control flow restructuring (e.g., inverted conditionals, deleted/added conditionals, change in the nesting level or scope of conditionals, merged/split conditionals) the diff should reflect the change as being behavior-preserving. For example, in the diff shown in Figure \ref{fig:control-flow-restructuring} an \texttt{if} statement is deleted and replaced with a conditional expression, but at the same time the condition of the \texttt{if} statement
	``\texttt{sf == null || sf.getParam() == null}''
	is inverted in the conditional expression as 
	``\texttt{sf != null \&\& sf.getParam() != null}''.
	Since, the condition is inverted, we see that the execution of the statements is inverted too. Initially, the body of the \texttt{if} statement included statement ``\texttt{return null;}'', which is now moved to the ``else'' part of the conditional expression ``\texttt{: null}''. On the other hand, the last statement of the method on the left side ``\texttt{return sf.getParam()[featureIndex];}'', which is executed only if the condition of the \texttt{if} statement is evaluated as \textit{false}, is moved to the ``then'' part of the conditional expression ``\texttt{? sf.getParam()[featureIndex]}''.
	The way the statements and sub-expressions are mapped in the diff preserves the program behavior after the control flow restructuring, and thus gives us the confidence to consider these mappings as correct.
	\item \textbf{Consistent identifier renaming}: The mappings should conform to consistent identifier renames. For example, method call renames within statement mappings should conform with the corresponding method declaration renames and field/variable reference renames within statement mappings should conform with the corresponding field/variable declaration renames.
	As discussed in Section \ref{sec:refactoring-awareness}, the diff shown in Figure \ref{fig:refactoring-aware-GTS} does not conform to \textit{consistent identifier renaming}, as the reference to variable \texttt{object} is matched with \texttt{itemKey}, while the declaration of variable \texttt{object} is renamed to \texttt{item}. On the other hand, the diff shown in Figure \ref{fig:refactoring-aware-RMiner} conforms to the \textit{consistent identifier renaming} criterion.
	\item \textbf{Consistent inline comment location}: In many cases, there exist similar statements that go through similar changes within the same method. 
	This introduces multiple potential matching candidates for approaches that generate mappings at statement level, such as RefactoringMiner. 
	Moreover, the statements around them might also change. Thus, the \textit{statement continuity} criterion is not helpful in that case. However, inline comments within the method might remain unchanged or be slightly changed.
	With the \textit{consistent inline comment location} criterion, we check whether statements with multiple matching candidates are surrounded by the same inline comments on the left and right side of the diff.
	For example, in the diff shown in Figure \ref{fig:inline-comments} there are two assignment statements for variable \texttt{text}, which are identical and change in exactly the same way on the right side of the diff. Moreover, the assertion statements that follow are also very similar and change as well. Therefore, there is some ambiguity about what is the right mapping for these statements. The way the statements are mapped in the diff makes them surrounded by the same unchanged inline comments, and thus gives us the confidence to consider these mappings as correct.
\end{enumerate}

\section{Evaluation}
Following the template defined by Briand et al.~\cite{Briand:GQM-templates} for the Goal-Question-Metric (GQM) paradigm, we define our study as:
We \textit{analyze} six AST diff tools for the \textit{purpose} of evaluating their accuracy \textit{with respect to} AST node mappings from the  \textit{viewpoint} of code reviewers and developers that want to understand the changes in a commit or pull request in the  \textit{context} of a benchmark including bug-fixing and refactoring commits from open-source projects.

More specifically, we investigate the following research questions:
\begin{enumerate}[wide,label=\bfseries RQ\arabic*., leftmargin=*,labelwidth=!, labelindent=0pt]
	\item What is the accuracy of each tool in multi-mappings?
	\item How many semantically incompatible mappings are generated by each tool?
	\item What is the accuracy of each tool in matching program elements (i.e., method, field declarations)?
	\item How do refactorings affect the accuracy of each tool?
	\item How many mappings are missed or mismatched by each tool due to the lack of commit-level change analysis?
	\item What is the overall accuracy of each tool?
	\item What is the percentage of perfect diffs generated by each tool?
	\item What is the execution time of each tool?
	\item What is the effect of the changes presented in Section~\ref{sec:changes-to-improve-statement-mapping-accuracy} on the accuracy of RefactoringMiner 3.0 over the prior version 2.0?
\end{enumerate}

In many of the subsections that follow we provide precision, recall, and F-score, which are computed as follows:
\begin{definition}
\vspace{-2mm}
Given the set of AST node mappings listed in the benchmark $S_B$ and the set of AST node mappings generated by a tool $S_T$, $\text{True Positives} = S_B \cap S_T$,
$\text{False Positives} = S_T \setminus S_B$ and $\text{False Negatives} = S_B \setminus S_T$.
Therefore, the precision and recall of a tool $S_T$ is computed as
$\text{Precision} = \frac{|TP|}{|TP| + |FP|}$ $\text{Recall} = \frac{|TP|}{|TP| + |FN|}$
$\text{F-score} = \frac{2\times|TP|}{2\times|TP| + |FP| + |FN|}$
\vspace{-2mm}
\end{definition}

Since all tools are able to match identical AST sub-trees with 100\% accuracy, including all AST node mappings in a file would artificially increase the computed precision and recall. As a result, we decided to exclude all AST node mappings nested under unchanged program elements from both $S_B$ and $S_T$ when computing precision and recall.

\subsection{RQ1: Multi-mappings}
\label{sec:multi-mappings}
\subsubsection{Experimental procedure}
For this research question, we focus our analysis on multi-mappings, i.e., AST nodes that have more than one corresponding nodes on the right side of the diff (one-to-many mappings), or the left side of the diff (many-to-one mappings).
More specifically, we filter all AST node mappings marked as multi-mappings in our benchmark $S_B$, and compute the accuracy of the AST diff tools on this particular subset of mappings.
RefactoringMiner is the only tool that supports multi-mappings. The other AST diff tools, typically match only one instance of the duplicated code, and the other duplication instances are either unmatched (i.e., appear as deleted/added code in the diff) or are mismatched with irrelevant code, as happened in the case shown in Figure~\ref{fig:multi-mappings-GTG}. With this research question, we want to highlight the negative impact that the lack of support for multi-mappings has on the accuracy of the tools.

\subsubsection{Results and discussion}

\begin{table}[ht]
	\vspace{-2mm}
	\caption{Precision and recall for multi-mappings in both datasets}
	\vspace{-4mm}
	\centering
	\fontsize{7pt}{10pt}\selectfont
	\setlength\tabcolsep{2px}
		\begin{tabular}{r l r l | r l r l | r l r l | r l r l | r l r l | r l r l}
			\toprule
			\multicolumn{4}{c}{\textbf{RM 3.0}} & 
			\multicolumn{4}{c}{\textbf{GT 3.0 greedy}} & 
			\multicolumn{4}{c}{\textbf{GT 3.0 simple}} &
			\multicolumn{4}{c}{\textbf{GT 2.1.0}} &
			\multicolumn{4}{c}{\textbf{IJM}} &
			\multicolumn{4}{c}{\textbf{MTDiff}} \\
			\multicolumn{2}{c}{\textbf{Precision}} &
			\multicolumn{2}{c}{\textbf{Recall}} &
			\multicolumn{2}{c}{\textbf{Precision}} &
			\multicolumn{2}{c}{\textbf{Recall}} &
			\multicolumn{2}{c}{\textbf{Precision}} &
			\multicolumn{2}{c}{\textbf{Recall}} &
			\multicolumn{2}{c}{\textbf{Precision}} &
			\multicolumn{2}{c}{\textbf{Recall}} &
			\multicolumn{2}{c}{\textbf{Precision}} &
			\multicolumn{2}{c}{\textbf{Recall}} &
			\multicolumn{2}{c}{\textbf{Precision}} &
			\multicolumn{2}{c}{\textbf{Recall}} \\ \midrule
			\sbar{1410}{4}{23} & \sbar{149}{168}{1284} & \sbar{147}{101}{1286} & \sbar{124}{156}{1309} & \sbar{37}{165}{1396} & \sbar{108}{257}{1325} \\
			\bottomrule
		\end{tabular}
	\label{table:multi-mapping-accuracy}
	\vspace{-2mm}
\end{table}

As shown in Table~\ref{table:multi-mapping-accuracy}, IJM is clearly the worst-performing tool on multi-mappings with the lowest precision and recall, as in many cases it fails to match even a single instance of the duplicated code.
This result is quite surprising as IJM is essentially an improved version of GumTree 2.1.0, which has a far better multi-mapping accuracy.
Therefore, we performed an in-depth investigation to find the reason behind IJM's poor performance. We discovered that IJM authors modified the structure of AST by merging \texttt{SimpleName} leaf AST nodes with their parent nodes. This was done to avoid the problem of semantically incompatible mappings, as explained in Section~\ref{sec:motivating-examples}.
However, this change made the depth of the AST shorter.
Normally, a method invocation has a minimum height of 2, as the name of the method is a leaf \texttt{SimpleName} node nested under the \texttt{MethodInvocation} parent node. After the merging, the height of the \texttt{MethodInvocation} subtree becomes 1.
As a result, the first phase of IJM that executes GumTree on the entire files to find the largest identical subtrees, fails to match statements of height equal to 1 (i.e., method invocation statements with a merged name node),
because GumTree employs a threshold for the hyperparameter $minHeight$
(i.e., length of the longest path from one leaf to the root of the subtree) equal to 2.
For this reason, many of the duplicated statements that are simple method invocations without arguments are not getting matched with their corresponding statements in the extracted methods.

From the remaining tools, RefactoringMiner has by far the highest accuracy (99.7\% precision, 98.4\% recall), which is expected as it is the only tool supporting multi-mappings. 
GumTree 3.0 (simple) is the second-best performing tool on multi-mappings with a precision of 59\% and a recall of 10\%. 
GumTree 3.0 (greedy) has a similar recall to GumTree 3.0 (simple), but it has a lower precision by -12\%, as due to its greedy nature, it tends to mismatch the remaining duplication instances with irrelevant (possibly newly added) code.
GumTree 2.1.0 has a very similar performance to GumTree 3.0 (greedy) with a slightly lower precision and recall (around -2\%).
\findings{RQ1 findings}{	
As expected all tools that do not support multi-mappings have a low recall within the range 2-10\%.
Moreover, the tools that are more greedy in nature (GumTree 3.0 greedy) have lower precision than their non-greedy counterparts (GumTree 3.0 simple), because they tend to force incorrect mappings for the unmatched multi-mappings. Based on their multi-mapping accuracy, the tools are ranked as follows:
\begin{enumerate}[leftmargin=*]
	\item \textbf{RefactoringMiner 3.0} (F-score=99\%)
	\item \textbf{GumTree 3.0 simple} (F-score=17.5\%)
	\item \textbf{GumTree 3.0 greedy} (F-score=17\%)
	\item \textbf{GumTree 2.1.0} (F-score=14.5\%)
	\item \textbf{MTDiff} (F-score=12\%)
	\item \textbf{IJM} (F-score=4.5\%)
\end{enumerate}
}

\subsection{RQ2: Semantically incompatible mappings}
\label{sec:RQ2}
\subsubsection{Experimental procedure}
For this research question, we investigate AST node types that can play different semantic roles in different contexts.
In most cases, the different semantic roles that can be played by an AST type are incompatible with each other.
However, language-agnostic diff tools match AST nodes of the same type, regardless of their semantic role in the program.
To partially address this limitation of language-agnostic diff tools, we extended GumTree 3.0 greedy and simple versions, so that they generate more fine-grained AST type labels for leaf AST nodes. In particular, for all leaf AST nodes, we change their AST type label by merging their actual type label with the type label of their parent AST node.
For example, a \texttt{SimpleName} node with a \texttt{SimpleType} parent node will have a type label \texttt{SimpleType-SimpleName}, while a \texttt{SimpleName} node with a \texttt{MethodInvocation} parent node will have a type label \texttt{MethodInvocation-SimpleName}.
In this way, we can avoid matching leaf AST nodes of the same AST type, which have parent nodes of different AST types. We name these variants of GumTree 3.0 as \textit{Fine-Grained} (FG).

Since all tools use the Eclipse JDT parser for generating ASTs, we analyzed the Eclipse JDT documentation to find AST types used in multiple contexts.
Below we list the AST types that are involved in multiple contexts and play a semantically incompatible role in each context:

\begin{itemize}[leftmargin=*]
	\item \header{Block} This AST type is used in 7 different contexts, namely as (1) the body of a method declaration, (2) the body of an initializer declaration, (3) the body of a \texttt{catch} clause, (4) the body of a \texttt{try} statement, (5) the body of a \texttt{synchronized} statement, (6) the body of a \texttt{finally} block, and (7) the body of control statements (\texttt{while}, \texttt{for}, \texttt{if}).
	\item \header{SingleVariableDeclaration} This AST type is used in 5 different contexts, namely as (1) \texttt{catch} clause exception, (2) \texttt{enhanced-for} formal parameter, (3) method declaration parameter, (4) lambda expression parameter, and (5) pattern variable in \texttt{instanceof} expression (Java 17 feature).
	\item \header{MethodDeclaration} This AST type is used to represent both method and constructor declarations. A constructor serves the semantic role of creating instances of a class and initializing its attributes, while a method serves the semantic role of providing the behavior for an instance of a class based on its current state.
	Thus, it makes no semantic sense to match a constructor with a method.
	\item \header{Modifier} This AST type is used to represent access modifiers (\texttt{public}, \texttt{private}, \texttt{protected}),
	\texttt{static},
	\texttt{abstract},
	\texttt{final},
	\texttt{native},
	\texttt{synchronized},
	\texttt{transient},
	\texttt{volatile},
	\texttt{strictfp},
	\texttt{default},
	\texttt{sealed}, and
	\texttt{non-sealed} modifiers.
	Among these modifiers, only the access modifiers are semantically compatible with each other, and the \texttt{final}, \texttt{sealed}, \texttt{non-sealed} modifiers (Java 17 feature) are semantically compatible with each other.
	All other modifiers have a unique semantic role in the program, and thus are not replaceable by another modifier (i.e., should not be matched with another modifier).
	\item \header{Type} This AST type is used in 22 different contexts, namely (1) method declaration return type, (2) receiver type and (3) list of thrown exception types, (4) \texttt{record} pattern type (Java 19 feature), (5) \texttt{instanceof} expression right operand, (6) type declaration super class type and (7) list of implemented interface types,
	(8) list of permitted types (Java 17 feature) (9) cast expression type, (10) class instance creation type, (11) creation reference type, (12) field declaration type, (13) variable declaration type, (14) method reference parameter type (in Javadoc), (15) type literal type, (16) type method reference type, (17) annotation type member declaration type,
	(18) wildcard-type bound, (19) array-type component and (20) element type, (21) qualified-type qualifier, and (22) parameterized-type type.
	\item \header{SimpleName} This AST type is used in 25 different contexts, namely (1) method declaration name and (2) receiver qualifier, (3) \texttt{record} pattern name (Java 19 feature), (4) variable declaration name, (5) type parameter name, (6-8) type/super/expression method reference name, (9) field access name, (10) super field access name, (11) method invocation name, (12) super method invocation name, (13) method reference parameter name (in Javadoc), (14) method reference name (in Javadoc), (15) member reference name (in Javadoc), (16) \texttt{enum} constant declaration name, (17) annotation type member declaration name, (18) type declaration name, (19) labeled statement label, (20) \texttt{continue} statement label, (21) \texttt{break} statement label, (22) member value pair name (in annotations), (23) qualified-type name, (24) qualified-name name, and (25) name-qualified type name.
\end{itemize}

\begin{definition}
	A mapping $M = (m_1, m_2) \in S_T$ (i.e., the mappings generated by tool $T$) is considered as \textit{semantically incompatible} if $m_1$ and $m_2$ have the same AST node type, the parents of $m_1$ and $m_2$ have a different AST node type (i.e., different context), and $M \notin S_B$, where $S_B$ is our benchmark of mappings.
\end{definition}

Based on this definition and the mappings generated by a tool $S_T$, we can compute the \textit{number of semantically incompatible mappings} for tool $T$.
Especially for modifiers, our definition is slightly modified as semantically incompatible modifiers can have parents of the same AST type.
\begin{definition}
	A modifier mapping $M = (m_1, m_2) \in S_T$ is considered as \textit{semantically incompatible} if $m_1$ and $m_2$ have the \texttt{Modifier} AST type, but belong to a different semantic group (i.e., access modifiers, seal modifiers, and each remaining modifier in its own group).
\end{definition}

\subsubsection{Results and discussion}
\begin{table}[ht]
	\caption{Number of semantically incompatible mappings in both datasets}
	\vspace{-4mm}
	\centering
	\setlength{\tabcolsep}{3pt}
	\begin{tabular}{l|r|r|r|r|r|r|r|r}
		\toprule
		\textbf{AST node Type} & 
		\makecell[c]{\textbf{RM}\\\textbf{3.0}} & 
		\makecell[c]{\textbf{GT 3.0}\\\textbf{greedy}} & 
		\makecell[c]{\textbf{GT 3.0}\\\textbf{simple}} & 
		\makecell[c]{\textbf{GT}\\\textbf{2.1.0}} & 
		\textbf{IJM} & \textbf{MTDiff} &
		\makecell[c]{\textbf{GT 3.0}\\\textbf{greedy (FG)}} & 
		\makecell[c]{\textbf{GT 3.0}\\\textbf{simple (FG)}}
		\\
		\toprule
		Block                     & 0 & 83 & 116 & 79 & 61 & 77 & 71 & 98 \\
		SingleVariableDeclaration & 0 & 4 & 3 & 1 & 0 & 5 & 3 & 3 \\
		MethodDeclaration         & 0 & 0 & 0 & 1 & 0 & 6 & 0 & 0 \\
		Modifier                  & 0 & 6 & 1 & 7 & 6 & 9 & 5 & 1 \\
		Type                      & 0 & 118 & 94 & 60 & 0 & 398 & 111 & 95 \\
		SimpleName                & 0 & 418 & 0 & 356 & 18 & 793 & 0 & 0 \\
		\bottomrule
		\textbf{Total}            & 0 & 629 & 214 & 504 & 85 & 1288 & 190 & 197 \\
		\bottomrule
	\end{tabular}
	\label{table:semantically-incompatible-mappings}
\end{table}

As we can observe from Table~\ref{table:semantically-incompatible-mappings}, RefactoringMiner is the only tool that does not generate semantically incompatible mappings for the six investigated AST node types.
This can be attributed to its semantic-aware algorithm design that uses language clues and refactoring information to guide the matching process, instead of randomly matching AST subtrees based on their similarity, regardless of their semantic context and relevance.
On the other hand, all other examined tools exhibit semantic violations to a greater or lesser extent.
Below we list the most common semantic violations for each AST node type, and discuss the individual behavior of each tool.
\begin{itemize}[leftmargin=*]
	\item \header{Block violations} The most common semantic violation for blocks is matching the body of a method declaration with the body of and \texttt{if}, \texttt{for}, or \texttt{try} statements. All tools are affected by this problem with GumTree 3.0 simple having the largest number of instances (116) and IJM having the smallest number of instances (61). This usually happens when the body of the control statement contains a large portion of the code inside the method body, and there is some control re-structuring (e.g., surrounding the body of a method within a newly added \texttt{try} block).
	\item \header{SingleVariableDeclaration violations} The most common semantic violation for single-variable-declarations is matching the formal parameter of an \texttt{enhanced-for} with a method parameter. All tools except for IJM and RefactoringMiner exhibit this behavior.
	MTDiff is the only tool matching a \texttt{catch} clause exception declaration with a method parameter or an \texttt{enhanced-for} formal parameter.
	\item \header{MethodDeclaration violations} GumTree 2.1.0 and MTDiff are the only tools matching a constructor with a method. MTDiff has the largest number of instances (6), while GumTree 2.1 has only one instance of this violation. 
	\item \header{Modifier violations} The most common semantic violation for modifiers is matching access modifiers (\texttt{public}, \texttt{private}, \texttt{protected}) with the
	\texttt{static} modifier. All tools except for RefactoringMiner exhibit this behavior with MTDiff having the largest number of instances (9) and GumTree 3.0 simple having only one instance.
	\item \header{Type violations} The most common semantic violation for types is matching the type of a field declaration with the return type of a method or the type of a local variable declaration. Moreover, matching the type of a method parameter with the type of a local variable declaration is another common semantic violation. MTDiff, which has the largest number of instances for this violation tends to match the type within type-literals (i.e., \texttt{\textbf{Type}.class}) with regular type references as the aforementioned ones.
	\item \header{SimpleName violations} The most common semantic violation for simple names (i.e., identifiers) is not matching the three components of a method invocation (i.e., receiver, name, arguments) accordingly, and also mismatching these three components with semantically irrelevant simple names, which are not even within method invocations. As already discussed in Section~\ref{sec:multi-mappings}, IJM modified the structure of AST by merging \texttt{SimpleName} leaf AST nodes with their parent nodes to avoid this semantic violation, which explains the reason it has only 18 instances of this violation. IJM violations mainly correspond to method invocation arguments being matched with a method invocation receiver, as the method names are merged with their parent method invocations, and thus they cannot be semantically mismatched.
\end{itemize}

A notable finding is that GumTree 3.0 simple has zero semantic violations for simple names (i.e., identifiers), while it has 94 for type references, in contrast to its greedy counterpart, which has semantic violations for both of these AST types.
Both versions of GumTree follow the same top-down phase to find the largest identical sub-trees with the \textit{minHeight} hyperparameter threshold configured equal to 2. As a result, while \texttt{SimpleName} nodes with identical values cannot be matched in this phase, as their height is equal to 1, \texttt{Type} nodes with identical values can be matched, as their height is at minimum equal to 2 (e.g., a \texttt{SimpleType} has a \texttt{SimpleName} node as leaf with the actual name of the type).
In the bottom-up phase that follows, the greedy version applies a non-conservative approach that matches AST subtrees of height equal to 1 (i.e., leaf AST nodes), while the simple version is more conservative and avoids this approach. This difference in the bottom-up phase of the GumTree simple version is responsible for having zero semantic violations for \texttt{SimpleName} nodes.
Another notable but expected finding is that the tools which are more greedy in nature (i.e., GumTree greedy and MTDiff) have the largest number of semantically incompatible mappings.

With respect to the GumTree variants using fine-grained AST type labels for leaf AST nodes (last two columns in Table~\ref{table:semantically-incompatible-mappings}), we can observe that this technique greatly benefits the GumTree greedy version by eliminating completely semantically incompatible mappings for \texttt{SimpleName} nodes, and also slightly reducing the number of semantically incompatible mappings for the other AST node types. 
On the other hand, the effect of this technique on the GumTree simple version is very small, as it only slightly reduces the the number of semantically incompatible mappings for \texttt{Block} nodes.
Despite the improvements, still none of the two GumTree variants can achieve having zero semantic violations.
This finding shows that the language-aware features embodied in RefactoringMiner (i.e., the program element matching algorithm shown in Figure~\ref{fig:workflow}) and IJM (i.e., the \textit{partial matching} technique) are more effective in eliminating semantically incompatible mappings.
\findings{RQ2 findings}{
	RefactoringMiner is the only tool with zero semantic violations for the six investigated AST node types.
	The tools that are more greedy in nature (i.e., GumTree greedy and MTDiff) tend to generate a larger number of mappings with semantic violations.
	On the contrary, the tools that utilize language-aware features (i.e., RefactoringMiner and IJM) tend to generate a smaller number of mappings with semantic violations.
	Based on their number of semantically incompatible mappings, the tools are ranked as follows:
	\begin{enumerate}[leftmargin=*]
		\item \textbf{RefactoringMiner 3.0} (0 semantic violations)
		\item \textbf{IJM} (85 semantic violations)
		\item \textbf{GumTree 3.0 greedy with fine-grained AST type labels} (190 semantic violations)
		\item \textbf{GumTree 3.0 simple with fine-grained AST type labels} (197 semantic violations)
		\item \textbf{GumTree 3.0 simple} (214 semantic violations)
		\item \textbf{GumTree 2.1.0} (504 semantic violations)
		\item \textbf{GumTree 3.0 greedy} (629 semantic violations)
		\item \textbf{MTDiff} (1288 semantic violations)
	\end{enumerate}
}

\subsection{RQ3: Program element mappings}
\label{sec:program-element-mappings}
\subsubsection{Experimental procedure}
For this research question, we investigate the ability of the tools to match correctly program elements (i.e., type, method, field declarations).
More specifically, 
we filter all AST node mappings corresponding to type, method, and field declarations in our benchmark $S_B$, and compute the accuracy of the AST diff tools on this particular subset of mappings.
As most AST diff tools are language-agnostic, they do not utilize any language clues in order to guide the matching process.
Instead, they try to ``blindly'' find the largest identical (or similar) AST sub-trees in a pair of files, and then match the corresponding method/field declarations containing these AST sub-trees.

However, this approach has some undesired implications.
When a relatively large part of a method is extracted into a new method, the original method declaration on the left side is matched with the extracted method declaration on the right side, and thus the remaining method declaration (calling the extracted method) appears as newly added on the right side of the diff.
This gives the code reviewers the wrong impression that the original method is ``renamed'' to the extracted method, which makes it more difficult to understand the actual change intended by the commit authors (i.e., \textsc{Extract Method} refactoring).
As we will see in the next subsection regarding RQ4, \textsc{Extract Method} is the most common refactoring in our dataset.
Moreover, there is empirical evidence showing that \textsc{Extract Method} is the most popular method-related refactoring applied by developers~\cite{Negara:ECOOP:2013}.
Silva et al. monitored 285 GitHub projects for a period of two months and found that \textsc{Extract Method} refactoring had the largest number of instances~\cite{WhyWeRefactor}.
Therefore, the implications of not handling properly \textsc{Extract Method} refactorings when generating AST diff are rather significant.

Regarding field declarations, RefactoringMiner is the only tool among the compared ones that uses field references to match renamed fields.
More specifically, in Phase 2 (Figure~\ref{fig:workflow}) it first establishes the corresponding method pairs within a class (including methods with changes in their signature), and then utilizes the statement mappings within the matched method pairs to find name-identifier replacements referring to a pair of field declarations.
If these replacements are consistent across all method pairs, then we have high confidence that the referenced pair of fields should be matched as renamed.
On the other hand, traditional AST diff tools rely only on the similarity of the field declaration AST sub-trees to match fields.
However, field declarations are in general small sub-trees as they consist only of a type, a name, a list of modifiers, and an optional initializer expression, and thus the pairs of matching field declarations are rather ambiguous when multiple fields have been renamed in a class.
In a previous work~\cite{RefactoringMiner2}, we showed that GumTree 2.1.2 had 78\% precision  and 70\% recall in the detection of renamed fields, and 75\% precision and 62\% recall in the detection of field type changes.
Negara et al.~\cite{Negara:ECOOP:2013} found that \textsc{Rename Field} is the second most popular refactoring manually applied by developers after \textsc{Rename Local Variable}.
Moreover, Ketkar et al.~\cite{10.1145/3368089.3409725} 416K commits from 129 open-source projects and found that field type changes are more commonly and frequently applied by developers than field renames. They also found that developers who changed the type of a
variable or attribute, they also renamed it in 55\% of the examined instances.
Therefore, field type changes and renames tend to co-occur, thus making the matching of field declarations even more ambiguous for traditional AST diff tools.

\subsubsection{Results and discussion}
\begin{table}[ht]
	\caption{Precision and recall for program element mappings in both datasets}
	\vspace{-4mm}
	\centering
	\fontsize{7pt}{10pt}\selectfont
	\setlength\tabcolsep{2px}
	\begin{tabular}{l r | r l r l | r l r l | r l r l | r l r l | r l r l | r l r l}
		\toprule
		\textbf{Prog} & & 
		\multicolumn{4}{c}{\textbf{RM 3.0}} & 
		\multicolumn{4}{c}{\textbf{GT 3.0 greedy}} & 
		\multicolumn{4}{c}{\textbf{GT 3.0 simple}} &
		\multicolumn{4}{c}{\textbf{GT 2.1.0}} &
		\multicolumn{4}{c}{\textbf{IJM}} &
		\multicolumn{4}{c}{\textbf{MTDiff}} \\
		\textbf{Elem} & \textbf{\#} &
		\multicolumn{2}{c}{\textbf{Precision}} &
		\multicolumn{2}{c}{\textbf{Recall}} &
		\multicolumn{2}{c}{\textbf{Precision}} &
		\multicolumn{2}{c}{\textbf{Recall}} &
		\multicolumn{2}{c}{\textbf{Precision}} &
		\multicolumn{2}{c}{\textbf{Recall}} &
		\multicolumn{2}{c}{\textbf{Precision}} &
		\multicolumn{2}{c}{\textbf{Recall}} &
		\multicolumn{2}{c}{\textbf{Precision}} &
		\multicolumn{2}{c}{\textbf{Recall}} &
		\multicolumn{2}{c}{\textbf{Precision}} &
		\multicolumn{2}{c}{\textbf{Recall}} \\ \midrule
		Type & 1432 &
		\sbarOnlyZero{1431}{0}{0} & \sbarOnlyZero{1412}{1}{19} & \sbarOnlyZero{1412}{1}{19} & \sbarOnlyZero{1415}{1}{16} & \sbarOnlyZero{1413}{2}{18} & \sbarOnlyZero{1406}{11}{25} \\
		Enum & 17 &
		\sbar{17}{0}{0} & \sbar{17}{0}{0} & \sbar{17}{0}{0} & \sbar{16}{0}{1} & \sbar{17}{0}{0} & \sbar{17}{0}{0} \\
		Method & 2289 &
		\sbar{2277}{1}{1} & \sbar{1710}{118}{568} & \sbar{1764}{147}{514} & \sbar{1730}{112}{548} & \sbar{1885}{11}{393} & \sbar{1717}{225}{561} \\
		Field & 245 &
		\sbar{242}{0}{3} & \sbar{128}{2}{117} & \sbar{129}{3}{116} & \sbar{145}{4}{100} & \sbar{109}{2}{136} & \sbar{141}{4}{104} \\
		\bottomrule
		Overall & 3983 &
		\sbar{3979}{1}{4} & \sbar{3267}{121}{704} & \sbar{3322}{151}{649} & \sbar{3306}{117}{665} & \sbar{3424}{15}{547} & \sbar{3281}{240}{690} \\
		\bottomrule
	\end{tabular}
	\label{table:program-element-mapping-accuracy}
\end{table}

Table~\ref{table:program-element-mapping-accuracy} shows the mapping accuracy of the tools per program element type.
The second column in Table~\ref{table:program-element-mapping-accuracy} shows the number of instances for each program element type that were considered in the computation of precision and recall, which corresponds to the number of program elements with changed AST trees.
We would like to emphasize that we excluded the program elements that have identical AST trees, since all tools can match them with 100\% accuracy, and thus their inclusion would artificially increase the computed precision and recall.
For the sake of providing complete information,
the total number of type, enum, method, and field declarations in our benchmark is 2745, 139, 47795, and 16438, respectively.
From the aforementioned numbers, we can infer that only a small portion of program elements changed in the benchmark commits (i.e., 52\% of the type declarations, 12\% of the enum declarations, 4.8\% of the method declarations, and 1.5\% of the field declarations).

Based on the results shown in Table~\ref{table:program-element-mapping-accuracy} all tools can match type and enum declarations with a high precision and recall. 
This is an expected outcome, as type declarations are typically large AST subtrees (i.e., they include methods and fields), and thus it is very uncommon to have different type declarations with highly similar AST trees (i.e., there is less ambiguity).
Moreover, we can observe that all GumTree-based tools have a very similar precision (between 99.2-99.9\%) and recall (between 98.3-98.9\%) in matching type/enum declarations.
However, RefactoringMiner is the only tool that achieves 100\% accuracy, as it relies on method and field signatures to match renamed/moved classes (Figure~\ref{fig:workflow}, Phase 3), and thus can match classes even if their method bodies have been extensively modified. 

Regarding method declarations, RefactoringMiner has an almost perfect precision and recall with only one false positive and one false negative.
The second-best performing tool is IJM, which has 99.4\% precision and 82.7\% recall.
This good performance can be attributed to \textit{partial matching}, which restricts the scope of matching to method declarations with the same signature (i.e., first, IJM attempts to match methods only by their signature, and then it attempts to match the
remaining methods by their signature and by their method body).
The reason RefactoringMiner has a better performance than IJM is because the process followed in Figures~\ref{fig:workflow} and~\ref{fig:overview} can additionally match methods with changes in their signatures, methods from/to which other methods have been extracted/inlined, and also methods that have been merged to a single method or split to multiple methods.
All other tools have a quite close performance with a precision ranging between 88-94\% and a recall ranging between 75-77\%.

Regarding field declarations, RefactoringMiner has a perfect precision with zero false positives and a recall of 98.8\% with three false negatives.
There is no clear second-best performing tool, as GumTree 3.0 greedy has the second highest precision with 98.5\% and GumTree 2.1.0 has the second highest recall with 59.2\%.
Notably, IJM does not perform as good as it does for method declarations, since it has the lowest recall among all tools (44.5\%), but it has the third-best precision (98.2\%).
IJM also applies \textit{partial matching} for field declarations, but it processes all field declaration subtrees together in a single iteration, without processing separately those that have identical signatures (i.e., same type and name).
After some investigation, we found that IJM's lower recall is due to some non-identical field declarations being moved to newly introduced inner classes within the same file.
The first phase of IJM that executes GumTree on the entire files to find the largest identical subtrees, fails to match the moved field declarations, since they are non-identical (e.g., they have differences in modifiers). In the phase where IJM performs partial matching for field declarations, it includes only the fields declared within the same class, and thus it is unable to match attributes moved to other classes, even if these classes are within the same file (i.e., inner classes).
The other GumTree-based tools search for similar sub-trees in the entire file, and thus are able to match these moved field declarations regardless of their class container.
Although, RefactoringMiner follows a similar approach to IJM by first matching field declarations within the same classes (Figure~\ref{fig:workflow}, Phase 2), it has an additional phase at the end (Figure~\ref{fig:workflow}, Phase 4) that detects moved field declarations between different classes.
Moreover, by utilizing field references, RefactoringMiner can match fields with co-occurring changes in their name, type, modifiers and initializer (i.e., when most or every part of the field declaration subtree has changed), while the other tools fail as the subtree similarity of such field declarations is very small, which explains their lower recall values. 
\findings{RQ3 findings}{	
	Our analysis shows that all tools have a relatively high accuracy in matching type/enum declarations, but RefactoringMiner is the only tool with 100\% precision and recall. However, when it comes to matching method and field declarations RefactoringMiner still has an almost perfect accuracy, while the recall of the other tools drops significantly. In particular, all AST diff tools, except for RefactoringMiner, have considerably low recall in matching field declarations (44-59\%).
	Therefore, we can conclude that utilizing programming language clues, such as the signatures and references of program elements, can improve significantly the mapping accuracy for method and field declarations.
	Based on their overall program element mapping accuracy, the tools are ranked as follows:
	\begin{enumerate}[leftmargin=*]
		\item \textbf{RefactoringMiner 3.0} (F-score=99.9\%)
		\item \textbf{IJM} (F-score=92.4\%)
		\item \textbf{GumTree 2.1.0} (F-score=89.4\%)
		\item \textbf{GumTree 3.0 simple} (F-score=89.3\%)
		\item \textbf{GumTree 3.0 greedy} (F-score=88.8\%)
		\item \textbf{MTDiff} (F-score=87.6\%)
	\end{enumerate}
}

\subsection{RQ4: Refactoring-related mappings}
\subsubsection{Experimental procedure}
For this research question, we investigate how different refactoring types affect the mapping accuracy of AST diff tools.
To isolate the refactoring effect, we carefully select the AST sub-trees that could be potentially affected by each particular refactoring transformation, and compute the accuracy of the AST diff tools on the subset of mappings included within the selected AST sub-trees. The AST sub-trees considered for each refactoring type are listed below:
\begin{itemize}[leftmargin=*]
	\item \header{Extract Method} The source method declaration from which the method is extracted on the left and right side of the diff, and the extracted method itself on the right side of the diff.
	\item \header{Inline Method} The target method declaration to which the method is inlined on the left and right side of the diff, and the inlined method itself on the left side of the diff.
	\item \header{Extract Variable} The pairs of statements that reference the extracted variable, and the extracted variable declaration itself on the right side of the diff.
	\item \header{Inline Variable} The pairs of statements that reference the inlined variable, and the inlined variable declaration itself on the left side of the diff.
	\item \header{Rename Variable/Parameter + Change Variable/Parameter Type + Parameterize Variable} The pairs of statements that reference the variable/parameter, and the variable/parameter declaration itself on the left and right side of the diff.
	\textsc{Parameterize Variable} is a refactoring that commonly occurs along with \textsc{Extract Method} and captures local variables that become parameters in the extracted method. In many cases, the name of the parameterized variable changes in the extracted method.
	There are also a few cases where \textsc{Parameterize Variable} occurs within the same method by eliminating a local variable and introducing a parameter in its place.
	\item \header{Replace Variable with Field} The pairs of statements that reference the variable/field, the variable declaration itself on the left side of the diff, and the field declaration on the right side of the diff.
	\item \header{Add Parameter} The method declaration to which the parameter is added on the left and right side of the diff, and the parameter declaration itself on the right side of the diff. The reason we include the complete method declaration is because the addition of a parameter is usually accompanied with newly added functionality affecting existing statements or adding new statements, which should be properly matched by the diff tool.
	\item \header{Rename Method} The renamed method declaration on the left and right side of the diff. The reason we include the complete method declaration is because renaming might involve some functionality change affecting existing statements or adding/deleting statements, which should be properly matched by the diff tool.
\end{itemize}

We included in our analysis only the refactoring types with ten or more instances, as having a small number of instances poses a threat to the generalizability of our findings.
Moreover, we excluded all refactoring types that move code between different files, such as \textsc{Move Method}, \textsc{Pull Up Method}, \textsc{Move Field}, because we investigate the accuracy of the tools on inter-file mappings separately in the next Section~\ref{sec:inter-file-mappings}.
Therefore, all examined refactoring types have a local effect within the same file.
Finally, in cases where different refactoring instances of the same type affect exactly the same AST subtrees, we keep only one instance in our results to avoid recounting the same effect.
For example, when duplicated code is extracted from the same method, all \textsc{Extract Method} instances affect the same source method and the same extracted method, and thus we keep only one instance in our results.

\subsubsection{Results and discussion}
\begin{table}[ht]
	\caption{Precision and recall for refactoring-related mappings in both datasets considering only statement mappings}
	\vspace{-4mm}
	\centering
	\fontsize{7pt}{10pt}\selectfont
	\setlength\tabcolsep{1px}
	\begin{tabular}{l r | r l r l | r l r l | r l r l | r l r l | r l r l | r l r l}
		\toprule
		\textbf{Refactoring} & & 
		\multicolumn{4}{c}{\textbf{RM 3.0}} & 
		\multicolumn{4}{c}{\textbf{GT 3.0 greedy}} & 
		\multicolumn{4}{c}{\textbf{GT 3.0 simple}} &
		\multicolumn{4}{c}{\textbf{GT 2.1.0}} &
		\multicolumn{4}{c}{\textbf{IJM}} &
		\multicolumn{4}{c}{\textbf{MTDiff}} \\
		\textbf{Type} & \textbf{\#} &
		\multicolumn{2}{c}{\textbf{Precision}} &
		\multicolumn{2}{c}{\textbf{Recall}} &
		\multicolumn{2}{c}{\textbf{Precision}} &
		\multicolumn{2}{c}{\textbf{Recall}} &
		\multicolumn{2}{c}{\textbf{Precision}} &
		\multicolumn{2}{c}{\textbf{Recall}} &
		\multicolumn{2}{c}{\textbf{Precision}} &
		\multicolumn{2}{c}{\textbf{Recall}} &
		\multicolumn{2}{c}{\textbf{Precision}} &
		\multicolumn{2}{c}{\textbf{Recall}} &
		\multicolumn{2}{c}{\textbf{Precision}} &
		\multicolumn{2}{c}{\textbf{Recall}} \\ \midrule
		\textsc{Extract Method}        & 178& \sbar{6021}{24}{27} & \sbar{4457}{498}{1591} & \sbar{4447}{455}{1601} & \sbar{4341}{526}{1707} & \sbar{4069}{323}{1979} & \sbar{4259}{626}{1789}\\
		\textsc{Rename Variable}       & 77 & \sbar{349}{1}{5} & \sbar{147}{126}{207} & \sbar{144}{127}{210} & \sbar{143}{137}{211}& \sbar{110}{75}{244} & \sbar{145}{139}{209}\\
		\textsc{Extract Variable}      & 75 & \sbar{135}{2}{2} & \sbar{94}{24}{43} & \sbar{98}{22}{39} & \sbar{94}{24}{43} & \sbar{85}{28}{52} & \sbar{105}{25}{32}\\
		\textsc{Change Var Type}       & 47 & \sbar{76}{0}{2} & \sbar{67}{3}{11} & \sbar{66}{3}{12} & \sbar{64}{6}{14} & \sbar{36}{6}{42} & \sbar{62}{13}{16} \\
		\textsc{Add Parameter}         & 45 & \sbarOnlyZero{1427}{1}{7} & \sbar{1053}{476}{381} & \sbar{1023}{460}{411} & \sbar{817}{678}{617} & \sbar{1053}{377}{381} & \sbar{1357}{83}{77} \\
		\textsc{Param Variable}        & 30 & \sbar{93}{0}{0} & \sbar{76}{14}{17} & \sbar{80}{11}{13} & \sbar{67}{12}{26} & \sbar{20}{10}{73} &  \sbar{58}{24}{35} \\
		\textsc{Inline Method}         & 28 & \sbar{492}{5}{6} & \sbar{447}{49}{51} & \sbar{452}{36}{46} & \sbar{421}{69}{77} & \sbar{431}{33}{67} & \sbar{435}{45}{63} \\
		\textsc{Inline Variable}       & 22 & \sbar{17}{0}{0} & \sbar{5}{5}{12} & \sbar{8}{4}{9} & \sbar{6}{5}{11} & \sbar{7}{6}{10} & \sbar{8}{5}{9} \\
		\textsc{Rename Parameter}      & 19 & \sbar{48}{0}{0} & \sbar{42}{6}{6} & \sbar{41}{5}{7} & \sbar{40}{6}{8} & \sbar{37}{6}{11} & \sbar{42}{5}{6} \\
		\textsc{Change Param Type}     & 18 & \sbar{69}{0}{0} & \sbar{57}{7}{12} & \sbar{60}{7}{9} & \sbar{54}{8}{15} & \sbar{58}{4}{11} & \sbar{62}{3}{7} \\
		\textsc{Rename Method}         & 15 & \sbarOnlyZero{109}{0}{1} & \sbar{105}{1}{5} & \sbar{107}{2}{3} & \sbar{108}{1}{2} & \sbar{105}{1}{5} & \sbar{105}{6}{5} \\
		\textsc{Replace Var w Field}   & 10 & \sbar{20}{0}{0} & \sbar{17}{2}{3} & \sbar{15}{0}{5} & \sbar{17}{4}{3} & \sbar{14}{1}{6} & \sbar{14}{3}{6} \\
		\bottomrule
		Overall & 564 &
		\sbar{8856}{33}{50} & \sbar{6567}{1211}{2339} & \sbar{6541}{1132}{2365} & \sbar{6172}{1476}{2734} & \sbar{6025}{870}{2881} & \sbar{6652}{977}{2254} \\
		\bottomrule
	\end{tabular}
	\label{table:refactoring-related-mapping-accuracy}
\end{table}
Based on the results shown in Table~\ref{table:refactoring-related-mapping-accuracy}, RefactoringMiner has consistently the highest precision and recall for all examined refactoring types.
The second-best performing tool differs for different sets of refactoring types.
For instance, MTDiff has the second-best accuracy (based on F-score) for parameter-related refactorings (\textsc{Add Parameter}, \textsc{Rename Parameter}, \textsc{Change Parameter Type}), as well as \textsc{Extract Variable}.
GumTree 3.0 simple has the second-best accuracy for \textsc{Extract Method}, \textsc{Inline Method}, \textsc{Parameterize Variable}, and \textsc{Inline Variable}.
GumTree 3.0 greedy has the second-best accuracy for variable-related refactorings (\textsc{Rename Variable}, \textsc{Change Variable Type}, \textsc{Replace Variable with Field}), but it is only slightly better than the third-best accuracy by GumTree 3.0 simple.
Notably, there is no refactoring type for which IJM has the second-best accuracy, and for the reasons explained in Sections~\ref{sec:multi-mappings} and~\ref{sec:program-element-mappings} it has the lowest recall value for \textsc{Extract Method}, \textsc{Extract Variable}, \textsc{Rename Variable}, \textsc{Change Variable Type}, \textsc{Parameterize Variable}, and \textsc{Rename Parameter} refactorings.
GumTree 2.1.0 has the second-best accuracy for only one refactoring type, namely \textsc{Rename Method}, but all tools have a quite similar performance on this refactoring type, and thus there is no clear advantage for GumTree 2.1.0.

Since some of the refactoring types listed in Table~\ref{table:refactoring-related-mapping-accuracy} involve sub-expression mappings (see Section~\ref{sec:refactoring-mechanics}),
we decided to further investigate the performance of the tools at sub-expression level.
However, due to the compatibility issues among the tools discussed in Section~\ref{sec:tools}, 
the tools that can provide comparable sub-expression mappings are those using the AST generator \texttt{gumtree.gen.jdt-3.0.0}, namely RefactoringMiner and the greedy/simple versions of GumTree 3.0.
Table~\ref{table:refactoring-related-mapping-accuracy-3.0} shows the precision and recall of the three tools at a more fine-grained level that additionally includes sub-expression mappings.

\begin{table}[ht]
	\caption{Precision and recall for refactoring-related mappings in both datasets considering statement and sub-expression mappings}
	\vspace{-4mm}
	\centering
	\fontsize{8pt}{10pt}\selectfont
	\setlength\tabcolsep{1px}
	\begin{tabular}{l r | r l r l | r l r l | r l r l }
		\toprule
		\textbf{Refactoring} & & 
		\multicolumn{4}{c}{\textbf{RM 3.0}} & 
		\multicolumn{4}{c}{\textbf{GT 3.0 greedy}} & 
		\multicolumn{4}{c}{\textbf{GT 3.0 simple}} \\
		\textbf{Type} & \textbf{\#} &
		\multicolumn{2}{c}{\textbf{Precision}} &
		\multicolumn{2}{c}{\textbf{Recall}} &
		\multicolumn{2}{c}{\textbf{Precision}} &
		\multicolumn{2}{c}{\textbf{Recall}} &
		\multicolumn{2}{c}{\textbf{Precision}} &
		\multicolumn{2}{c}{\textbf{Recall}} \\ \midrule
		\textsc{Extract Method}        & 178& \sbar{7726}{53}{58} & \sbar{4965}{1661}{2819} & \sbar{4934}{1298}{2850} \\
		\textsc{Rename Variable}       & 77 & \sbar{745}{7}{13} & \sbar{162}{254}{596} & \sbar{157}{160}{601} \\
		\textsc{Extract Variable}      & 75 & \sbar{276}{9}{7} & \sbar{147}{188}{136} & \sbar{143}{116}{140} \\
		\textsc{Change Variable Type}  & 47 & \sbar{188}{3}{6} & \sbar{97}{77}{97} & \sbar{90}{24}{104} \\
		\textsc{Parameterize Variable} & 30 & \sbarOnlyZero{193}{0}{1} & \sbar{110}{78}{84} & \sbar{118}{41}{76} \\
		\textsc{Inline Method}         & 28 & \sbar{685}{6}{8} & \sbar{572}{215}{121} & \sbar{581}{124}{112} \\
		\textsc{Inline Variable}       & 22 & \sbarOnlyZero{49}{1}{1} & \sbar{20}{34}{30} & \sbar{21}{18}{29} \\
		\bottomrule
	\end{tabular}
	\label{table:refactoring-related-mapping-accuracy-3.0}
\end{table}
By comparing Tables~\ref{table:refactoring-related-mapping-accuracy-3.0} and~\ref{table:refactoring-related-mapping-accuracy}, we can observe that RefactoringMiner maintains high accuracy levels with some small decrease in its precision and recall. 
However, GumTree 3.0 greedy has a considerable drop in its precision, and has consistently a larger drop in precision than GumTree 3.0 simple.
This result shows that the greedy version of GumTree 3.0 often forces incorrect mappings at the sub-expression level that harm considerably its precision.
On the other hand, the same greedy nature helps to slightly increase the recall over its non-greedy counterpart for some refactoring types, but this small improvement in recall comes with a much larger cost in its precision.
Overall, we can conclude that GumTree simple has a clearly better performance than GumTree greedy when we take into consideration sub-expression mappings.
\findings{RQ4 findings}{	
	RefactoringMiner has consistently the highest precision and recall for all examined refactoring types. Regarding the second-best performance, each tool has an advantage over a different set of refactoring types. MTDiff has an advantage in parameter-related refactorings and extract variable. GumTree 3.0 simple has an advantage in extract/inline method. GumTree 3.0 greedy has an advantage in local-variable-related refactorings. 
	Based on their overall refactoring mapping accuracy, the tools are ranked as follows:
	\begin{enumerate}[leftmargin=*]
		\item \textbf{RefactoringMiner 3.0} (F-score=99.5\%)
		\item \textbf{MTDiff} (F-score=80.5\%)
		\item \textbf{GumTree 3.0 simple} (F-score=78.9\%)
		\item \textbf{GumTree 3.0 greedy} (F-score=78.7\%)
		\item \textbf{IJM} (F-score=76.3\%)
		\item \textbf{GumTree 2.1.0} (F-score=74.6\%)
	\end{enumerate}
}
\subsection{RQ5: Inter-file mappings}
\label{sec:inter-file-mappings}
\subsubsection{Experimental procedure}
For this research question, we focus our analysis on inter-file mappings, i.e., AST nodes that have a corresponding node in a file other than the one they are located in.
More specifically, we filter all AST node mappings marked as inter-file mappings in our benchmark $S_B$, and compute the accuracy of the AST diff tools on this particular subset of mappings.
Typically, inter-file mappings are caused by refactorings moving methods and fields between existing files (i.e., \textsc{Move Method}, \textsc{Move Field}), or to new files (i.e., \textsc{Extract Class}, \textsc{Extract Superclass}).

As none of the compared tools is designed to support inter-file mappings, we implemented
the Staged Tree Matching (STM) algorithm proposed by Fujimoto et al.~\cite{StagedTreeMatching} for the GumTree 3.0 greedy and simple versions. 
Staged Tree Matching first applies the standard GumTree algorithm on the pairs of files modified in a commit. Then, it constructs a project-level AST for each version by connecting the AST roots corresponding to each file to a pseudo-project-root node, and runs another matching phase on the project-level ASTs by considering only the remaining unmatched AST subtrees.

\subsubsection{Results and discussion}
\begin{table}[ht]
	\caption{Precision and recall for inter-file mappings in both datasets considering statement and sub-expression mappings}
	\vspace{-4mm}
	\centering
	\fontsize{8pt}{10pt}\selectfont
	\setlength\tabcolsep{2px}
	\begin{tabular}{r l r l | r l r l | r l r l}
		\toprule
		\multicolumn{4}{c}{\textbf{RM 3.0}} & 
		\multicolumn{4}{c}{\textbf{GT 3.0 greedy (STM)}} &
		\multicolumn{4}{c}{\textbf{GT 3.0 simple (STM)}} \\
		\multicolumn{2}{c}{\textbf{Precision}} &
		\multicolumn{2}{c}{\textbf{Recall}} &
		\multicolumn{2}{c}{\textbf{Precision}} &
		\multicolumn{2}{c}{\textbf{Recall}} &
		\multicolumn{2}{c}{\textbf{Precision}} &
		\multicolumn{2}{c}{\textbf{Recall}} \\ \midrule
		\sbar{1425}{0}{6} & \sbar{1385}{139}{46} & \sbar{1393}{150}{38} \\
		\bottomrule
	\end{tabular}
	\label{table:inter-file-mapping-accuracy}
\end{table}
As shown in Table~\ref{table:inter-file-mapping-accuracy}, RefactoringMiner has a perfect precision, as all moves it detected were correct, but it missed one method move, thus leading to 6 false negative statement mappings and a recall of 99.6\%.
With respect to the Staged Tree Matching algorithm, we can observe that GumTree simple has a higher recall, while GumTree greedy has a higher precision.
The explanation for this finding is that GumTree greedy tends to forcefully match some moved statements with newly added statements within the same file, as shown in the example of Figure~\ref{fig:commit-GTG}, and thus these statements are not considered in the second matching phase of the algorithm, which eventually results in a lower recall.
For the same reason, GumTree greedy generally has less statements considered in the second phase compared to GumTree simple, and thus the chances of generating false mappings are less, which eventually results in a higher precision.
This trade-off in precision and recall between the two GumTree versions actually shows that the language-aware approach followed by RefactoringMiner to find moved code by analyzing how method calls and field accesses are updated within matched statements and binding those references to the corresponding method/field declarations is more effective in terms of accuracy, as it allows to reason about the code that remained in the same file and the code that has been moved to a different file.
\findings{RQ5 findings}{	
	RefactoringMiner has a perfect precision in inter-file mappings, which is higher by 9-10\% than the precision achieved by the Staged Tree Matching algorithm implemented for the two GumTree 3.0 versions. Moreover, RefactoringMiner has a higher recall by 2-3\% compared to the two GumTree 3.0 versions.
	Based on their inter-file mapping accuracy, the tools are ranked as follows:
	\begin{enumerate}[leftmargin=*]
		\item \textbf{RefactoringMiner 3.0} (F-score=99.8\%)
		\item \textbf{GumTree 3.0 greedy} (F-score=93.74\%)
		\item \textbf{GumTree 3.0 simple} (F-score=93.68\%)
	\end{enumerate}
}

\subsection{RQ6: Overall accuracy}
\label{sec:RQ6}
\subsubsection{Experimental procedure}
For this research question, we compute the overall accuracy of the tools by including all AST node mappings in our benchmark $S_B$.
Moreover, given the compatibility issues among the tools discussed in Section~\ref{sec:tools},
we compute the accuracy at two levels of granularity.
At the coarse-grained level, we include mappings between statements and program elements, as all tools can provide this information.
At the fine-grained level, we include sub-expression mappings in addition to the mappings considered at the coarse-grained level.
The tools that can provide comparable sub-expression mappings are those using the AST generator \texttt{gumtree.gen.jdt-3.0.0}, namely RefactoringMiner and the greedy/simple versions of GumTree 3.0.
Moreover, we include in the evaluation, our implementation of the GumTree greedy and simple version with the techniques discussed in Sections~\ref{sec:RQ2} and~\ref{sec:inter-file-mappings}, namely \textit{Fine-Grained AST types} (FG) and \textit{Staged Tree Matching} (STM) combined.

\subsubsection{Results and discussion}
\begin{table}[ht]
	\caption{Overall mapping accuracy considering only statement mappings}
	\vspace{-4mm}
	\centering
	\fontsize{7pt}{10pt}\selectfont
	\setlength\tabcolsep{2px}
	\begin{tabular}{l | r l r l | r l r l | r l r l | r l r l | r l r l | r l r l}
		\toprule
		&
		\multicolumn{4}{c}{\textbf{RM 3.0}} & 
		\multicolumn{4}{c}{\textbf{GT 3.0 greedy}} & 
		\multicolumn{4}{c}{\textbf{GT 3.0 simple}} &
		\multicolumn{4}{c}{\textbf{GT 2.1.0}} &
		\multicolumn{4}{c}{\textbf{IJM}} &
		\multicolumn{4}{c}{\textbf{MTDiff}} \\
		\textbf{Dataset} &
		\multicolumn{2}{c}{\textbf{Precision}} &
		\multicolumn{2}{c}{\textbf{Recall}} &
		\multicolumn{2}{c}{\textbf{Precision}} &
		\multicolumn{2}{c}{\textbf{Recall}} &
		\multicolumn{2}{c}{\textbf{Precision}} &
		\multicolumn{2}{c}{\textbf{Recall}} &
		\multicolumn{2}{c}{\textbf{Precision}} &
		\multicolumn{2}{c}{\textbf{Recall}} &
		\multicolumn{2}{c}{\textbf{Precision}} &
		\multicolumn{2}{c}{\textbf{Recall}} &
		\multicolumn{2}{c}{\textbf{Precision}} &
		\multicolumn{2}{c}{\textbf{Recall}} \\ \midrule
		Defects4J &
		\sbar{34689}{57}{145} & \sbar{34260}{273}{574} & \sbar{34313}{306}{521} & \sbar{34326}{385}{508} & \sbar{34332}{356}{502} & \sbar{34211}{550}{623} \\
		Refactoring &
		\sbar{14411}{38}{68} & \sbar{10715}{990}{3752} & \sbar{10684}{971}{3783} & \sbar{10252}{1115}{4215} & \sbar{10341}{803}{4126} & \sbar{10952}{790}{3515} \\
		\bottomrule
		Overall &
		\sbar{49100}{95}{213} & \sbar{44975}{1263}{4326} & \sbar{44997}{1277}{4304} & \sbar{44578}{1500}{4723} & \sbar{44673}{1159}{4628} & \sbar{45163}{1340}{4138} \\
		\bottomrule
	\end{tabular}
	\label{table:overall-mapping-accuracy-2.1}
\end{table}
As we can observe from Table~\ref{table:overall-mapping-accuracy-2.1}, at the coarse-grained level all tools exhibit a very high accuracy in the Defects4J dataset with RefactoringMiner having the highest precision and recall.
However, in the Refactoring dataset the differences in accuracy are considerably larger with RefactoringMiner having a +6.4\% precision and +24\% recall than the second-best performing tool, which is MTDiff.
This result proves our argument that bug fixing commits are a low barrier for assessing the accuracy of AST diff tools, and thus more challenging datasets should be used, such as the Refactoring benchmark we introduce in this paper.
Our analysis in the previous research questions has shown that the lack of support for multi-mappings and inter-file mappings contributes a considerable number of false negatives for all tools other than RefactoringMiner. Moreover, the lack of language-awareness contributes false negatives in method and field declaration mappings, and the lack of refactoring-awareness contributes false negatives in statement mappings, especially for variable-related refactorings.
Although MTDiff seems to have the second-best performance based on F-score, it is hard to make a safe conclusion about which tool is actually the second-best performing one, as all tools (with the exception of RefactoringMiner) have a quite similar overall F-score at statement level ranging between 93.5-94.3\%.
However, there is some indication that MTDiff has a slightly better performance in the Refactoring dataset with +1.6\% precision and recall than the third-best performance.

\begin{table}[ht]
	\caption{Overall mapping accuracy considering statement and sub-expression mappings}
	\vspace{-4mm}
	\centering
	\fontsize{7pt}{10pt}\selectfont
	\setlength\tabcolsep{2px}
	\begin{tabular}{l | r l r l | r l r l | r l r l | r l r l | r l r l}
		\toprule
		&
		\multicolumn{4}{c}{\textbf{RM 3.0}} & 
		\multicolumn{4}{c}{\textbf{GT 3.0 greedy}} & 
		\multicolumn{4}{c}{\textbf{GT 3.0 simple}} &
		\multicolumn{4}{c}{\makecell{\textbf{GT 3.0 greedy}\\\textbf{STM+FG}}} &
		\multicolumn{4}{c}{\makecell{\textbf{GT 3.0 simple}\\\textbf{STM+FG}}} \\
		\textbf{Dataset} &
		\multicolumn{2}{c}{\textbf{Precision}} &
		\multicolumn{2}{c}{\textbf{Recall}} &
		\multicolumn{2}{c}{\textbf{Precision}} &
		\multicolumn{2}{c}{\textbf{Recall}} &
		\multicolumn{2}{c}{\textbf{Precision}} &
		\multicolumn{2}{c}{\textbf{Recall}} &
		\multicolumn{2}{c}{\textbf{Precision}} &
		\multicolumn{2}{c}{\textbf{Recall}} &
		\multicolumn{2}{c}{\textbf{Precision}} &
		\multicolumn{2}{c}{\textbf{Recall}} \\ \midrule
		Defects4J &
		\sbar{46922}{140}{324} & \sbar{43998}{1145}{3248} & \sbar{46206}{765}{1040} & \sbar{43708}{1007}{3538} & \sbar{46018}{797}{1228} \\
		Refactoring &
		\sbar{20586}{78}{161} & \sbar{14573}{2756}{6174} & \sbar{15016}{2304}{5731} & \sbar{15884}{2616}{4863} & \sbar{16383}{2368}{4364} \\
		\bottomrule
		Overall &
		\sbar{67508}{218}{485} & \sbar{58571}{3901}{9422} & \sbar{61222}{3069}{6771} & \sbar{59592}{3623}{8401} & \sbar{62401}{3165}{5592} \\
		\bottomrule
	\end{tabular}
	\label{table:overall-mapping-accuracy-3.0}
\end{table}

At the fine-grained level that additionally includes sub-expression mappings, we can observe by comparing Table~\ref{table:overall-mapping-accuracy-3.0} with Table~\ref{table:overall-mapping-accuracy-2.1} that the precision and recall of all tools drops.
However, GumTree 3.0 greedy has the largest accuracy drop among the three tools, which means that it tends to force more incorrect sub-expression mappings, but also miss more sub-expression mappings due to its greedy nature.
RefactoringMiner has still a remarkable overall precision and recall of 99.7\% and 99.3\%, respectively,
showing that it performs very well at the sub-expression level too.
More specifically, in the Refactoring dataset it has a +13\% precision and +27\% recall than the second-best performing tool, which is GumTree 3.0 simple.
GumTree 3.0 simple has consistently better precision and recall than GumTree 3.0 greedy in both datasets, when considering sub-expression mappings, while the accuracy differences between these tools are very small, when considering only statement mappings.

Regarding the \textit{Fine-Grained AST types} (FG) and \textit{Staged Tree Matching} (STM) techniques, we can observe from Table~\ref{table:overall-mapping-accuracy-3.0} that their effect is positive on both precision and recall in the Refactoring dataset, which includes more inter-file mappings and more complex changes (i.e., changes prone to semantically incompatible mappings) than the Defects4J dataset.
More specifically, the techniques contributed in GumTree greedy and simple version an increased precision by 1.8\% and 0.7\%, respectively, and an increased recall by 6.4\% and 6.6\%, respectively.
However, in the Defects4J dataset, the techniques contributed a small decrease in recall by 0.6\% and 0.4\%, respectively, while the precision remained almost unchanged.
Further investigation revealed that this small drop in recall is mainly due to the \textit{Fine-Grained AST types} (FG) technique, which may also have a negative effect.
For example, if a variable (i.e., \texttt{SimpleName} AST type), which is part of an infix expression, gets wrapped into a method invocation as an argument within the same infix expression, the FG technique will block the mapping of this variable, since the variable's \textit{fine-grained} AST type originally was \texttt{InfixExpression-SimpleName} and then changed to \texttt{MethodInvocation-SimpleName}.
This finding shows that the FG technique can be potentially improved by allowing the mapping of leaf AST nodes with parents of different AST types in some certain scenarios, as the aforementioned one.
In conclusion, despite the potential accuracy improvements brought by these language-independent techniques, still none of the two GumTree variants can achieve the accuracy of RefactoringMiner. 
Moreover, our experiment shows that these language-independent techniques may also have a negative effect on accuracy under certain scenarios.
The language-aware features embodied in RefactoringMiner (i.e., the program element matching algorithm shown in Figure~\ref{fig:workflow}, the detection of inter-file move refactorings, the use of language clues, such as method calls and field references, to reason about and guide the matching process)
are more effective than the language-independent solutions that have been explored in the literature to overcome some of the limitations of the state-of-the-art AST diff tools.
\findings{RQ6 findings}{	
	All tools except for RefactoringMiner, have a quite similar F-score at statement level ranging between 93.5-94.3\%.
	RefactoringMiner brings significant accuracy improvements, especially in the Refactoring dataset and when considering sub-expression mappings, with +13\% in precision and +27\% in recall than the second-best performing tool.
	GumTree 3.0 simple has consistently better precision and recall than GumTree 3.0 greedy in both datasets, when considering sub-expression mappings.
	Based on their overall mapping accuracy, the tools are ranked as follows:
	\begin{enumerate}[leftmargin=*]
		\item \textbf{RefactoringMiner 3.0} (F-score=99.7\% at statement level and 99.5\% at sub-expression level)
		\item \textbf{MTDiff} (F-score=94.3\% at statement level)
		\item \textbf{GumTree 3.0 simple} (F-score=94.2\% at statement level and 92.6\% at sub-expression level)
		\item \textbf{GumTree 3.0 greedy} (F-score=94.2\% at statement level and 89.8\% at sub-expression level)
		\item \textbf{IJM} (F-score=93.9\% at statement level)
		\item \textbf{GumTree 2.1.0} (F-score=93.5\% at statement level)	
	\end{enumerate}
}

\subsection{RQ7: Perfect diff rate}
\subsubsection{Experimental procedure}
\vspace{-1mm}
For this research question, we investigate the ability of the tools to generate perfect diffs. A diff generated by a tool $T$ is considered perfect, if the set of AST node mappings in $S_T$ is identical to the benchmark mappings $S_B$ (i.e., $S_T = S_B$).
Therefore, \textit{perfect diff rate} is the percentage of commits for which a tool generated the perfect diff.
Since even small diff imperfections can cause confusion to the developer trying to understand the changes intended by the commit author, \textit{perfect diff rate} essentially measures the ability of the tools to generate \textit{noise-free} or \textit{confusion-free} diffs, as the perfect diff reflects the commit author intentions in the most accurate way.

\subsubsection{Results and discussion}
\begin{table}[ht]
	\caption{Perfect diff rate considering only statement mappings}
	\vspace{-4mm}
	\centering
	\fontsize{8pt}{10pt}\selectfont
	\setlength\tabcolsep{3px}
	\begin{tabular}{l | r l | r l | r l | r l | r l  | r l }
		\toprule
		\textbf{Dataset} &
		\multicolumn{2}{c}{\textbf{RM 3.0}} & 
		\multicolumn{2}{c}{\textbf{GT 3.0 greedy}} & 
		\multicolumn{2}{c}{\textbf{GT 3.0 simple}} &
		\multicolumn{2}{c}{\textbf{GT 2.1.0}} &
		\multicolumn{2}{c}{\textbf{IJM}} &
		\multicolumn{2}{c}{\textbf{MTDiff}} \\
		\toprule
		Defects4J &
		\pbar{715}{800} & \pbar{603}{800} & \pbar{579}{800} & \pbar{605}{800} & \pbar{605}{800} & \pbar{546}{800} \\
		Refactoring &
		\pbar{155}{188} & \pbar{28}{188} & \pbar{26}{188} & \pbar{21}{188} & \pbar{26}{188} & \pbar{11}{188} \\
		\bottomrule
		Overall &
		\pbar{870}{988} & \pbar{631}{988} & \pbar{605}{988} & \pbar{626}{988} & \pbar{631}{988} & \pbar{557}{988} \\
		\bottomrule
	\end{tabular}
	\label{table:perfect-diff-rate-2.1}
	\vspace{-5mm}
\end{table}

\begin{table}[ht]
	\vspace{-3mm}
	\caption{Perfect diff rate considering statement and sub-expression mappings}
	\vspace{-4mm}
	\centering
	\fontsize{8pt}{10pt}\selectfont
	\setlength\tabcolsep{3px}
	\begin{tabular}{l | r l | r l | r l }
		\toprule
		\textbf{Dataset} &
		\multicolumn{2}{c}{\textbf{RM 3.0}} & 
		\multicolumn{2}{c}{\textbf{GT 3.0 greedy}} & 
		\multicolumn{2}{c}{\textbf{GT 3.0 simple}} \\
		\toprule
		Defects4J &
		\pbar{687}{800} & \pbar{145}{800} & \pbar{506}{800} \\
		Refactoring &
		\pbar{132}{188} & \pbar{9}{188} & \pbar{16}{188} \\
		\bottomrule
		Overall &
		\pbar{819}{988} & \pbar{154}{988} & \pbar{522}{988} \\
		\bottomrule
	\end{tabular}
	\label{table:perfect-diff-rate-3.0}
	\vspace{-4mm}
\end{table}
Table~\ref{table:perfect-diff-rate-2.1} shows the perfect diff rate of the tools considering only statement mappings (i.e., coarse-grained level).
As we can observe from Table~\ref{table:perfect-diff-rate-2.1}, all tools are able to generate the perfect diff for a relatively large percentage of cases from the Defects4J dataset, with RefactoringMiner having the highest rate with 89.4\% and the remaining tools ranging between 68-75\%.
Regarding the Refactoring dataset, with the exception of RefactoringMiner that achieves a rate of 82.4\%, the rates of the other tools decrease dramatically and range between 6-15\%.
This result is another proof that bug fixing commits are a low barrier for assessing the accuracy of AST diff tools, and thus more challenging datasets should be used, such as the Refactoring benchmark we introduce in this paper.
Table~\ref{table:perfect-diff-rate-3.0} shows the perfect diff rate of the tools considering statement and sub-expression mappings (i.e., fine-grained level).
By comparing Table~\ref{table:perfect-diff-rate-3.0} with Table~\ref{table:perfect-diff-rate-2.1}, we can observe that all tools have a lower perfect diff rate.
This is reasonable as the tools need to correctly match all fine-grained sub-expression mappings to achieve the perfect diff.
However, GumTree 3.0 greedy is the tool that has by far the largest decrease in its perfect diff rate. This result is consistent with our findings in RQ4 and RQ6, showing that the greedy version of GumTree 3.0 very often forces incorrect mappings at the sub-expression level.
On the other hand, RefactoringMiner achieves a perfect diff rate of 70.2\% in the Refactoring dataset, and 85.9\% in the Defects4J dataset, which is a remarkable achievement, given the challenge of matching correctly all fine-grained sub-expression mappings.

\findings{RQ7 findings}{	
	RefactoringMiner has the highest perfect diff rate in both levels of granularity (i.e., with and without including sub-expression mappings).
	All tools, except for RefactoringMiner, exhibit a very low perfect diff rate for the Refactoring dataset ranging between 6-15\%.
	Based on their overall perfect diff rate, the tools are ranked as follows:
	\begin{enumerate}[leftmargin=*]
		\item \textbf{RefactoringMiner 3.0} (87.9\%)
		\item \textbf{IJM} (63.9\%)
		\item \textbf{GumTree 3.0 greedy} (63.9\%)
		\item \textbf{GumTree 2.1.0} (63.4\%)
		\item \textbf{GumTree 3.0 simple} (61.2\%)
		\item \textbf{MTDiff} (56.4\%)	
	\end{enumerate}
}
\subsection{RQ8: Execution time}
\subsubsection{Experimental procedure}
For this research question, we measure the execution time of each tool on each dataset in our benchmark.
We executed each tool separately on each commit of the Defects4J and Refactoring datasets, 
and recorded the time taken to generate the AST diff using the \texttt{System.nanoTime} Java method to get the start time and end time of the diff generation process for each commit.
All tools were executed on the same machine with the following specifications:
Apple M2 @ 3.49GHz, 16GB LPDDR5-6400 RAM, 512GB SSD, MacOS Sonoma 14.2, Java 17.0.6 x64.

\subsubsection{Results and discussion}
\begin{figure}[ht]
	\centering
	\vspace{-1.5mm}
	\includegraphics[width=\linewidth]{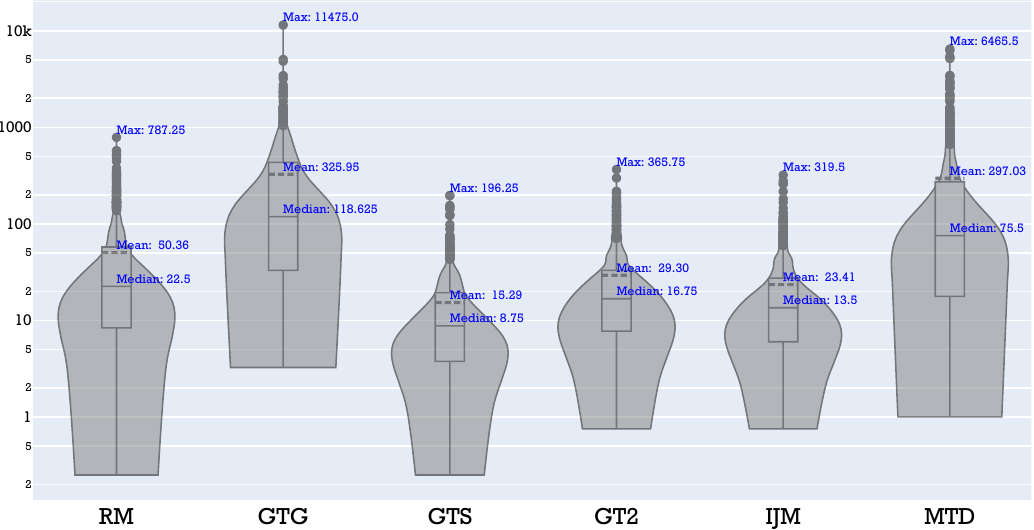}
	\vspace{-7mm}
	\caption{Execution time per Defects4J commit in milliseconds.}
	\label{fig:runtimes-d4j}
	\vspace{-3mm}
\end{figure}

\begin{figure}[ht]
	\centering
	\includegraphics[width=\linewidth]{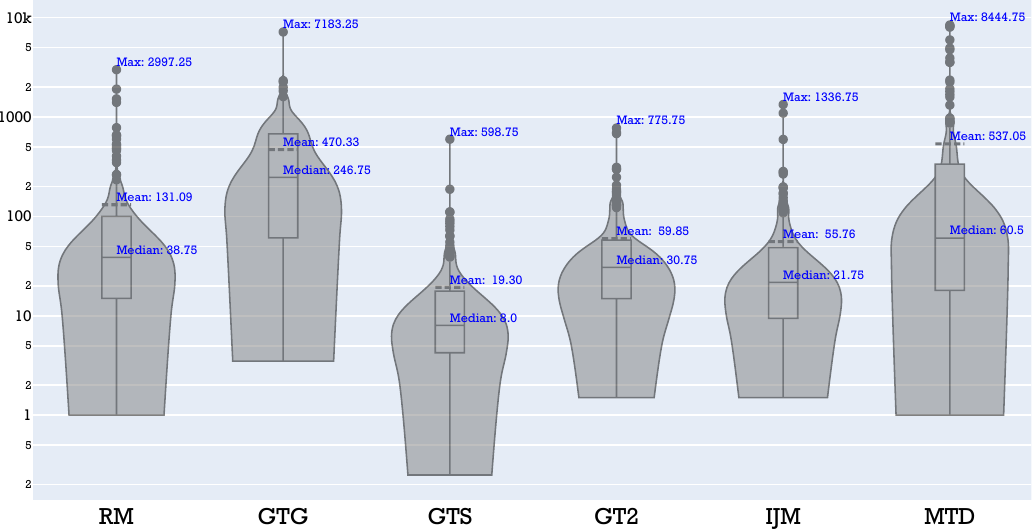}
	\vspace{-7mm}
	\caption{Execution time per Refactoring commit in milliseconds.}
	\label{fig:runtimes-ref}
	\vspace{-3mm}
\end{figure}

Figures \ref{fig:runtimes-d4j} and \ref{fig:runtimes-ref} show the execution time distribution of each tool on the Defects4J and Refactoring datasets, respectively.
Clearly, GumTree 3.0 simple is the fastest tool, followed by IJM and GumTree 2.1, which have a very similar performance, as IJM is an extension of GumTree 2.1.
RefactoringMiner comes in fourth place with an execution time that is 2.5 times slower on median and 3.3 times slower on average than GumTree (simple).
Still, it has execution times within the same order of magnitude as the three faster tools, and it can process each bug fixing commit in the Defects4J dataset in less than 800 ms.
MTDiff comes in the fifth place, and
GumTree greedy (i.e., the default algorithm in GumTree 3.0 configuration) has the slowest execution time among all tools, and is an order of magnitude slower than the GumTree simple (13.5 times slower on median and 21.3 times slower on average).

As explained before, the Refactoring benchmark includes inter-file mappings, and RefactoringMiner is the only tool supporting such mappings with the cost of additional processing.
Therefore, its execution time differences with the other tools are expected to be larger than those observed in the Defects4J dataset.
Still, we can observe from Figure \ref{fig:runtimes-ref} that the tools are ranked in the same order in terms of their execution time as in the Defects4J dataset,
but RefactoringMiner is now 4.8 times slower on median and 6.8 times slower on average than GumTree 3.0 simple.
We inspected the 5 commits in which RefactoringMiner has an execution time above 1 second, and they either involved inter-file mappings, or very large files (i.e., 15K LOC).

To create a ranking of the tools based on their execution time, for each pairwise combination of tools, we applied the Wilcoxon signed-rank test on their execution time distributions with the following null hypothesis:
``the execution time of tool \textit{X} is smaller than the execution time of tool \textit{Y}''.
We selected this test, because the execution time distributions are matched or dependent samples, meaning that we have execution times for the same commits in all distributions.
Table~\ref{table:execution-times} shows all 15 pair combinations of tools that the statistical test was applied for.
The rows of the table represent tool \textit{X}, while the columns represent tool \textit{Y} in the statistical test.
The symbol $<$ indicates that the null hypothesis was accepted with significance at 95 percent confidence level, and thus we can conclude that tool \textit{X} is faster than \textit{Y}.
\begin{table}[ht]
	\caption{Results of the Wilcoxon signed-rank test on the execution time distributions of the tools in both datasets} 
	\vspace{-4mm}
	\centering 
	\fontsize{8pt}{10pt}\selectfont
	\setlength\tabcolsep{3px}
		\begin{tabular}{l | c c c c c}
			\toprule
			\backslashbox{X}{Y} & \textbf{IJM} & \textbf{GumTree 2.1.0} & \textbf{RefactoringMiner} & \textbf{MTDiff} & \textbf{GumTree 3.0 greedy}\\ \midrule
			\textbf{GumTree 3.0 simple}  & $<$              & $<$            & $<$       & $<$    & $<$ \\
			\textbf{IJM}                 &                  & $<$            & $<$       & $<$    & $<$ \\
			\textbf{GumTree 2.1.0}       &                  &                & $<$       & $<$    & $<$ \\ 
			\textbf{RefactoringMiner}    &                  &                &           & $<$    & $<$ \\
			\textbf{MTDiff}              &                  &                &           &        & $<$ \\
			\bottomrule
		\end{tabular}
		\label{table:execution-times}
		\vspace{-2mm}
\end{table}

\findings{RQ8 findings}{
	RefactoringMiner is slower than GumTree 3.0 simple, GumTree 2.1 and IJM, but still its execution time is within the same order of magnitude.
	GumTree 3.0 simple is an order of magnitude faster than GumTree 3.0 greedy, and given it has also higher precision and recall, it should be preferred over GumTree greedy.
	Based on their execution times, the tools are ranked as follows:
	\begin{enumerate}[leftmargin=*]
		\item \textbf{GumTree 3.0 simple}
		\item \textbf{IJM}
		\item \textbf{GumTree 2.1.0}
		\item \textbf{RefactoringMiner 3.0}
		\item \textbf{MTDiff}
		\item \textbf{GumTree 3.0 greedy}
	\end{enumerate}
}

\subsection{RQ9: Accuracy comparison between RefactoringMiner 3.0 and 2.0}
\subsubsection{Experimental procedure}
As discussed in Section~\ref{sec:changes-to-improve-statement-mapping-accuracy}, we made some changes in the statement mapping process of RefactoringMiner 2.0~\cite{RefactoringMiner2}, which is the core of the tool.
With this research question, we want to investigate what is the effect of these changes on the accuracy of RefactoringMiner 3.0 over the prior version 2.0.
Since RefactoringMiner 2.0 does not have AST diff generation capabilities, we focused our comparison on features that both versions 2.0 and 3.0 support, namely the matching of program element declarations (i.e., type, method, field declarations) and statements.
Therefore, we utilized our benchmark at the appropriate level of granularity (i.e., \texttt{gumtree.gen.jdt-2.1.0}, see Section~\ref{sec:tools}) to compute the accuracy of versions 2.0 and 3.0 on program element declaration and statement mappings, multi-mappings, inter-file-move mappings, and refactoring-related mappings, respectively.
Moreover, we computed and compared the number of semantically incompatible mappings and perfect diff rate for versions 2.0 and 3.0.

\subsubsection{Results and discussion}
\begin{table}[ht]
	\vspace{-3mm}
	\caption{Precision and recall for statement mappings and program element declaration mappings}
	\vspace{-4mm}
	\centering
	\fontsize{8pt}{10pt}\selectfont
	\setlength\tabcolsep{2px}
	\begin{tabular}{l | l | r l r l | r l r l}
		\toprule
		& &
		\multicolumn{4}{c}{\textbf{RM 3.0}} & 
		\multicolumn{4}{c}{\textbf{RM 2.0}} \\
		& \textbf{Dataset} &
		\multicolumn{2}{c}{\textbf{Precision}} &
		\multicolumn{2}{c}{\textbf{Recall}} &
		\multicolumn{2}{c}{\textbf{Precision}} &
		\multicolumn{2}{c}{\textbf{Recall}} \\ \midrule
		\multirow{2}{*}{\textbf{Statement}}
		& Defects4J &
		\sbar{34689}{57}{145} & \sbar{30124}{296}{4710}  \\
		\multirow{2}{*}{\textbf{mappings}}
		& Refactoring &
		\sbar{14411}{38}{68} & \sbar{13244}{525}{1235}  \\
		& Overall &
		\sbar{49100}{95}{213} & \sbar{43368}{821}{5945}  \\
		\midrule
		\multirow{2}{*}{\textbf{Declaration}}
		& Defects4J &
		\sbar{2480}{0}{0} & \sbar{2434}{1}{46}  \\
		\multirow{2}{*}{\textbf{mappings}}
		& Refactoring &
		\sbar{1499}{1}{4} & \sbar{1409}{3}{94}  \\
		& Overall &
		\sbar{3979}{1}{4} & \sbar{3843}{4}{140}  \\
		\midrule
		\multirow{1}{*}{\textbf{Multi-mappings}}
		& Overall &
		\sbar{1410}{4}{23} & \sbar{1035}{253}{398} \\
		\midrule
		\multirow{1}{*}{\textbf{Refactoring mappings}}
		& Overall &
		\sbar{10670}{40}{67} & \sbar{9744}{639}{993} \\
		\midrule
		\multirow{1}{*}{\textbf{Inter-file mappings}}
		& Overall &
		\sbar{1438}{0}{6} & \sbar{1357}{6}{87} \\
		\midrule
		\multirow{1}{*}{\textbf{Perfect diff rate}}
		& Overall &
		& \pbar{870}{988} & & & \pbar{529}{988} & \\
		\bottomrule
	\end{tabular}
	\vspace{-2mm}
	\label{table:RM2-comparison}
\end{table}
As we can observe from Table~\ref{table:RM2-comparison}, RefactoringMiner 3.0 has a slightly higher precision and considerably higher recall in comparison to version 2.0 with respect to statement mappings.
The differences are more evident in the Refactoring dataset, where RefactoringMiner 3.0 has a +3.5\% precision and +8\% recall than version 2.0,
while in the Defects4J dataset version 3.0 has a +0.8\% precision and +13.1\% recall than version 2.0.
We performed a thorough investigation to understand why version 2.0 performs worse than version 3.0 in statement mappings, and we found four main reasons.
First, there are some program elements that contain statements, but are not processed by version 2.0.
Such program elements are initializer blocks and the initializers of field declarations and \texttt{enum} constants, which are used to initialize fields and are executed before class constructors. RefactoringMiner supported the processing of initializer blocks in version 2.3,
the diff of field declaration initializers in version 2.2, and the diff of \texttt{enum} constant initializers in version 2.3.2.
In particular, some commits from project JSoup in the Defects4J dataset have \texttt{enum} constants intialized with anonymous classes containing a large number of statements, which explains the lower recall observed in the Defects4J dataset compared to the Refactoring dataset.
Second, there are some refactoring types that contribute statement mappings, but are not supported by version 2.0.
For example, \textsc{Replace Anonymous with Class} is a refactoring that contains statement mappings between the methods of the anonymous class and the methods of the introduced class, and was supported in version 3.0.
Third, we found cases in version 2.0 diffs, where repeated statements within the body of a method were not accurately matched. The improved sorting criteria discussed in Section~\ref{sec:candidate-sorting} in cases of ties (i.e., when there are multiple possible matches for a given statement), reduced both false positives and negatives, thus contributing to improved precision and recall.
The fourth reason is related to the replacement algorithm used by RefactoringMiner to match non-identical statements.
As explained in Section~\ref{sec:background} (Phase 2) and in \cite{RefactoringMiner2}, RefactoringMiner applies
syntactically valid AST node replacements within a pair of non-identical input statements, and matches these statements only if they become textually identical after the application of replacements.
This design choice avoids the use of arbitrary code similarity thresholds to match non-identical statements.
Throughout the evolution of the tool from version 2.0 to 3.0, the kinds of supported replacements increased, thus enabling the matching of statement pairs that could not be matched before, as the replacement algorithm could not make them textually identical.
For example, version 2.3 supports the replacement of variables with boolean literals and the replacement of variables with class instance creations, while these kinds of replacements were not supported in previous versions of the tool.
In conclusion, the first, second and fourth reasons mainly affect the recall of the tool, while the third reason affects both precision and recall.

Regarding program element declaration mapping accuracy, we can observe from  Table~\ref{table:RM2-comparison} that the precision of versions 2.0 and 3.0 is very close, as version 3.0 has a +0.2\% precision in the Refactoring dataset, and a +0.1\% precision in the Defects4J dataset.
However, the differences in recall are larger, especially in the Refactoring dataset, where version 3.0 has a +6\% recall than version 2.0.
Again, we performed a thorough investigation to explain the observed differences in recall, and we found two main reasons.
The first reason is that version 2.0 is not processing the initializers of field declarations and \texttt{enum} constants to find differences, as already explained in the previous paragraph. 
However, in some cases field declarations and \texttt{enum} constants are initialized with an anonymous class, and thus version 2.0 is missing to match the method declarations within the anonymous classes in the field/enum-constant initializers. 
The second reason is related to an inherent limitation of version 2.0, which is not able to match method declarations with empty bodies, as it requires to find at least one statement mapping in order to match a pair of methods.
Developers commonly instantiate interfaces with anonymous classes (e.g., when implementing listeners for various events) and leave some methods with empty implementation.
Version 2.0 is unable to match such methods with an empty body within anonymous class declarations.
This issue was fixed in version 3.0 of RefactoringMiner by matching the empty methods within anonymous classes based on their signature similarity.
Both of these reasons affect the recall of the tool in matching method declarations within anonymous classes.

Regarding multi-mapping accuracy, we can observe from  Table~\ref{table:RM2-comparison} that version 2.0 has a considerably lower precision (-19.3\%) and recall (-26.2\%) than version 3.0.
The lower recall in multi-mappings is an expected finding, as version 2.0 does not support several refactorings that contribute multi-mappings, such as the \textsc{Consolidate Duplicate Conditional Fragments} refactoring (first supported in version 3.0, see Section ~\ref{sec:multi-mapping support}) and the \textsc{Merge/Split Method}, \textsc{Parameterize Test} and \textsc{Merge Catch} refactorings (supported in version 2.4, see Table~\ref{table:refactorings}). 
Version 2.0 essentially supports the multi-mappings established from \textsc{Extract Method} refactorings, where duplicated code fragments from the same or different methods are merged into a single method, and \textsc{Inline Method} refactorings, where a single code fragment is duplicated in all call sites of the inlined method.
The interesting finding is that many of these missed multi-mappings contribute false positive mappings (i.e., decrease precision), as they tend to be mismatched with other irrelevant statements. These findings show that supporting refactoring types eliminating or introducing duplicate code is essential for improving both precision and recall in statement mappings.

Regarding refactoring-related mappings, as shown in Table~\ref{table:RM2-comparison}, version 2.0 has a lower recall (-8.6\%) and precision (-5.8\%) than version 3.0.
This can be attributed to two main reasons. Version 2.0 supports considerably less refactoring types (40 in total) compared to version 3.0, which supports over 100 refactoring types, API changes and language migrations.
These unsupported refactoring types contribute negatively in the recall of version 2.0.
Moreover, the overall improvement in statement mapping accuracy achieved in version 3.0, also improved the precision and recall in refactoring-related statement mappings, and thus the overall refactoring detection accuracy.

Regarding inter-file mappings, as shown in Table~\ref{table:RM2-comparison}, version 2.0 has a slightly lower recall (-5.6\%) than version 3.0, while its precision is very similar to that of version 3.0. 
We performed a thorough investigation to explain the observed difference in recall, and we found two main reasons behind the false negatives.
The first reason is that version 2.0 fails to detect some moved and renamed attributes, while version 3.0 is utilizing the field references found in statement mappings within moved methods to infer \textsc{Move and Rename Attribute} refactorings.
The second reason is that version 2.0 is not processing the pairs of anonymous class declarations within moved methods, and thus the methods declared in the anonymous classes are not getting matched.

In terms of semantically incompatible mappings, version 2.0 has six such mappings, while version 3.0 has zero. Four of these mappings were between anonymous class method body blocks and composite statement body blocks, while two mappings were incompatible type mappings due to incorrectly matched variable declaration statements. Still compared to the other AST diff tools that were evaluated in Section~\ref{sec:RQ2}, RefactoringMiner 2.0 has a considerably smaller number of semantically incompatible mappings, which shows that the core algorithm driving the matching process (Figure~\ref{fig:workflow}) was already very effective in generating semantically compatible mappings since version 2.0.

Finally, version 2.0 has a much lower \textit{perfect diff rate} (-35\%) compared to version 3.0, and actually has the lowest \textit{perfect diff rate} among all tools. This finding is rather expected, as RefactoringMiner 2.0 is not processing some parts of the AST (e.g., initializer blocks, field declaration and enum constant initializers, and import declarations), and thus in many cases it is not able to generate a complete diff covering the entire ASTs.
\findings{RQ9 findings}{
	Our analysis has shown that RefactoringMiner 2.0 was not processing parts of the AST, as its matching process is fully guided based on language-dependent features and clues. In other words,  RefactoringMiner is not trying to randomly match code fragments between the left and right ASTs based on their similarity, but it is utilizing the program structure, the program element declaration types, and language clues (e.g., method calls, method signatures, field references) to reason about and guide the matching process on specific parts of the ASTs.
	In version 3.0, we enhanced the aforementioned reasoning/guiding algorithm to process parts of the AST that were previously not handled (e.g., initializer blocks, field declaration initializers), thus improving the recall of the tool by 8-13\% in statement mappings and by 2-6\% in program element declaration mappings.
	Moreover, the changes in the statement mapping process of RefactoringMiner discussed in Section~\ref{sec:changes-to-improve-statement-mapping-accuracy} improved both precision and recall of the tool, as RefactoringMiner 3.0 makes less mistakes when breaking ties for statements with multiple possible matches.
	Moreover, in version 3.0, we supported several refactoring types that contribute multi-mappings, which further improved the statement mapping accuracy of the tool.
}

\subsection{Discussion}
In this section, we provide insights based on our experiment findings.

\header{How to eliminate semantically incompatible mappings} The fact that AST diff tools can match semantically incompatible AST nodes is something already known in the literature.
As a matter of fact, IJM~\cite{IJM} attempted to overcome the semantic violations related to name identifiers by modifying the structure of AST (i.e, by merging \texttt{SimpleName} leaf AST nodes with their parent nodes). However, as discussed in RQ1, this change in the AST structure made the depth of the AST shorter, and affected negatively the accuracy of IJM on matching statements of height equal to 1 (i.e., method invocation statements with a merged name node), as
GumTree employs a default threshold for the hyperparameter $minHeight$ (i.e., length of the longest path from one leaf to the root of the subtree), which is equal to 2.

Based on our findings, changing the AST structure is not the correct approach to eliminate semantically incompatible AST node mappings.
A better approach is to introduce the semantic role of the AST nodes in the program (e.g., method call name, variable name, type name) as a node property, and match two nodes of the same AST type, only if they have the same or a compatible semantic role property. Alternatively, semantic compatibility can be forced for AST nodes playing multiple semantic roles (e.g., \texttt{SimpleName}) by checking whether their direct parents have the same AST type.

\header{Greed is bad} GumTree greedy and MTDiff represent algorithms trying to maximize the number of generated AST node mappings.
GumTree greedy applies a non-conservative approach that matches AST subtrees of height equal to 1 in the bottom-up phase of the algorithm.
MTDiff uses 5 optimizations to generate additional \textit{move} edit operations.
Based on our findings, GumTree greedy has consistently lower accuracy than its non-greedy counterpart (i.e., GumTree simple) in all examined research questions.
More specifically, when compared to the simple version, GumTree greedy has lower accuracy in multi-mappings (RQ1), more semantically incompatible mappings (RQ2), lower accuracy in program element mappings (RQ3), lower accuracy in refactoring-related mappings (RQ4), more false positives in inter-file mappings (RQ5), lower overall accuracy (RQ6), and much lower perfect diff rate when considering sub-expression mappings (RQ7).
The same findings hold when comparing MTDiff with its less-greedy counterpart, which is GumTree 2.1.0.
More specifically, when compared to GumTree 2.1.0, MTDiff has lower accuracy in multi-mappings (RQ1), more semantically incompatible mappings (RQ2), lower accuracy in program element mappings (RQ3), more false positives in inter-file mappings (RQ5), and lower perfect diff rate (RQ7).
In addition, their greedy nature has considerable runtime cost, as GumTree greedy and MTDiff are an order of magnitude slower than all other AST diff tools in both Defects4J and Refactoring datasets (RQ8).

GumTree 3.0 is a very established tool, widely used by researchers and practitioners in downstream tasks.
In GumTree 3.0 released version, the default algorithm is the greedy one.
We strongly recommend to the GumTree development team to change the default algorithm to the simple version, as the majority of researchers and practitioners are using this tool with its default configuration.

\header{Partial matching is a good idea}
Partial matching is essentially a divide-and-conquer approach introduced by IJM to divide the AST into different sections of declarations based on their kind (e.g., import declarations, method declarations, field declarations), and then apply the tree matching process separately on each section of declarations.
Based on the discussion of RQ3 results, this approach improved significantly the program element mapping accuracy for method declarations compared to GumTree 2.1.0 (i.e., the tool extended by IJM).
However, we also found that partial matching was not so effective for field declarations, as it fails to match slightly modified field declarations moved to inner classes within the same file.
Thus, IJM requires an additional post-processing step to match field declarations between different classes declared within the same file.

Although partial matching requires some basic language-awareness as input to the algorithm (i.e., the main program declarations in each programming language), as well as some criteria to match declarations of the same kind (i.e., signature-based matching), it is a very effective approach to narrow-down the matching process to different sections of the AST instead of the whole AST.


\header{Language-awareness is helpful to guide the matching process}
GumTree and all tools based on it, apply a language-independent AST node matching algorithm,
which  tries to ``blindly'' find the largest identical (or similar) AST sub-trees in a pair of files without relying on language clues to guide the matching process.
The premise is that being language-independent, it is possible to devise a solution
that can support any programming language (i.e., silver bullet).
However, based on our findings, being 100\% language-independent may negatively affect the accuracy of the generated AST node mappings, due to the ``blind'' matching process.
Programming language clues, such as method calls, field accesses, and refactoring transformations can guide the matching process to more accurate mappings.
For example, when in the place of a deleted piece of code, there is a call to a newly added method in a file, it is very likely that the deleted piece of code has been moved to this newly added method (i.e., \textsc{Extract Method} refactoring).
Therefore, instead of ``blindly'' trying to match the deleted piece of code within the entire file, the matching process can focus on the code within the body of this newly added method. This approach can avoid false mappings with newly added code in other parts of the AST that is coincidentally similar to the deleted code, and guide the matching process towards the correct mappings.

Obviously, computing bindings between references (i.e., method calls and field accesses) and their corresponding declarations (i.e., method and field declarations), as well as detecting refactoring operations, are rather challenging tasks for some programming languages, especially at commit (or pull request) level, where we have a small portion of the project's code available for analysis and it is not possible to build the code and get these bindings from the compiler.
If we are able to overcome these challenges, then the design and development of \textit{hybrid} AST diff tools that combine language-independent and language-specific features seems a very promising direction.

\header{We need more challenging benchmarks}
Based on our analysis of RQ6 and RQ7 results, all examined tools achieve a higher accuracy in the Defects4J dataset compared to the Refactoring dataset.
Moreover, the differences in their accuracy are rather small in the Defects4J dataset compared to the Refactoring dataset.
These results show that bug-fixing commits, which typically include small (a few lines of code), local (typically located within a single method), and rather simple changes,
are not suitable for assessing the accuracy of AST diff tools, as they offer AST matching problem instances of rather low complexity and challenge.
However, the majority of AST diff tools have been evaluated in bug fixing datasets, which makes the evaluation findings unreliable and hard to interpret.
Moreover, these evaluations were not objective, as they were based on the number of generated mappings and the
size of the resulting edit script, without validating whether the mappings and edit actions are correct and reflect the actual changes intended by the commit author. 

We call the AST differencing community to follow our lead and contribute new and more challenging benchmarks, as these can help advance the capabilities of our tools, and reveal limitations that have not been discovered yet. To this end, we have developed a benchmarking infrastructure and tooling \cite{DiffBenchmark} that can automate parts of this labour-intensive process.

\subsection{Limitations}
To provide an overview of the limitations of our work, we examined the AST diffs for 15 commits in which our tool has some of its lowest accuracy values, despite having higher accuracy than the other tools.
We found four commonly occurring issues:
\begin{enumerate}[leftmargin=*]
	\item Our tool may generate inaccurate mappings in test code, because unit tests share a lot of identical or similar statements (e.g., assertions, test object creations), and thus there is more ambiguity in the matching process, as a given statement may have multiple candidate matching statements.
	Moreover, developers apply some test-specific refactorings in test code~\cite{9769994}, which are not supported by RefactoringMiner yet.
	As this is a common issue that all AST diff tools exhibit to some extent, the AST diff community should develop special techniques for matching test code, and create benchmarks with commits including test files to evaluate the accuracy of the tools on test changes.
	
	\item The algorithm described in Section~\ref{sec:multi-mapping support}
	for generating multi-mappings when duplicated code is moved out of or moved into conditionals does not cover scenarios that follow a different nesting pattern than the expected one. For example, in a commit from project apache/hbase~\cite{hbase} the statements in lines L164-168 are duplicated 3 times in R152-159, R163-169 and R174-181, but the conditional statements do not form an \texttt{if-else-if} chain as expected, but are sequential \texttt{if} statements. Thus, our algorithm can be improved to cover more scenarios.
	
	\item 
	Despite the improvements described in Section~\ref{sec:leaf-mapping-sorting-criteria} for handling ties in leaf statement mappings, we still observe some false mappings for \textit{jump} statements (e.g., \texttt{break}, \texttt{continue}, \texttt{return}), especially when the control flow of the program is heavily restructured, as in the case of a commit from project Nutz~\cite{Nutz}.
	
	\item
	A limitation that is inherited by GumTree simple (used in Algorithm~\ref{algo:SubTreeMatcher}) is its inability to match string literals with small variations (i.e., only identical string literals can be matched). However, greedy algorithms, such as GumTree (greedy) and MTDiff can overcome this limitation with the risk of having more false mappings. For pairs of statements where we have high confidence that they correspond to a correct mapping, we could utilize algorithms that can match AST nodes with different values, especially for string literals with small variations.
\end{enumerate}

Arguably, someone could claim that the main limitation of this work is \textit{language specificity}, as it builds upon RefactoringMiner (a tool supporting Java programs) to obtain the statement mappings.
Our counter-argument and main point of this work is that AST diff should not be treated in a language-agnostic manner, as language-awareness provides very useful information for improving the AST mapping accuracy, thus leading to a diff output that reflects more accurately the changes intended by the commit author.
Through illustrative examples and experiment results, we have shown that by utilizing \textit{language clues} (i.e., method calls and signatures, field references), the \textit{semantic role} of AST nodes in the program, \textit{refactoring} information, and by supporting \textit{multi-mappings} and \textit{inter-file} mappings, we can achieve considerable improvements in accuracy.

\subsection{Threats to Validity}
\header{External validity} The benchmark used for our experiments includes nearly 1,000 cases (800 bug fixes from 17 different projects and 188 refactoring commits from 105 different projects). This dataset ensures a certain level of diversity with respect to the domain of the projects and the characteristics of the changes (i.e., refactorings vs. bug fixes).
\header{Internal validity} Although we did our best effort to reduce bias in the construction of our benchmark by incorporating the input of six different AST diff tools, we cannot claim the oracle is completely unbiased, as all validators are co-authors of this work.
The vast majority of the AST mappings in the benchmark were straightforward to validate using well-defined diff assessment criteria (Section \ref{sec:diff-assessment-criteria}).
Only around 3\% of the examined cases included debatable mappings, which were resolved after extensive discussions between the two authors of the paper, and by asking the opinion of our lab members, when the co-authors were not able to reach a consensus.
\header{Verifiability}
We make the source code of RefactoringMiner \cite{RefactoringMiner} and our benchmark \cite{DiffBenchmark} publicly
available to enable the replication of our experiments and
facilitate future research on AST diff techniques.
\section{Related Work}
\label{sec:related-work}
The most significant and influential work on AST differencing is GumTree by Falleri et al. \cite{gumtree}.
GumTree is a language-independent abstract syntax tree matcher, and generates the AST node mappings between two trees in two phases.
In the first (top-down) phase, it detects the largest identical subtrees.
In order to find the largest identical subtrees, GumTree must compare all sub-tree combinations within two files.
Due to the recursive nature of trees, the runtime complexity is prone to grow very fast. To achieve a better runtime, GumTree exploits a hashing technique. There are two hash values associated with every tree. The first one is calculated by considering the label and the value of the tree and the second one is calculated based on the label only (\textit{structure hash}). 
The structure hash is useful to find structurally isomorphic trees excluding the actual node values. 
Hash values for non-leaf nodes are dependent on the values of their children. 
In the first phase, both labels and values are considered, 
as GumTree aims to find the largest identical subtrees.
GumTree starts with the highest trees available in both files and performs a comparison by checking their hash values. This comparison goes step by step deeper into the inner nodes, until a minimum height equal to 2 is reached.
The tuning of the minimum height parameter is essential to avoid matching small trees that are leaves, but have identical values.
The second (bottom-up) phase aims to match the trees that are not matched previously, but a fair amount of their children are matched.
More specifically, two trees are matched if the ratio of their common descendants is greater or equal to 0.5.

Iterative Java Matcher (IJM) \cite{IJM} is built upon GumTree and aims to improve the quality of the generated \textit{move} and \textit{update} actions, specifically for Java projects.
The first enhancement (\textit{partial matching}) restricts the scope of matching to selected parts of the source code to decrease the chance of mismatches. 
Instead of applying the GumTree algorithm on the entire Java files, IJM provides partial input to GumTree by pruning specific parts of the Java file, such as the import statements, type declarations, enumeration declarations, method declarations, and field declarations.
The second enhancement (\textit{merging name nodes}) addresses another limitation of GumTree, which is prone to match nodes whose parents correspond to different AST types. Since this limitation mostly occurs for the \texttt{SimpleName} AST type, IJM combines \texttt{SimpleName} nodes with their parent nodes to prevent inaccurate mappings. The last enhancement (\textit{name-aware matching}) improves the bottom-up phase of GumTree by considering the string similarity of the names of the nodes in addition to their node types.

Move-optimized Tree Differencing (MTDiff) \cite{MTDiff} is based on ChangeDistiller \cite{ChangeDistiller} and uses 5  optimizations, namely \textit{identical subtree}, \textit{longest common subsequence}, \textit{unmapped leaves}, \textit{inner node}, and \textit{leaf move}, to shorten the resulting edit scripts.
When applied to Gumtree, RTED, JSync, and ChangeDistiller, they led to shorter scripts for 18-98\% of the changes in the histories of 9 open-source software repositories.

CLDiff \cite{ClDiff} uses GumTree (without any changes in its implementation) to obtain AST mappings and edit actions, and generates concise code differences by grouping fine-grained code differences at or above the statement level.
Finally, it links the related concise code differences according to five predefined links, namely \textit{Def-Use} linking changes in declarations with changes in their references, \textit{Abstract-Method} linking changes in abstract methods with changes in their sub-class implementations, \textit{Override-Method} linking changes in overridden methods with changes in their overriding methods, \textit{Implement-Method} linking changes in interface methods with changes in their implementing methods, and \textit{Systematic-Change} linking similar changes caused by systematic edits (e.g., refactoring).
\section{Conclusions}
In this work, we presented a novel AST diff tool based on RefactoringMiner \cite{RefactoringMiner2} that resolves all limitations of the current tools:
(1) it supports multi-mappings, 
(2) it avoids matching semantically incompatible AST nodes, 
(3) it uses language clues to guide the matching process, 
(4) it is refactoring-aware, 
(5) it supports diff at the commit level.
To evaluate our tool and compare it with the state-of-the-art tools, we created the first benchmark of AST node mappings, including 800 bug-fixing commits from Defects4J \cite{Defects4J} and 188 refactoring commits from the Refactoring Oracle \cite{RefactoringMiner2}, following a very meticulous and well-defined process.
Our evaluation showed that our tool achieved a considerably higher precision and recall, especially for refactoring commits, it does not generate semantically incompatible AST node mappings, and has an execution time that is comparable with that of the faster tools.
More specifically, our tool has the best performance in 7 out of 8 diff quality dimensions covered by our evaluation framework, including multi-mapping accuracy (RQ1), semantically compatible mappings (RQ2), program element mapping accuracy (RQ3), refactoring-related mapping accuracy (RQ4), inter-file mapping accuracy (RQ5), overall mapping accuracy (RQ6), and perfect diff rate (RQ7).
The only category our tool does not have the best performance is execution time (RQ8), where it ranks fourth out of 6 tools.

\header{Implications for researchers}
The accurate AST node mappings generated by RefactoringMiner can improve the performance of several downstream tasks that rely on AST diff information.
For instance, automated program repair \cite{10.1007/s10664-019-09780-z, 10.1007/s10664-021-10087-1, 10.5555/2486788.2486855, BugBuilder} and library/API migration \cite{9079197, 10.1145/3387905.3388608} techniques mine the change history of open-source projects to extract bug-fixing patches and API usage migration scenarios, respectively, based on AST edit scripts. The AST diff accuracy affects the quality of the extracted change patterns.
Finally, our benchmark of AST node mappings can be used to evaluate the accuracy of novel AST diff tools.

\header{Implications for practitioners}
The exceptional accuracy of RefactoringMiner's AST diff can enable fine-grained and refactoring-aware code evolution analysis \cite{CodeTracker}, such as tracking a specific statement (e.g., code block) in the commit change history of a project, and finding the commits it changed.
Codoban et al. \cite{7332446} surveyed 217 developers and found that 85\% of them consider software history important to their development activities and 61\% need to refer to history at least several times a day.
An interesting direction is to develop an intelligent AST diff visualization UI based on the output of RefactoringMiner, which can overlay semantic and refactoring information on top of the changes \cite{spike, ClDiff}.
Software developers can use such an intelligent change visualization tool to conduct their code reviews, saving considerable effort and time in understanding the changes,
and thus improving their overall productivity.

\section{Future Research}
We believe that a diff reflecting with high accuracy the changes intended by the commit author is an important foundation upon which intelligent code review assistants could be built, as having accurate mappings between two versions of code establishes what portion of the code remained the same (even if it was refactored) and what portion of the code is truly deleted or added.
However, there are still missing components to enable intelligent assistants.
The first component is navigation assistance.
\begin{itemize}[leftmargin=*]
\item In what order and granularity should we navigate the code reviewer over the changes to help her understand better the ``story'' and motivation behind the changes?
\item Should we start from behavior-preserving changes (i.e., refactorings) to give a higher-level view of the design changes intended by the commit author, and then proceed with non-behavior-preserving changes to give more insights about the new features or bug fixes intended by the commit author?
\item How should we deal with overlapping behavior-preserving and non-behavior-preserving changes?
\item Should we group relevant changes and present them as one change to the code reviewer?
\end{itemize}
The second component is explanation assistance.
Ideally, the code reviewer would like to know why a particular change happened.
Some of these explanations can come directly from diff-related information.
For example, multi-mappings can be associated with the intention of eliminating duplicated code.
Inversion, splitting and merging of conditional statements can be associated with the intention of re-structuring the control-flow of the program.
However, for changes that cannot be explained by refactorings, and especially for changes implementing a new functionality or fixing a bug, the explanation is not straightforward.
For such changes we could utilize commit message information, linked issues in the commit message, or discussions from a Pull Request involving the commit.
If the aforementioned information is not helpful, we could utilize Large Language Models (LLMs) to provide an explanation.
This would require to investigate what is the best way to provide the required input and context to the LLM, what is the best approach to prompt the LLM, and how should we deal with known LLM limitations, such as hallucination and limited ability to control and reproduce the model's responses.

Another research direction is to extend RefactoringMiner, so that it is able to support more programming languages.
RefactoringMiner's core statement mapping algorithm (Figure~\ref{fig:overview}) is mostly language-independent, as it uses a string replacement algorithm to match statements that are textually different, and a set of heuristics to match method calls and object creations with differences in their arguments. For example, it already abstracts method invocations, object creations, and method references into objects of the same class, namely \texttt{AbstractCall}, and thus similar constructs from other programming languages could be represented by the same abstraction.
\begin{acks}
The authors would like to thank the members of the Refactoring research group at Concordia University, Victor Veloso, Tayeeb Hasan, Pedram Nouri, Palash Borhan Uddin, Diptopol Dam, and Mosabbir Khan Shiblu, for providing their opinion on debatable AST mappings whenever needed.
This research was partially supported by NSERC grant RGPIN-2018-05095.
\end{acks}

\bibliographystyle{ACM-Reference-Format}
\bibliography{sample-base}

\appendix

\end{document}